\documentclass[12pt]{article}
\usepackage{amsmath}
\usepackage{amssymb}
\usepackage{amsfonts}
\newcommand{\pgh}{\mathrm{pgh}}
\newcommand{\agh}{\mathrm{agh}}
\newcommand{\gh}{\mathrm{gh}}
\allowdisplaybreaks

\begin{document}

\author{Constantin Bizdadea\thanks{%
E-mail: bizdadea@central.ucv.ro}, Solange-Odile Saliu\thanks{%
E-mail: osaliu@central.ucv.ro}\\
Department of Physics, University of Craiova\\
13 Al. I. Cuza Str., Craiova 200585, Romania}
\title{Gauge-invariant massive BF models}
\date{}
\maketitle

\begin{abstract}
Consistent interactions that can be added to a free, Abelian gauge theory comprising a BF model and a finite set of massless real scalar fields are constructed from the deformation of the solution to the master equation based on specific cohomological techniques. Under the hypotheses of analyticity in the coupling constant, Lorentz covariance, spacetime locality, Poincar\'{e} invariance, supplemented with the requirement on the preservation of the number of derivatives on each field with respect to the free theory, we obtain that the deformation procedure leads to two classes of gauge-invariant interacting theories with a mass term for the BF vector field $A_{\mu }$ with $U(1)$ gauge invariance. In order to derive this result we have not used the Higgs mechanism based on spontaneous symmetry breaking.

\end{abstract}

\section{Introduction\label{intro}}

Topological BF theories~\cite{birmingham91} are important due to the fact that some interacting, non-Abelian versions are related to a Poisson structure algebra~\cite{stroblspec} characteristic to Poisson sigma models, which, in turn, are useful tools at the study of two-dimensional gravity. It is well known that pure gravity in $D=3$ is just a BF theory. Moreover, higher-dimensional General Relativity and  Supergravity in Ashtekar formalism may also be formulated as topological BF theories in the presence of some extra constraints~\cite{ezawa,freidel,smolin,ling}. In view of these results, it is relevant to construct the self-interactions in BF theories~\cite{bf1,bf2,bf3} as well as the couplings between BF models and other gauge or matter theories~\cite{bf4,bf5,bf6,bf7}.

The aim of this paper is to investigate the consistent interactions in four spacetime dimensions between an Abelian BF theory and a set of massless real scalar fields by means of the deformation of the solution to the master equation~\cite{PLB1993,CM1998} with the help of local BRST cohomology~\cite{CMP1995a,CMP1995b,PR2000}. The field sector of the four-dimensional BF model consists in one scalar field $\varphi $, two vector fields $\{A_{\mu }$, $H^{\mu }\}$, and a two-form $B^{\mu \nu }$. We work under the hypotheses of analyticity in the coupling constant, Lorentz covariance, spacetime locality, Poincar\'{e} invariance, supplemented with the requirement on the preservation of the number of derivatives on each field with respect to the free theory. As a consequence of our procedure, we are led to two classes of gauge invariant interacting theories with a mass term for the BF vector field $A_{\mu }$ with $U(1)$ gauge invariance. The derivation of the above classes of gauge invariant massive theories represents the main result of this paper. We remark that in order to derive the previously mentioned massive models we have not used in any way the Higgs mechanism based on spontaneous symmetry breaking~\cite{Englert,Higgs1,Higgs2,Kibble}. Thus, our main result reveals a novel mass generation mechanism that deserves to be further investigated with respect to a collection of Maxwell vector fields and a set of real massless scalar fields.

Our strategy goes as follows. Initially, we determine in Section \ref{model} the antifield-BRST symmetry of the free model, which splits as the sum between the Koszul--Tate differential and the longitudinal exterior derivative, $s=\delta +\gamma $. In Section \ref{deform} we briefly present the reformulation of the problem of constructing consistent interactions in gauge field theories in terms of the deformation of the solution to the master equation. Next, in Section \ref{first} we determine the first-order deformation of the solution to the master equation for the model under consideration. The first-order deformation belongs to the local cohomology $H^{0}(s|d)$, where $d$ is the exterior spacetime derivative. We find that the first-order deformation is parameterized by five types of smooth functions of the undifferentiated scalar fields form the theory. Section \ref{high} is devoted to the investigation of higher-order deformations. The consistency of the first-order deformation restricts the above  mentioned functions to fulfill two kinds of equations (consistency equations). Based on these equations we prove that the higher-order deformations can be taken to stop at order three in the coupling constant. The identification of the interacting model is developed in Section \ref{inter}. Initially, we infer the general form of the Lagrangian action and its gauge symmetries. Next, we emphasize two types of solutions to the consistency equations, which lead to the previously mentioned gauge invariant massive theories. Section \ref{concl} closes the paper with the main conclusions and some comments. The present paper contains also two appendices, in which the concrete form of the first-order deformation used in the main body of the paper as well as some formulas concerning the gauge structure of the interacting model are derived.

\section{Starting model\label{model}}

We start from a free model in $D=4$ spacetime dimensions describing a topological BF theory with a maximal field spectrum (a scalar $\varphi$, two sorts of vector fields denoted by $\{A_{\mu}, H^{\mu}\}$, and a two-form $B^{\mu\nu}$) plus a finite set of massless real scalar fields $\phi ^{A}$ ($A=\overline{1,N}$). For notational ease we designate the entire collection of massless real scalar fields by $\phi $ and the full field spectrum by $\Phi^{\alpha_{0}}$
\begin{equation}
\phi \equiv \big\{\phi ^{A}\big\}_{A=\overline{1,N}}, \qquad \Phi^{\alpha_{0}} \equiv \{ \varphi ,A_{\mu},H^{\mu},B^{\mu\nu},\phi \}. \label{1}
\end{equation}
The Lagrangian action underlying this model reads as
\begin{align}
S^{\mathrm{L}} [\Phi^{\alpha_{0}}]&=\int d^4x \Big[ H^{\mu} \partial_{\mu }\varphi + \tfrac{1}{2} B^{\mu\nu} \partial_{[\mu } A_{\nu ]} + \tfrac{1}{2} k_{AB} \big( \partial_{\mu } \phi ^{A} \big) \big( \partial ^{\mu } \phi ^{B} \big) \Big] \nonumber \\
&\equiv S^{\mathrm{L,BF}} [\varphi , A _{\mu}, H ^{\mu}, B ^{\mu\nu}] + S^{\mathrm{L,scalar}} [\phi]. \label{2}
\end{align}
We work with a mostly negative metric in a Minkowski spacetime of dimension $D=4$, $\sigma ^{\mu \nu }=\sigma _{\mu \nu }=( +---)$ and a metric tensor $k _{AB}$ with respect to the matter field indices (constant, symmetric, invertible, and positively defined), $\phi _{A}= k _{AB} \phi ^{B}$. In this context, the elements of its inverse will be symbolized by $k ^{AB}$. Everywhere in this paper the notation $[\mu \ldots \nu ]$ signifies complete antisymmetry with respect to the (Lorentz) indices between brackets, with the conventions that the minimum number of terms is always used and the result is never divided by the number of terms. For instance, the expression $\partial_{[\mu }A_{\nu ]}$ from action (\ref{2}) means $\partial_{\mu }A_{\nu }-\partial_{\nu }A_{\mu }$.

The BF action is invariant under the nontrivial (infinitesimal) gauge transformations
\begin{align}
\delta _{\Omega^{\alpha_{1}}} \varphi &=0, & \delta _{\Omega^{\alpha_{1}}} A_{\mu} &= \partial_{\mu } \epsilon, \label{3} \\
\delta _{\Omega^{\alpha_{1}}} H^{\mu} &= -2\partial _{\lambda} \xi^{\lambda\mu}, & \delta _{\Omega^{\alpha_{1}}} B^{\mu\nu} &= -3 \partial _{\lambda} \epsilon^{\lambda\mu\nu}, \label{4}
\end{align}
while the action of the matter fields possesses no nontrivial gauge symmetries of its own
\begin{equation}
\delta _{\Omega^{\alpha_{1}}} \phi ^{A} =0 ,\qquad A=\overline{1,N}, \label{5}
\end{equation}
such that (\ref{3})--(\ref{5}) actually represent a generating set of (infinitesimal) gauge transformations with respect to the overall free action (\ref{2}). The notation $\Omega^{\alpha_{1}}$ collects all the gauge parameters
\begin{equation}
\Omega^{\alpha_{1}}\equiv \{ \epsilon, \xi^{\lambda\mu}, \epsilon^{\lambda\mu\nu}\}, \label{6}
\end{equation}
which are bosonic, completely antisymmetric (where appropriate), and otherwise arbitrary tensors of definite orders defined on the spacetime manifold. The above set of gauge transformations, written in a compact form as $\delta_{\Omega^{\alpha_{1}}}\Phi^{\alpha_{0}}$, is off-shell reducible of order two. Indeed, if we transform the gauge parameters like
\begin{equation}
\Omega ^{\alpha_{1}}=\Omega ^{\alpha_{1}} (\Omega ^{\alpha_{2}}) \Leftrightarrow \left\{
\begin{array}{l}
\epsilon (\Omega ^{\alpha_{2}}) =0,\\
\xi^{\mu\nu} (\Omega ^{\alpha_{2}}) =-3 \partial _{\lambda} \xi^{\lambda\mu\nu},\\
\epsilon ^{\mu\nu\rho} (\Omega ^{\alpha_{2}}) =-4 \partial _{\lambda} \epsilon^{\lambda\mu\nu\rho},
\end{array}
\right. \label{7}
\end{equation}
in terms of the first-order reducibility parameters
\begin{equation}
\Omega ^{\alpha_{2}} \equiv \{ \xi ^{\lambda\mu\nu}, \epsilon ^{\lambda\mu\nu\rho}\}, \label{8}
\end{equation}
that are bosonic, completely antisymmetric, and otherwise arbitrary tensors on the Minkowski spacetime, then the gauge transformations of all fields vanish everywhere on the space of field histories (off-shell)
\begin{equation}
\delta _{\Omega ^{\alpha_{1}}(\Omega ^{\alpha_{2}})} \Phi ^{\alpha_{0}}=0. \label{9}
\end{equation}
The last identities cover all the first-order reducibility relations of the set (\ref{3})--(\ref{5}) of gauge transformations. Next, if we transform the first-order reducibility parameters as
\begin{equation}
\Omega ^{\alpha_{2}} = \Omega ^{\alpha_{2}} (\Omega ^{\alpha_{3}}) \Leftrightarrow
\left\{
\begin{array}{l}
\xi^{\mu\nu\rho} (\Omega ^{\alpha_{3}}) =-4 \partial _{\lambda} \xi^{\lambda\mu\nu\rho},\\
\epsilon ^{\lambda\mu\nu\rho} (\Omega ^{\alpha_{3}})=0,
\end{array}
\right. \label{10}
\end{equation}
in terms of the second-order reducibility parameters
\begin{equation}
\Omega ^{\alpha_{3}} \equiv \{\xi ^{\lambda\mu\nu\rho}\}, \label{11}
\end{equation}
with $\xi ^{\lambda\mu\nu\rho}$ a bosonic, completely antisymmetric, and otherwise arbitrary tensor, then all the transformed gauge parameters (\ref{7}) vanish also off-shell
\begin{equation}
\Omega ^{\alpha_{1}} (\Omega ^{\alpha_{2}} (\Omega ^{\alpha_{3}}))=0. \label{12}
\end{equation}
The above identities stand for all the second-order reducibility relations corresponding to the gauge transformations of action (\ref{2}). The reducibility order of this gauge theory in $D=4$ is equal to two since the transformed first-order reducibility parameters (\ref{10}) vanish if and only if the second-order reducibility parameters also vanish (actually if they are constant, but, due to their explicit dependence on the spacetime coordinates, these constants can be taken to vanish)
\begin{equation}
\Omega ^{\alpha_{2}} (\Omega ^{\alpha_{3}})=0 \Leftrightarrow \Omega ^{\alpha_{3}} \equiv \xi ^{\lambda\mu\nu\rho}=0. \label{13}
\end{equation}
Actually, the BF sector, given its topological character, carries no physical degrees of freedom, and hence all the physical degrees of freedom of this free model are provided by the presence of the matter scalars. As an issue of terminology, if we write transformations (\ref{3})--(\ref{5}), (\ref{7}), and (\ref{10}) in condensed De Witt notations (where any discrete index is understood to contain a continuous, spacetime one as well, say $\alpha _{0}\equiv (\alpha _{0},x)$, and a sum taken over any such index automatically includes a spacetime integral with respect to the corresponding continuous one) like
\begin{equation*}
\delta_{\Omega^{\alpha_{1}}}\Phi^{\alpha_{0}} = Z _{\hphantom{\alpha _{0}}\alpha _{1}}^{\alpha _{0}} \Omega^{\alpha_{1}}, \qquad \Omega ^{\alpha_{1}} (\Omega ^{\alpha_{2}}) = Z _{\hphantom{\alpha _{1}}\alpha _{2}}^{\alpha _{1}} \Omega^{\alpha_{2}}, \qquad \Omega ^{\alpha_{2}} (\Omega ^{\alpha_{3}}) = Z _{\hphantom{\alpha _{2}}\alpha _{3}}^{\alpha _{2}} \Omega^{\alpha_{3}},
\end{equation*}
then $Z _{\hphantom{\alpha _{0}}\alpha _{1}}^{\alpha _{0}}$ are known as gauge generators, while $Z _{\hphantom{\alpha _{1}}\alpha _{2}}^{\alpha _{1}}$ and $Z _{\hphantom{\alpha _{2}}\alpha _{3}}^{\alpha _{2}}$ are called first-order and respectively second-order reducibility functions.

Moreover, all the commutators among the gauge transformations of the fields vanish off-shell, $[\delta_{\Omega^{\alpha_{1}}},\delta_{\Omega^{\prime\alpha_{1}}}]\Phi^{\alpha_{0}}=0$, such the associated gauge algebra is Abelian. The previous properties combined with the linearity of the field equations following from action (\ref{2}) in all fields allow us to conclude that the overall free model under consideration is a linear gauge theory with a definite Cauchy order, equal to four.

Next, we construct the BRST differential algebra for the free model under study in the context of the antifield-antibracket formalism~\cite{BVPLB81,BVPLB83,BVPRD83,BVNPB84,CMP1990a,NPPB1990,Princeton1992,PR1995,IJGMMP1995,NPPB1997}. Related to the BF sector, we use the notations and results exposed in~\cite{AP2003}. We introduce the BRST generators as the original fields $\Phi^{\alpha_{0}}$ from relation (\ref{1}), the ghosts as dynamical variables respectively associated with both gauge and reducibility parameters displayed in (\ref{6}), (\ref{8}), and (\ref{11}), together with their corresponding antifields (denoted by star variables)
\begin{align}
\Phi ^{\alpha_{0}} & \equiv \big\{ \varphi , A _{\mu}, H ^{\mu}, B ^{\mu\nu}, \phi ^{A} \big\}, & \Phi ^{\ast} _{\alpha_{0}} & \equiv \{ \varphi ^{\ast} ,A ^{\ast\mu}, H ^{\ast}_{\mu}, B ^{\ast} _{\mu\nu}, \phi ^{\ast} _{A} \}, \label{14}\\
\Omega ^{\alpha_{1}} \rightarrow \eta ^{\alpha_{1}} & \equiv \{ \eta, C ^{\mu\nu}, \eta ^{\mu\nu\rho}\}, & \eta ^{\ast} _{\alpha_{1}} & \equiv \{ \eta ^{\ast}, C ^{\ast}_{\mu\nu}, \eta ^{\ast}_{\mu\nu\rho}\}, \label{15}\\
\Omega ^{\alpha_{2}} \rightarrow \eta ^{\alpha_{2}} & \equiv \{ C ^{\mu\nu\rho}, \eta ^{\lambda\mu\nu\rho} \}, & \eta ^{\ast} _{\alpha_{2}} & \equiv \{ C ^{\ast} _{\mu\nu\rho}, \eta ^{\ast} _{\lambda\mu\nu\rho}\}, \label{16}\\
\Omega ^{\alpha_{3}} \rightarrow \eta ^{\alpha_{3}} & \equiv \{ C ^{\lambda\mu\nu\rho} \}, & \eta ^{\ast} _{\alpha_{3}} & \equiv \{ C ^{\ast} _{\lambda\mu\nu\rho} \}. \label{17}
\end{align}
For notational ease, it is convenient to organize the fields/ghosts and respectively antifields into
\begin{equation}
\chi ^{\Delta} \equiv \{ \Phi ^{\alpha_{0}}, \eta ^{\alpha_{1}}, \eta ^{\alpha_{2}}, \eta ^{\alpha_{3}} \}, \qquad \chi ^{\ast} _{\Delta} \equiv \{ \Phi ^{\ast} _{\alpha_{0}}, \eta ^{\ast} _{\alpha_{1}}, \eta ^{\ast} _{\alpha_{2}}, \eta ^{\ast} _{\alpha_{3}} \}. \label{18}
\end{equation}
The $\mathbb{Z}_2$ grading of the BRST algebra in terms of the Grassmann parity ($\varepsilon $) is inferred from the observation that all the original fields together with the accompanying gauge and reducibility parameters are bosonic, so, according to the general rules of the antifield formalism, we take
\begin{equation}
\varepsilon(\eta^{\alpha_{k}})=k \, \mathrm{mod} \, 2, \quad k=1,2,3, \qquad \varepsilon(\chi^{\ast}_{\Delta})= (\varepsilon(\chi^{\Delta})+1) \, \mathrm{mod} \,2. \label{19}
\end{equation}
The Grassmann parity is then lifted to the BRST algebra by means of its additive action modulo 2 against multiplication. In agreement with the usual prescriptions of the BRST method, the BRST algebra is endowed with three more gradings (correlated with the main derivatives/differentials acting on this algebra): two $\mathbb{N}$-gradings along the antifield number $\agh$ and respectively the pure ghost number $\pgh$ and a total $\mathbb{Z}$-grading in terms of the ghost number $\gh$. These are instated by setting the values of the corresponding degrees at the level of the BRST generators
\begin{align}
\agh (\chi^{\Delta}) & =0, & \agh (\Phi^{\ast}_{\alpha_{0}}) & =1, & \agh (\eta^{\ast}_{\alpha_{k}}) & =k+1, & k & =1,2,3, \label{20}\\
\pgh (\chi^{\ast}_{\Delta}) & =0, & \pgh (\Phi^{\alpha_{0}}) & =0, & \pgh (\eta^{\alpha_{k}}) & =k, & k & =1,2,3, \label{21}
\end{align}
and by further using their additive behaviour with respect to multiplication. Finally, the (total) ghost number of any object with definite pure ghost and antifield numbers is defined like $\gh (a) = \pgh (a) - \agh (a)$.

Due to the fact that the right-hand sides of both gauge transformations (\ref{3})--(\ref{5}) and relations (\ref{7}) and (\ref{10}) do not depend on the fields $\Phi^{\alpha_{0}}$ (or, in other words, all gauge generators and reducibility functions --- of order one and respectively two --- are field independent), it follows that the BRST differential $s$ reduces to a sum of two fermionic derivations
\begin{equation}
s=\delta + \gamma , \label{22}
\end{equation}
with $\delta $ the Koszul--Tate differential, graded in terms of $\agh$ ($\agh (\delta )=-1$), and $\gamma $ the longitudinal exterior derivative (in this case a true differential), graded by $\pgh$ ($\pgh (\gamma )=1$). These two degrees do not interfere ($\agh (\gamma )=0$, $\pgh (\delta )=0$), such that the total degree of the BRST differential (and of each of its components), namely $\gh$, becomes equal to $1$: $\gh (s)=\gh (\delta )=\gh (\gamma ) =1$. One of the major requirements of the BRST setting, namely the second-order nilpotency of $s$, becomes equivalent to three separate equations
\begin{equation}
s^2=0 \Leftrightarrow (\delta ^2=0, \, \delta \gamma + \gamma \delta =0, \, \gamma ^2=0), \label{23}
\end{equation}
which confirms that $\gamma$ can indeed be constructed as a true differential in the case of the free model under study. The actions of $\delta $ and $\gamma $ on the BRST generators that enforce (\ref{23}) as well as the fundamental cohomological requirements of the antifield BRST theory~\cite{CMP1990a,NPPB1990,Princeton1992,PR1995,IJGMMP1995,NPPB1997} are given by
\begin{align}
\delta \chi^{\Delta} &=0, & \delta \varphi ^{\ast} &= \partial _{\lambda }H^{\lambda }, & \delta A^{\ast\mu} &= \partial _{\lambda }B^{\lambda \mu}, \label{24}\\
\delta H^{\ast}_{\mu} &= - \partial _{\mu }\varphi , & \delta B^{\ast}_{\mu\nu} &= -\tfrac{1}{2} \partial _{[\mu }A _{\nu ]}, & \delta \phi ^{\ast}_{A} &= k _{AB} \Box \phi ^{B}, \label{25}\\
\qquad \delta \eta^{\ast} &= - \partial _{\lambda } A^{\ast \lambda}, & \delta C^{\ast}_{\mu\nu} &= \partial _{[\mu }H ^{\ast } _{\nu ]}, & \delta \eta^{\ast}_{\mu\nu\rho} &= \partial _{[\mu } B^{\ast}_{\nu \rho ]}, \label{26}\\
\delta C^{\ast}_{\mu\nu\rho} &= - \partial _{[\mu } C^{\ast}_{\nu \rho ]}, & \delta \eta^{\ast}_{\lambda\mu\nu\rho} &= - \partial _{[\lambda } \eta^{\ast}_{\mu \nu \rho]}, & \delta C^{\ast}_{\lambda\mu\nu\rho} &= \partial _{[\lambda } C^{\ast}_{\mu \nu \rho]}, \label{27}\\
\gamma \chi^{\ast}_{\Delta} &=0, & \gamma \varphi &=0, & \gamma A_{\mu} &= \partial _{\mu }\eta, \label{28}\\
\gamma H^{\mu} &= -2 \partial _{\lambda } C^{\lambda\mu }, & \gamma B^{\mu\nu} &= -3 \partial _{\lambda } \eta ^{\lambda\mu \nu}, & \gamma \phi ^{A} &=0, \label{29}\\
\gamma \eta &=0, & \gamma C^{\mu\nu} &= -3 \partial _{\lambda } C ^{\lambda\mu \nu}, & \gamma \eta^{\mu\nu\rho} &= -4 \partial _{\lambda } \eta ^{\lambda\mu \nu\rho}, \label{30}\\
\gamma C^{\mu\nu\rho} &= -4 \partial _{\lambda } C ^{\lambda\mu \nu\rho}, & \gamma \eta^{\lambda\mu\nu\rho} &= 0, & \gamma C^{\lambda\mu\nu\rho} &= 0, \label{31}
\end{align}
where both operators are assumed to act like right derivations and $\Box \equiv \partial _{\mu} \partial ^{\mu}$ symbolizes the d'Alembertian. We notice that the actions of $\gamma $ on all fields/ghosts can be obtained in this particular situation simply by replacing all gauge or reducibility parameters from the right-hand sides of relations (\ref{3})--(\ref{5}), (\ref{7}), (\ref{10}), and (\ref{13}) with the corresponding ghosts introduced in (\ref{15})--(\ref{17}).

A striking feature of the antifield approach, in spite of its essentially Lagrangian origins, resides in the (anti)canonical action of the BRST differential~\cite{BVPLB81,BVPLB83,BVPRD83,BVNPB84,CMP1990a,NPPB1990,Princeton1992,PR1995,IJGMMP1995,NPPB1997}, $s\cdot = (\cdot ,S)$, where its (anti)canonical generator $S$ is a bosonic functional of ghost number equal to $0$ ($\varepsilon (S)=0$, $\gh (S)=0$) that is solution to the classical master equation $(S,S)=0$, where $(,)$ symbolizes the antibracket. This (anti)canonical structure is obtained by postulating that each antifield is respectively conjugated to the corresponding field/ghost, $(\chi^{\Delta},\chi^{\ast}_{\Delta ^{\prime}})=\delta ^{\Delta} _{\Delta ^{\prime}}$, and is shown to display properties that are fully complementary to the generalized Poisson bracket from the Hamiltonian formalism for theories with both bosonic and fermionic degrees of freedom. The classical master equation is completely equivalent to the second-order nilpotency of $s$ and its solution also implements the main cohomological requirements at the level of the BRST differential. In the case of the free gauge theory under study the solution to the classical master equation takes a simple form, expressed by
\begin{align}
S=\int d^4x\Big[ & H^{\mu}\partial_{\mu }\varphi+\tfrac{1}{2}B^{\mu\nu}\partial_{[\mu }A_{\nu ]}+\tfrac{1}{2}k_{AB} \big( \partial_{\mu }\phi ^{A} \big) \big(\partial^{\mu }\phi ^{B}\big) \nonumber \\
&+A^{\ast\mu} \partial _{\mu }\eta -2 H^{\ast}_{\mu} \partial _{\lambda } C^{\lambda\mu } -3 B^{\ast}_{\mu\nu} \partial _{\lambda } \eta ^{\lambda\mu \nu} \nonumber \\
&-3 C^{\ast}_{\mu\nu}\partial _{\lambda } C ^{\lambda\mu \nu} -4 \eta^{\ast}_{\mu\nu\rho} \partial _{\lambda } \eta ^{\lambda\mu \nu\rho} -4 C^{\ast}_{\mu\nu\rho} \partial _{\lambda } C ^{\lambda\mu \nu\rho}\Big]. \label{32}
\end{align}
The solution to the classical master equation is constructed in such a way to encode the entire gauge structure of a given theory. Relation (\ref{32}) (and also formula (\ref{22})) must be viewed like a decomposition of the canonical generator of the BRST differential (or respectively of the BRST differential itself) along the antifield number $\agh$. Thus, its component of $\agh $ equal to $0$ is nothing but the Lagrangian action of the starting gauge model, while its projection on $\agh $ equal to $1$ is written as the antifields of the original fields times the gauge transformations of the corresponding fields with the gauge parameters $\Omega^{\alpha_{1}}$ replaced by the ghosts $\eta^{\alpha_{1}}$ (of pure ghost number $1$). The structure of the remaining terms, of antifield numbers strictly greater than $1$, reveals all the remaining tensor properties of the gauge algebra and reducibility of the chosen generating set of gauge transformations. In our case there appear only terms of $\agh$ equal to $2$ and respectively $3$, that are linear in both antifields and ghosts, whose origin is due to the first- and respectively second-order reducibility of the gauge transformations (\ref{3})--(\ref{5}). The absence of terms at least quadratic in ghosts or respectively in antifields is directly correlated with the abelianity of the associated gauge algebra and the off-shell behaviour of the accompanying reducibility relations. Moreover, we observe that all the properties of the Lagrangian formulation of the free model (\ref{2}), such as spacetime locality~\cite{CMP1990b}, Lorentz covariance, Poincar\'{e} invariance and so on, are preserved by the solution to the classical master equation.

\section{Deformation procedure\label{deform}}

The long standing problem of constructing consistent interactions in gauge field theories has been solved in an elegant and yet economic fashion by reformulating it as a problem of deforming the classical solution to the master equation~\cite{PLB1993,CM1998} in the framework of the local BRST cohomology~\cite{CMP1995a,CMP1995b,PR2000}. Thus, if consistent interactions can be constructed for a given ``free'' gauge theory, then the associated solution to the classical master equation, $S$, can be deformed along a coupling constant (deformation parameter) $g$ to another functional $\bar{S}$, which is precisely the solution to the master equation for the interacting gauge theory
\begin{equation}
S\rightarrow \bar{S}=S +gS_1 +g^2 S_2 +g^3 S_3 + g^4 S_4 +\cdots , \qquad (\bar{S},\bar{S})=0. \label{33}
\end{equation}
The consistency of deformations requires that the deformed gauge theory preserves the number of physical degrees of freedom of the starting ``free'' system (the field content and the number of independent gauge symmetries are the same via keeping the maximum reducibility order and also the number of independent reducibility relations at each order for both theories), and hence the field/ghost and antifield spectra are unchanged. It is understood that $\bar{S}$ should satisfy all the other standard properties required by the BRST-antifield formalism (and already assumed to be verified by $S$), namely, to be a bosonic functional of ghost number $0$ ($\varepsilon (\bar{S})=0$, $\gh (\bar{S})=0$) of fields, ghosts, and antifields. From the expansion of $\bar{S}$ we find that the main equation of the deformation procedure, $(\bar{S},\bar{S})=0$, becomes equivalent to the following chain of equations obtained by projection on the various powers in the coupling constant (and also accounting for the canonical action of the BRST differential of the initial gauge theory, $s\cdot = (\cdot , S)$)
\begin{align}
g^0&:(S,S)=0, \label{34}\\
g^1&:sS_1=0, \label{35}\\
g^2&:\tfrac{1}{2}(S_1,S_1)+sS_2=0,\label{36}\\
g^3&:(S_1,S_2)+sS_3=0, \label{37}\\
g^4&:\tfrac{1}{2}(S_2,S_2)+(S_1,S_3)+sS_4=0, \label{38}\\
& \vdots \nonumber
\end{align}
Equation (\ref{34}) is satisfied by assumption (since $S$ is the generator of the BRST symmetry for the ``free'' gauge theory). The remaining ones are to be solved recursively, from lower to higher orders, such that each equation corresponding to a given order of perturbation theory, say $i$ ($i\geq 1$), contains a single unknown functional, namely, the deformation of order $i$, $S_i$. Equation (\ref{35}) demands that the first-order deformation is $s$-closed. Nevertheless, we discard the class of $s$-exact solutions since these can be shown to correspond to trivial gauge interactions of the Lagrangian action~\cite{PLB1993,CM1998} and can be eliminated by some (possibly nonlinear) field redefinitions. In view of this, it follows that the nontrivial first-order deformations of the solution to the classical master equation are constrained by equation (\ref{35}) to be the (nontrivial) equivalence classes of the cohomology of the BRST differential $s$ in ghost number $0$ computed in the space of all (local and nonlocal) functionals of fields, ghosts, and antifields. This specific cohomology is nonempty since it contains all Lagrangian physical observables of the initial ``free'' gauge theory, so we can admit that equation (\ref{35}) possesses nontrivial solutions. The existence of solutions to the remaining higher-order equations is shown by means of the triviality of the antibracket map in the BRST cohomology computed in the space of all functionals~\cite{PLB1993}. Unfortunately, this procedure does not guarantee the spacetime locality of the deformed solution $\bar{S}$, and thus neither the locality of the interacting Lagrangian action.

Under the hypothesis of spacetime locality of deformations, if we make the notations
\begin{gather}
S_1 =\int a \, d^Dx , \quad S_2=\int b \, d^Dx , \quad S_3=\int c \, d^Dx , \quad S_4=\int d \, d^Dx , \label{39}\\
\tfrac{1}{2}(S_1,S_1) = \int \Delta \, d^Dx , \quad (S_1,S_2)= \int \Lambda \, d^Dx , \label{40}\\
\tfrac{1}{2}(S_2,S_2)+(S_1,S_3)=\int \Gamma \, d^Dx , \quad \cdots \label{41}
\end{gather}
where the nonintegrated densities of the deformations of various orders, $a$, $b$, $c$, $d$, and so on, are all bosonic and of ghost number equal to $0$ (such that $\Delta $, $\Lambda $, $\Gamma $, etc. become fermionic and of total ghost number equal to $1$), then equations (\ref{35})--(\ref{38}) etc. take the local form (in dual language)
\begin{align}
g^1&: sa = \partial _{\mu} j ^{\mu}, \label{42}\\
g^2&: sb = -\Delta + \partial _{\mu} k ^{\mu}, \label{43}\\
g^3&: sc = -\Lambda + \partial _{\mu} l ^{\mu}, \label{44}\\
g^4&: sd = -\Gamma + \partial _{\mu} m ^{\mu}, \label{45}\\
&\vdots \nonumber
\end{align}
where all the currents ($j ^{\mu}$, $k ^{\mu}$, etc.) are local, fermionic, and of ghost number $1$. The above chain of equations should be solved recursively, starting from lower to higher orders of perturbation theory. Equation (\ref{42}) stipulates that the nonintegrated density of the first-order deformation is a local BRST co-cycle of ghost number $0$. Its solution is unique up to addition of trivial quantities, i.e., of $s$-exact terms modulo divergences at the level of $a$ and respectively of $s$ acting on the corresponding currents modulo the divergence of a two-form in relation with $j ^{\mu}$
\begin{equation}
a\rightarrow a^{\prime} = a+s\bar{a}+\partial _{\mu} \bar{j} ^{\mu}, \qquad j ^{\mu} \rightarrow j ^{\prime \mu} = j ^{\mu} + s\bar{j} ^{\mu}+ \partial _{\nu} k ^{\nu\mu}, \label{46}
\end{equation}
(with $\bar{a}$ local, fermionic, and of ghost number $-1$, the current $\bar{j} ^{\mu}$ also local, but bosonic and of ghost number $0$, and the two-tensor $k ^{\nu\mu}$ local, fermionic, of $\gh $ equal to $1$, and antisymmetric, $k ^{\nu\mu}=-k ^{\mu\nu}$) in the sense that
\begin{equation}
sa - \partial _{\mu} j ^{\mu} \equiv sa^{\prime} - \partial _{\mu} j ^{\prime\mu} =0. \label{47}
\end{equation}
In other words, $a$ is constrained now to belong to a nontrivial class of the local BRST cohomology (cohomology of $s$ modulo $d$ --- with $d$ the exterior differential in spacetime) in $\gh =0$ computed in the algebra of (local) nonintegrated densities, $H^0(s|d)$. In this context, there is no warranty that there exist nontrivial solutions to the first-order deformation equation in this algebra. Moreover, assuming one actually finds such solutions, $a$, it is possible that there are still no local solutions with respect to the second- or higher-order deformations. Indeed, $\Delta $ introduced in the former relation from (\ref{40}) is local now due to the previous assumption on $a$ since the antibracket of two local functionals is always local. Nevertheless, it may not read as an $s$-exact object modulo a divergence, with both the object and the corresponding current some local quantities, in which case the second-order deformation equation, (\ref{43}), possesses no solutions with respect to $b$ in the algebra of local nonintegrated densities. If one imposes further restrictions on the deformations of the solution to the classical master equation, such as Poincar\'{e} invariance, Lorentz covariance, etc., then equations (\ref{42})--(\ref{45}), etc. remain valid, but the first-order deformation $a$ should be regarded as a nontrivial element of $H^0(s|d)$ computed now in an even more restricted algebra, of nonintegrated densities that are also Poincar\'{e} invariant, Lorentz covariant, etc.

\section{Setting the problem. First-order deformation\label{first}}

In the sequel we apply the deformation procedure exposed previously with the purpose of generating consistent interacting gauge theories in $D=4$ whose free limit is precisely the gauge theory described by relations (\ref{2})--(\ref{5}). We are interested only in (nontrivial) deformations that comply with the standard hypotheses of field theory: analyticity in the coupling constant, Lorentz covariance, spacetime locality, and Poincar\'{e} invariance. By analyticity in the coupling constant we mean that the deformed solution to the classical master equation, $\bar{S}$ as in (\ref{33}), is an analytic function of $g$ and reduces to the canonical BRST generator (\ref{32}) of the starting model in the free limit ($g=0$). The other requirements are translated at the level of its nonintegrated densities at all orders of perturbation theory to be expressed by (nontrivial) bosonic functions of ghost number $0$ that are: (A) (background) Lorentz scalars; (B) smooth functions of the undifferentiated original fields $\Phi^{\alpha_{0}}$ (see notation (\ref{1})); (C) polynomials in the derivatives of the original fields up to a finite order; (D) polynomials in the ghosts, antifields, and their spacetime derivatives up to a finite order (items (B), (C), and (D) ensure the spacetime locality); (E) without an explicit dependence of the spacetime coordinates (Poincar\'{e} invariance). In addition, we impose the conservation of the number of derivatives on each field with respect to the free limit and call it the derivative order assumption. This means that: (1) the interacting Lagrangian density may contain at most two derivatives of the fields at each order in the coupling constant; (2) all the vertices containing two derivatives are limited to terms that are quadratic in the first-order derivatives of the matter fields; (3) the other vertices, with one or no derivatives, are not restricted. In this way, the derivative order of the equations of motion for each field is the same in the free and respectively interacting theory. We recall that the interacting Lagrangian densities in various perturbative orders are obtained by projecting the corresponding nonintegrated densities of the deformed solution to the master equation on antifield number $0$. Actually, we show that it is possible to relax the derivative order assumption, work with the weaker requirement that the maximum number of derivatives allowed to enter each Lagrangian density is equal to two (without limiting the fields on which the derivatives may act), and then recover precisely the initial, stronger condition at the level of each nontrivial deformation.

By virtue of the discussion from the previous section, the nonintegrated density of the first-order deformation, $a$, should be a nontrivial element of the local BRST cohomology in ghost number equal to $0$, $H^0(s|d)$. In addition, all such solutions for $a$ will be selected such as to comply with the working hypotheses mentioned in the above. The nonintegrated density of the first-order deformation splits naturally into two components
\begin{equation}
a=a^{\mathrm{BF}}+a^{\mathrm{int}}, \label{48}
\end{equation}
where $a^{\mathrm{BF}}$ is responsible for the selfinteractions among the BF fields and $a^{\mathrm{int}}$ governs the couplings between the BF field spectrum and the matter scalar fields as well as the selfinteractions among the scalar fields. The two components display different contents of BRST generators ($a^{\mathrm{BF}}$ involves only the BF fields, ghosts, and antifields, while $a^{\mathrm{int}}$ mixes the BF and matter sectors), such that equation $sa =\partial _{\mu} j ^{\mu}$ becomes equivalent to two equations, one for each piece
\begin{align}
sa^{\mathrm{BF}} &= \partial _{\mu} j _\mathrm{BF}^{\mu}, \label{49}\\
sa^{\mathrm{int}} &= \partial _{\mu} j _\mathrm{int}^{\mu}. \label{50}
\end{align}

Equation (\ref{49}) was addressed in~\cite{AP2003} under some equivalent working hypotheses and in the framework of the same deformation approach. Employing all the results, notations, and conventions therein, it follows that $a^{\mathrm{BF}}$ can be taken to decompose as a sum between pieces with the antifield number ranging from $0$ to $4$
\begin{equation}
a^{\mathrm{BF}}=\sum _{j=0} ^{4} a _{j}^{\mathrm{BF}},\qquad \varepsilon(a _{j}^{\mathrm{BF}})=0, \qquad \gh(a _{j}^{\mathrm{BF}})=0, \qquad \agh(a _{j}^{\mathrm{BF}})=j, \label{51}
\end{equation}
where the expressions of its components take the form
\begin{align}
a _{0}^{\mathrm{BF}} = & -W(\varphi)H^{\mu}A_{\mu} , \label{52}\\
a _{1}^{\mathrm{BF}} = & \bigg( \varphi ^{\ast} W - H _{\mu} ^{\ast} H^{\mu} \frac{dW}{d\varphi } \bigg)\eta
+ \bigg( \frac{dW}{d\varphi } H _{[\mu} ^{\ast} A _{\nu ]} + 2W B _{\mu\nu} ^{\ast} \bigg) C ^{\mu\nu} , \label{53}\\
a _{2}^{\mathrm{BF}} = & - \bigg[ \bigg( \frac{dW}{d\varphi } C _{[\mu\nu} ^{\ast} + \frac{d^{2} W}{d\varphi ^{2}} H _{[\mu} ^{\ast} H _{\nu} ^{\ast} \bigg) A _{\rho ]} + 2 \bigg( \frac{dW}{d\varphi } H _{[\mu} ^{\ast} B _{\nu\rho ]} ^{\ast} + W \eta _{\mu\nu\rho} ^{\ast}\bigg) \bigg] C ^{\mu\nu\rho} \nonumber \\
&  +\bigg( \frac{dW}{d\varphi } C _{\mu\nu} ^{\ast} + \frac{d^{2} W}{d\varphi ^{2}} H _{\mu} ^{\ast} H _{\nu} ^{\ast} \bigg)\eta C ^{\mu\nu}, \label{54}\\
a _{3}^{\mathrm{BF}} = & \bigg\{ \bigg( \frac{dW}{d\varphi } C _{[\mu\nu\rho} ^{\ast} + \frac{d^{2} W}{d\varphi ^{2}} H _{[\mu} ^{\ast} C _{\nu\rho} ^{\ast} + \frac{d^{3} W}{d\varphi ^{3}} H _{[\mu} ^{\ast} H _{\nu} ^{\ast} H _{\rho} ^{\ast} \bigg) A _{\lambda ]} \nonumber \\
&+ 2 \bigg[ \bigg( \frac{dW}{d\varphi } C _{[\mu\nu} ^{\ast} + \frac{d^{2} W}{d\varphi ^{2}} H _{[\mu} ^{\ast} H _{\nu} ^{\ast} \bigg) B _{\rho \lambda ]} ^{\ast} + \frac{dW}{d\varphi } H _{[\mu}^{\ast} \eta _{\nu\rho\lambda ]} ^{\ast} +W \eta _{\mu \nu \rho \lambda } ^{\ast} \bigg] \bigg\} C ^{\mu\nu\rho\lambda} \nonumber \\
&- \bigg( \frac{dW}{d\varphi } C _{\mu\nu\rho} ^{\ast} + \frac{d^{2} W}{d\varphi ^{2}} H _{[\mu} ^{\ast} C _{\nu\rho ]} ^{\ast} + \frac{d^{3} W}{d\varphi ^{3}} H _{\mu} ^{\ast} H _{\nu} ^{\ast} H _{\rho} ^{\ast} \bigg) \eta C ^{\mu\nu\rho} , \label{55}\\
a _{4}^{\mathrm{BF}} = & \bigg[ \frac{dW}{d\varphi } C _{\mu\nu\rho\lambda} ^{\ast} + \frac{d^{2} W}{d\varphi ^{2}} \bigg( H _{[\mu} ^{\ast} C _{\nu\rho\lambda ]} ^{\ast} + C _{[\mu\nu} ^{\ast} C _{\rho\lambda ]} ^{\ast} \bigg) \nonumber \\
&+ \frac{d^{3} W}{d\varphi ^{3}} H _{[\mu} ^{\ast} H _{\nu} ^{\ast} C _{\rho\lambda ]} ^{\ast} + \frac{d^{4} W}{d\varphi ^{4}} H _{\mu} ^{\ast} H _{\nu} ^{\ast} H _{\rho} ^{\ast} H _{\lambda} ^{\ast} \bigg] \eta C ^{\mu\nu\rho\lambda}. \label{56}
\end{align}
Everywhere in the sequel the lower numeric index of any quantity serves as the value of its antifield number (like, for instance, the lower index $j$ in $a _{j}^{\mathrm{BF}}$). In the above $W=W(\varphi)$ is a smooth function depending only on the undifferentiated BF scalar field $\varphi $. In~\cite{AP2003} it was considered a second possible solution to equation (\ref{49}), that decomposes again in pieces of antifield number ranging between $0$ and $4$ and produces a Lagrangian at order one in the coupling constant which is quadratic in the components of the two-form $B ^{\mu\nu}$, $M(\varphi)\varepsilon _{\mu\nu\rho\lambda}B ^{\mu\nu} B^ {\rho\lambda}$, with $M(\varphi)$ another smooth function of $\varphi $. Here, we discard this last solution, although it is nontrivial and verifies all the working hypotheses, for a simple reason. The consistency of the first-order deformation (i.e., the existence of $b$ as solution to equation (\ref{43})) constrains the two functions $W$ and $M$ to satisfy the equation $W(\varphi)M(\varphi)=0$, so the associated BF selfinteractions cannot coexist in $D=4$. Therefore, one should analyze separately the complementary cases where a single function is nonvanishing. The case $W=0$ and $M$ arbitrary has been shown in~\cite{AP2003} to produce purely trivial couplings between the BF and an arbitrary set of matter fields and therefore we avoid it since our aim is to unveil all nontrivial couplings between the BF fields and a collection of real massless scalars. This argues the choice (\ref{52})--(\ref{56}) with respect to the first-order deformation in the BF sector.

Our next task is to establish the general form of the cross-coupling first-order deformation as solution to equation (\ref{50}). Although the Cauchy order of the overall model is equal to $4$, it is the BF sector alone, governed by the Lagrangian action $S^{\mathrm{L,BF}}[\varphi ,A_{\mu}, H^{\mu},B^{\mu\nu}]$ from (\ref{2}) and gauge transformations (\ref{3})--(\ref{4}), which is a linear gauge theory of Cauchy order $4$. The massless real scalar fields are separately described by a linear theory with the Lagrangian action $S^{\mathrm{L,scalar}}[\phi]$ from (\ref{2}) and without (nontrivial) gauge symmetries (see (\ref{5})), so its Cauchy order is equal to $1$. On these grounds, the matter sector can be shown to be able to contribute nontrivially to the first-order deformation $a^{\mathrm{int}}$ earliest in antifield number $1$ (see~\cite{AP2003} for a more detailed argument). Consequently, it is enough to expand $a^{\mathrm{int}}$ and $j _{\mathrm{int}} ^{\mu}$ along the antifield number like
\begin{align}
a^{\mathrm{int}} &= a _{1}^{\mathrm{int}} + a _{0}^{\mathrm{int}}, & j _{\mathrm{int}} ^{\mu} &= j _{\mathrm{int},1} ^{\mu} + j _{\mathrm{int},0} ^{\mu}, \label{57}\\
\agh(a _{1}^{\mathrm{int}}) &=\agh(j _{\mathrm{int},1} ^{\mu}) = 1, & \agh(a _{0}^{\mathrm{int}}) &=\agh(j _{\mathrm{int},0} ^{\mu}) = 0, \label{58}
\end{align}
being understood that in addition both components of $a^{\mathrm{int}}$ should be bosonic and of ghost number $0$, while both currents should be fermionic and of ghost number $1$. Of course, since both degrees $\gh $ and $\agh $ are now fixed, the third one ($\pgh $) is also completely known
\begin{equation}
\pgh(a _{0}^{\mathrm{int}}) =0, \qquad \pgh(a _{1}^{\mathrm{int}}) = \pgh(j _{\mathrm{int},0} ^{\mu}) =1, \qquad \pgh(j _{\mathrm{int},1} ^{\mu}) = 2. \label{59}
\end{equation}
Taking into account formula (\ref{57}) and decomposition (\ref{22}) of the BRST differential, equation (\ref{50}) becomes equivalent to two separate equations
\begin{align}
\gamma a _{1}^{\mathrm{int}} & = \partial _{\mu} j _{\mathrm{int},1} ^{\mu}, \label{60} \\
\delta a _{1}^{\mathrm{int}} + \gamma a _{0}^{\mathrm{int}} & = \partial _{\mu} j _{\mathrm{int},0} ^{\mu}. \nonumber
\end{align}
Since the antifield number of both hand sides of equation (\ref{60}) is strictly positive (equal to $1$), it can be safely replaced by its homogeneous version without loss of nontrivial terms, namely, one can always take $j _{\mathrm{int},1} ^{\mu} =0$ in (\ref{57}). The proof of this result is done in a standard manner (for instance, see~\cite{CMP1995b,NPB2001,PRD2003,IJMPA2004,IJGMMP2004,JHEP2006,PRD2006}) and enables the equivalence between (\ref{50}) and the simpler equations
\begin{align}
\gamma a _{1}^{\mathrm{int}} & = 0, \label{61} \\
\delta a _{1}^{\mathrm{int}} + \gamma a _{0}^{\mathrm{int}} & = \partial _{\mu} j _{\mathrm{int},0} ^{\mu}. \label{62}
\end{align}

Equation (\ref{61}) shows that $a _{1}^{\mathrm{int}}$ can be taken as a $\gamma $-closed object of pure ghost number $1$ (see (\ref{59})). All $\gamma $-exact quantities may be factored out from $a _{1}^{\mathrm{int}}$ as they eventually provoke trivial terms in $a ^{\mathrm{int}}$, which is translated into the fact that $a _{1}^{\mathrm{int}}$ is a nontrivial element of the cohomology of $\gamma $ in $\pgh = 1$ computed in the algebra of local nonintegrated densities in the framework of the above hypotheses, $H^{1}(\gamma )$. According to the detailed analysis from Appendix \ref{A}, we conclude that the general, nontrivial expression of the nonintegrated density of the  first-order deformation that mixes the BF and the matter sectors and, essentially, meets all the imposed requirements, can be written as
\begin{align}
a ^{\mathrm{int}} =& a _{0} ^{\mathrm{int}} + a _{1} ^{\mathrm{int}}, \label{174}\\
a _{0} ^{\mathrm{int}} = & - k _{AB} \big( \partial ^{\mu } \phi ^{A} \big) \bar{n} ^{B}(\varphi, \phi) A_{\mu} - \mathcal{V} (\varphi, \phi ) \nonumber\\
&+ \tfrac{1}{2} \mu _{AB} (\varphi, \phi ) \big( \partial _{\mu } \phi ^{A} \big) \big( \partial ^{\mu } \phi ^{B} \big) , \label{175}\\
a _{1} ^{\mathrm{int}} =& \bigg[ - H ^{\ast} _{\mu} k _{AB} \big( \partial ^{\mu } \phi ^{A} \big) \frac{\partial \bar{n} ^{B} (\varphi, \phi)}{\partial \varphi } + \phi ^{\ast} _{A}\bar{n} ^{A}(\varphi, \phi) \bigg] \eta , \label{176}
\end{align}
where
\begin{gather}
\bar{n} ^{A} (\varphi, \phi) = n ^{A} (\varphi) + T ^{AB} (\varphi) k _{BC} \phi ^{C}, \qquad T ^{AB} (\varphi) = - T ^{BA} (\varphi), \label{177} \\
\mu _{AB} (\varphi, \phi ) = \mu _{BA} (\varphi, \phi ), \qquad \mu _{AB} (\varphi, \phi ) \neq \frac{\partial u _{A} (\varphi, \phi )}{\partial \phi ^{B}} + \frac{\partial u _{B} (\varphi, \phi )}{\partial \phi ^{A}}. \label{178}
\end{gather}
We remark that the nonintegrated density of the first-order deformation is parameterized by three kinds of arbitrary smooth functions depending only on the undifferentiated BF scalar field $\varphi$ (a scalar $W$, an $N$-dimensional vector of components $n ^{A}$, and a skew-symmetric quadratic matrix of order $N$ of elements $T ^{AB}$) and by two types of arbitrary smooth functions of all the undifferentiated scalar fields from the theory $\{ \varphi,\phi \}$ (a scalar $\mathcal{V}$ and a symmetric, nontrivial quadratic matrix of order $N$ of elements $\mu _{AB}$). It is easy to see that $a ^{\mathrm{int}}$ contains also nontrivial selfinteractions among the  matter fields, as it has been permitted from the start (see the discussion following formula (\ref{48})). They follow from (\ref{175}) allowing the functions $\mathcal{V}$ and $\mu _{AB}$ to include terms that depend solely on $\phi$.

This completes the problem of obtaining the general form of the nontrivial nonintegrated density of the first-order deformation $a$ as solution to (\ref{42}) under some specific assumptions. It splits like in (\ref{48}), with $a ^{\mathrm{BF}}$ and $a ^{\mathrm{int}}$ governed by relations (\ref{51})--(\ref{56}) and respectively (\ref{174})--(\ref{178}).

\section{Higher-order deformations\label{high}}

Next, we pass to inferring the nonintegrated density of the second-order deformation $b$ as solution to equation (\ref{43}) or, in other words, to the consistency of the first-order deformation at order two in the coupling constant. By direct computation, from (\ref{48}) where we use (\ref{51})--(\ref{56}) and (\ref{174})--(\ref{176}), we find that
\begin{align}
\Delta = & - \bigg( k _{AC} \frac{\partial \bar{n} ^{C} (\varphi, \phi)}{\partial \phi ^{B}} + k _{BC} \frac{\partial \bar{n} ^{C} (\varphi, \phi)}{\partial \phi ^{A}} \bigg) \big( \partial ^{\mu} \phi ^{A} \big) \bar{n} ^{B} A _{\mu} \eta \nonumber \\
& + \bigg[ \tfrac{1}{2} \bigg( \mu _{AC} (\varphi, \phi) \frac{\partial \bar{n} ^{C} (\varphi, \phi)}{\partial \phi ^{B}} + \mu _{BC} (\varphi, \phi) \frac{\partial \bar{n} ^{C} (\varphi, \phi)}{\partial \phi ^{A}} \nonumber\\
& + \frac{\partial \mu _{AB} (\varphi, \phi)}{\partial \phi ^{C}} \bar{n} ^{C} (\varphi, \phi) + \frac{\partial \mu _{AB} (\varphi, \phi)}{\partial \varphi} W(\varphi ) \bigg) \big( \partial _{\mu } \phi ^{A} \big) \big( \partial ^{\mu } \phi ^{B} \big) \nonumber \\
& - \bigg( \frac{\partial \mathcal{V} (\varphi, \phi )}{\partial \phi ^{A}} \bar{n} ^{A} (\varphi, \phi) + \frac{\partial \mathcal{V} (\varphi, \phi )}{\partial \varphi} W(\varphi ) \bigg) \bigg] \eta \nonumber \\
&+ s \bigg\{ H _{\mu} ^{\ast} \big[ \mu _{AB} (\varphi, \phi) \big( \partial ^{\mu} \phi ^{A} \big) - k _{AB} A ^{\mu} \bar{n} ^{A} (\varphi, \phi) \big] \frac{\partial \bar{n} ^{B} (\varphi, \phi)}{\partial \varphi} \eta \nonumber \\
&+ \big[  \mu _{AB} (\varphi, \phi) \big( \partial ^{\mu} \phi ^{A} \big) - \tfrac{1}{2} k _{AB} A ^{\mu} \bar{n} ^{A} (\varphi, \phi) \big] \bar{n} ^{B} (\varphi, \phi) A _{\mu} \bigg\} . \label{179}
\end{align}
The quantities appearing on the first line from the right-hand side of (\ref{179}) are identically vanishing because $\bar{n} ^{A}$ of the form (\ref{177}) are by construction the solutions to the latter set of equations in (\ref{131}) (see the paragraph containing formula (\ref{144}) from Appendix \ref{A}). With this observation at hand and making the supplementary notations
\begin{align}
&\frac{\partial \mathcal{V} (\varphi, \phi )}{\partial \phi ^{A}} \bar{n} ^{A} (\varphi, \phi) + \frac{\partial \mathcal{V} (\varphi, \phi )}{\partial \varphi} W(\varphi ) \equiv \mathcal{V} ^{\prime} (\varphi, \phi ), \label{180}\\
&\mu _{AC} (\varphi, \phi) \frac{\partial \bar{n} ^{C} (\varphi, \phi)}{\partial \phi ^{B}} + \mu _{BC} (\varphi, \phi) \frac{\partial \bar{n} ^{C} (\varphi, \phi)}{\partial \phi ^{A}} + \frac{\partial \mu _{AB} (\varphi, \phi)}{\partial \phi ^{C}} \bar{n} ^{C} (\varphi, \phi) \nonumber \\
&+ \frac{\partial \mu _{AB} (\varphi, \phi)}{\partial \varphi} W(\varphi ) \equiv \mu _{AB} ^{\prime} (\varphi, \phi), \qquad \mu _{AB} ^{\prime} (\varphi, \phi) = \mu _{BA} ^{\prime} (\varphi, \phi), \label{181}
\end{align}
from (\ref{179}) and (\ref{43}) we further deduce that the existence of a local $b$ as solution to (\ref{43}) is \emph{equivalent} to the (consistency) condition
\begin{align}
\big[ \tfrac{1}{2} \mu _{AB} ^{\prime} (\varphi, \phi) \big( \partial _{\mu } \phi ^{A} \big) \big( \partial ^{\mu } \phi ^{B} \big) - \mathcal{V} ^{\prime} (\varphi, \phi ) \big] \eta= -s b ^{\prime} + \partial _{\mu} k ^{\prime \mu} , \label{182}
\end{align}
with $b ^{\prime}$ a bosonic, local nonintegrated density of ghost number $0$ and $k ^{\prime \mu}$ a local fermionic current of ghost number $1$. Due to the fact that $\eta $ is $s$-closed and $s A_\mu = \partial _{\mu} \eta$, we get that (\ref{182}) implies the \emph{necessary} requirement
\begin{equation}
\tfrac{1}{2} \mu _{AB} ^{\prime} (\varphi, \phi) \big( \partial _{\mu } \phi ^{A} \big) \big( \partial ^{\mu } \phi ^{B} \big) - \mathcal{V} ^{\prime} (\varphi, \phi ) = s \bar{a} ^{\prime } +  \partial _{\mu} \bar{j} ^{\prime \mu}, \label{183}
\end{equation}
where $\bar{a} ^{\prime }$ denotes a fermionic,  local nonintegrated density of ghost number $-1$ and $\bar{j} ^{\prime \mu}$ a local, bosonic current of ghost number $0$. Inspecting equation (\ref{183}), on the one hand we remark that its left-hand side is a bosonic, local nonintegrated density of ghost number $0$ that is $s$-closed since it depends only on the scalar fields from the theory and their spacetime derivatives, so an element of $H^{0} (s|d)$. On the other hand, (\ref{183}) asks that it is precisely in a trivial class from $H^{0} (s|d)$. As it has been argued in Appendix \ref{A} (see the discussion following formula (\ref{170})), equation (\ref{183}) is satisfied \emph{if and only if} $\mathcal{V} ^{\prime}$ is vanishing \emph{and} $\mu _{AB} ^{\prime}$ is trivial, i.e. of the form (\ref{171})
\begin{equation}
\mathcal{V} ^{\prime} (\varphi, \phi ) =0, \qquad \mu _{AB} ^{\prime} (\varphi, \phi) = \frac{\partial \lambda _{A} (\varphi, \phi )}{\partial \phi ^{B}} + \frac{\partial \lambda _{B} (\varphi, \phi )}{\partial \phi ^{A}}, \label{184}
\end{equation}
with $\{ \lambda _{A},A=\overline{1,N}\}$ some smooth functions of the undifferentiated scalar fields from the theory. It is easy to see that equation (\ref{183}) is also \emph{sufficient} for the existence of $b$ in the sense that its solutions ensure that (\ref{182}) displays at least one local solution with respect to $b ^{\prime}$. Indeed, by means of (\ref{184}) and using formula (\ref{172}), we obtain that
\begin{align}
&\big[ \tfrac{1}{2} \mu _{AB} ^{\prime} (\varphi, \phi) \big( \partial _{\mu } \phi ^{A} \big) \big( \partial ^{\mu } \phi ^{B} \big) - \mathcal{V} ^{\prime} (\varphi, \phi ) \big] \eta \nonumber \\
= & s \bigg\{ - \big( \partial ^{\mu } \phi ^{A} \big) \lambda _{A}(\varphi, \phi) A_{\mu}  +\bigg[ k ^{AB} \phi ^{\ast} _{A} \lambda _{B}(\varphi, \phi) - H ^{\ast} _{\mu} \big( \partial ^{\mu } \phi ^{A} \big) \frac{\partial \lambda _{A} (\varphi, \phi)}{\partial \varphi } \bigg] \eta \bigg\} \nonumber \\
&+ \partial _{\mu} \big[ \lambda _{A} (\varphi, \phi ) \big( \partial ^{\mu} \phi ^{A} \big) \eta \big] , \label{185}
\end{align}
so comparing (\ref{182}) with (\ref{185}) it follows that we can take for instance
\begin{equation}
b ^{\prime} = \big( \partial ^{\mu } \phi ^{A} \big) \lambda _{A}(\varphi, \phi) A_{\mu} + \bigg[ H ^{\ast} _{\mu} \big( \partial ^{\mu } \phi ^{A} \big) \frac{\partial \lambda _{A} (\varphi, \phi)}{\partial \varphi } - k ^{AB} \phi ^{\ast} _{A} \lambda _{B}(\varphi, \phi) \bigg] \eta . \label{186}
\end{equation}
Until now we emphasized that the existence of local solutions to equation (\ref{43}), which controls the nonintegrated density of the second-order deformation, is equivalent to the fact the functions that parameterize the first-order deformation are no longer arbitrary, but subject to conditions (\ref{184}). Indeed, by means of (\ref{180})--(\ref{181}) and recalling (\ref{177}), it is clear that (\ref{184}) constrain precisely the parameterizing functions $W(\varphi)$, $n ^{A} (\varphi)$, $T ^{AB} (\varphi)$, $\mathcal{V}(\varphi, \phi )$, and $\mu _{AB} (\varphi, \phi)$.

Next, we point out that we can still simplify the second set of relations from (\ref{184}) without loss of nontrivial terms at the level of the nonintegrated density of the first-order deformation, (\ref{48}). To this aim we act as mentioned in Appendix \ref{A} at the end of the paragraph containing formula (\ref{172}), namely, we add to (\ref{48}) some specific purely trivial terms and choose to work with
\begin{align}
a \rightarrow a &+ s \bigg[ H _{\mu} ^{\ast} \big( \partial ^{\mu } \phi ^{A} \big) \frac{\partial v _{A} (\varphi, \phi )}{\partial \varphi} - k ^{AB} \phi _{A} ^{\ast} v _{B} (\varphi, \phi ) \bigg] \nonumber \\
&+ \partial _{\mu} \big( v _{A} (\varphi, \phi ) \partial ^{\mu} \phi ^{A} \big) , \label{187}
\end{align}
which is the same (due to (\ref{172})) with adding to $\mu _{AB}$ from (\ref{175}) a trivial part, of the form (\ref{171})
\begin{equation}
\mu _{AB} (\varphi, \phi ) \rightarrow \mu _{AB} (\varphi, \phi ) + \frac{\partial v _{A} (\varphi, \phi )}{\partial \phi ^{B}} + \frac{\partial v _{B} (\varphi, \phi )}{\partial \phi ^{A}}. \label{188}
\end{equation}
Consequently, the functions $\mu _{AB} ^{\prime} (\varphi, \phi)$  introduced in (\ref{181}) transform like
\begin{equation}
\mu _{AB} ^{\prime} (\varphi, \phi) \rightarrow \mu _{AB} ^{\prime} (\varphi, \phi) + \frac{\partial \Theta _{A} (\varphi, \phi )}{\partial \phi ^{B}} + \frac{\partial \Theta _{B} (\varphi, \phi )}{\partial \phi ^{A}}, \label{189}
\end{equation}
with
\begin{align}
\Theta _{A} (\varphi, \phi ) \equiv & \frac{\partial v _{A} (\varphi, \phi )}{\partial \phi ^{C}} \bar{n} ^{C} (\varphi, \phi) + v _{C} (\varphi, \phi ) \frac{\partial \bar{n} ^{C} (\varphi, \phi)}{\partial \phi ^{A}} \nonumber \\
&+ \frac{\partial v _{A} (\varphi, \phi )}{\partial \varphi} W(\varphi ). \label{190}
\end{align}
Comparing the second set of relations appearing in (\ref{184}) with (\ref{189}), we conclude that we can absorb the terms depending on $\lambda _{A}$ into $\mu _{AB}$ by an appropriate trivial transformation (\ref{188}). Therefore, from now on we work with the purely homogeneous conditions
\begin{equation}
\mathcal{V} ^{\prime} (\varphi, \phi ) =0, \qquad \mu _{AB} ^{\prime} (\varphi, \phi) = 0, \label{191}
\end{equation}
written in explicit form (with the help of (\ref{180}) and (\ref{181})) like
\begin{align}
\frac{\partial \mathcal{V} (\varphi, \phi )}{\partial \phi ^{A}} \bar{n} ^{A} (\varphi, \phi) + \frac{\partial \mathcal{V} (\varphi, \phi )}{\partial \varphi} W(\varphi ) &=0, \label{192}\\
\mu _{AC} (\varphi, \phi) \frac{\partial \bar{n} ^{C} (\varphi, \phi)}{\partial \phi ^{B}} + \mu _{BC} (\varphi, \phi) \frac{\partial \bar{n} ^{C} (\varphi, \phi)}{\partial \phi ^{A}}& \nonumber \\
+ \frac{\partial \mu _{AB} (\varphi, \phi)}{\partial \phi ^{C}} \bar{n} ^{C} (\varphi, \phi) + \frac{\partial \mu _{AB} (\varphi, \phi)}{\partial \varphi} W(\varphi ) &=0, \label{193}
\end{align}
with $\bar{n} ^{A}$ given in (\ref{177}). We call (\ref{192}) and (\ref{193}) consistency equations since, in agreement with the analysis from the previous paragraph, they ensure the existence of solutions $b$ to the second-order deformation equation in local form, (\ref{43}), and thus the consistency of the nonintegrated density of the overall deformation of the solution to the master equation at order two of perturbation theory.

Assuming there exist nonvanishing and nontrivial solutions to the consistency conditions with respect to the parameterizing functions, where nontrivial refers strictly to the fact that the functions $\mu _{AB}$ are subject to the latter requirement from (\ref{178}), it follows from (\ref{179}) that equation (\ref{43}) takes the form
\begin{align}
s \bigg\{ b + H _{\mu} ^{\ast} \big[ \mu _{AB} (\varphi, \phi) \big( \partial ^{\mu} \phi ^{A} \big) - k _{AB} A ^{\mu} \bar{n} ^{A} (\varphi, \phi) \big] \frac{\partial \bar{n} ^{B} (\varphi, \phi)}{\partial \varphi} \eta & \nonumber \\
+ \big[  \mu _{AB} (\varphi, \phi) \big( \partial ^{\mu} \phi ^{A} \big) - \tfrac{1}{2} k _{AB} A ^{\mu} \bar{n} ^{A} (\varphi, \phi) \big] \bar{n} ^{B} (\varphi, \phi) A _{\mu} \bigg\} &= \partial _{\mu} k ^{\mu} . \label{194}
\end{align}
In consequence, we find that
\begin{align}
b = & - H _{\mu} ^{\ast} \big[ \mu _{AB} (\varphi, \phi) \big( \partial ^{\mu} \phi ^{A} \big) - k _{AB} A ^{\mu} \bar{n} ^{A} (\varphi, \phi) \big] \frac{\partial \bar{n} ^{B} (\varphi, \phi)}{\partial \varphi} \eta \nonumber \\
&- \big[ \mu _{AB} (\varphi, \phi) \big( \partial ^{\mu} \phi ^{A} \big) - \tfrac{1}{2} k _{AB} A ^{\mu} \bar{n} ^{A} (\varphi, \phi) \big] \bar{n} ^{B} (\varphi, \phi) A _{\mu} , \label{195}
\end{align}
and therefore the nonintegrated density of the second-order deformation splits into a sum of components with the antifield number equal to $0$ and respectively $1$
\begin{align}
b =& b_{0} + b_{1}, \label{196} \\
b_{0} =& - \mu _{AB} (\varphi, \phi) \big( \partial ^{\mu} \phi ^{A} \big) \bar{n} ^{B} (\varphi, \phi) A _{\mu} + \tfrac{1}{2} k _{AB} \bar{n} ^{A} (\varphi, \phi) \bar{n} ^{B} (\varphi, \phi) A ^{\mu} A _{\mu}, \label{197} \\
b_{1} =& H _{\mu} ^{\ast} \bigg[ - \mu _{AB} (\varphi, \phi) \big( \partial ^{\mu} \phi ^{A} \big) \frac{\partial \bar{n} ^{B} (\varphi, \phi)}{\partial \varphi} \nonumber \\
&+ \tfrac{1}{2} k _{AB} \frac{\partial \big(\bar{n} ^{A} (\varphi, \phi)\bar{n} ^{B} (\varphi, \phi)\big)}{\partial \varphi} A ^{\mu} \bigg]  \eta . \label{198}
\end{align}
From the previous expressions it follows that the second order of perturbation theory contributes only to the deformation of the gauge transformations corresponding to the one-form $H ^{\mu}$ from the BF sector and adds two kinds of vertices, which couple the BF to the matter fields and, essentially, meet all the requirements, including the derivative order assumption.

Next, we solve equation (\ref{44}), responsible for the nonintegrated density of the third-order deformation, $c$. In terms of notations (\ref{39})--(\ref{41}) and employing the results contained in formulas (\ref{48}), (\ref{51})--(\ref{56}), (\ref{174})--(\ref{176}), and (\ref{196})--(\ref{198}), by direct computation we arrive at
\begin{align}
\Lambda = & \tfrac{1}{2} \bigg( k _{AC} \frac{\partial \bar{n} ^{C} (\varphi, \phi)}{\partial \phi ^{B}} + k _{BC} \frac{\partial \bar{n} ^{C} (\varphi, \phi)}{\partial \phi ^{A}} \bigg) \bar{n} ^{A} (\varphi, \phi) \bar{n} ^{B} (\varphi, \phi) A^{\mu} A _{\mu} \eta \nonumber \\
& - \bigg( \mu _{AC} (\varphi, \phi) \frac{\partial \bar{n} ^{C} (\varphi, \phi)}{\partial \phi ^{B}} + \mu _{BC} (\varphi, \phi) \frac{\partial \bar{n} ^{C} (\varphi, \phi)}{\partial \phi ^{A}} \nonumber\\
& + \frac{\partial \mu _{AB} (\varphi, \phi)}{\partial \phi ^{C}} \bar{n} ^{C} (\varphi, \phi) + \frac{\partial \mu _{AB} (\varphi, \phi)}{\partial \varphi} W(\varphi ) \bigg) \big( \partial ^{\mu} \phi ^{A} \big) \bar{n} ^{B} (\varphi, \phi) A _{\mu} \eta \nonumber \\
& + s \bigg\{ - \tfrac{1}{2} \mu _{AB} (\varphi, \phi) A ^{\mu} \bigg[ H _{\mu} ^{\ast} \frac{\partial \big( \bar{n} ^{A} (\varphi, \phi) \bar{n} ^{B} (\varphi, \phi) \big)}{\partial \varphi} \eta \nonumber \\
& + \bar{n} ^{A} (\varphi, \phi) \bar{n} ^{B} (\varphi, \phi) A _{\mu} \bigg] \bigg\} . \label{199}
\end{align}
The terms contained in the first and respectively second and third line from the right-hand side of (\ref{199}) are identically vanishing on account of formulas (\ref{177}) and respectively (\ref{193}), such that equation (\ref{44}) becomes equivalent to
\begin{align}
s \bigg\{ c - \tfrac{1}{2} \mu _{AB} (\varphi, \phi) A ^{\mu} \bigg[ H _{\mu} ^{\ast} \frac{\partial \big( \bar{n} ^{A} (\varphi, \phi) \bar{n} ^{B} (\varphi, \phi) \big)}{\partial \varphi} \eta &\nonumber \\
+ \bar{n} ^{A} (\varphi, \phi) \bar{n} ^{B} (\varphi, \phi) A _{\mu} \bigg] \bigg\} &= \partial _{\mu} l ^{\mu}, \label{200}
\end{align}
from which we arrive at
\begin{equation}
c = \tfrac{1}{2} \mu _{AB} (\varphi, \phi) A ^{\mu} \bigg[ H _{\mu} ^{\ast} \frac{\partial \big( \bar{n} ^{A} (\varphi, \phi) \bar{n} ^{B} (\varphi, \phi) \big)}{\partial \varphi} \eta + \bar{n} ^{A} (\varphi, \phi) \bar{n} ^{B} (\varphi, \phi) A _{\mu} \bigg] . \label{201}
\end{equation}
In conclusion, the nonintegrated density of the third-order deformation decomposes, according to the distinct values of the antifield number, into a sum between two pieces
\begin{align}
c &= c _{0} + c _{1}, \label{202} \\
c _{0} &= \tfrac{1}{2} \mu _{AB} (\varphi, \phi) \bar{n} ^{A} (\varphi, \phi) \bar{n} ^{B} (\varphi, \phi) A ^{\mu} A _{\mu} , \label{203} \\
c _{1} &= \tfrac{1}{2} H _{\mu} ^{\ast} A ^{\mu} \mu _{AB} (\varphi, \phi) \frac{\partial \big( \bar{n} ^{A} (\varphi, \phi) \bar{n} ^{B} (\varphi, \phi) \big)}{\partial \varphi} \eta , \label{204}
\end{align}
where the functions $\bar{n} ^{A}$ of the form (\ref{177}) together with $\mu _{AB}$ are assumed to satisfy the consistency conditions (\ref{192}) and (\ref{193}). Analyzing (\ref{202})--(\ref{204}), we notice that the third order of perturbation theory deforms again only the gauge transformations of the BF one-form $H ^{\mu}$ and produces a single kind of cross-coupling vertices, while respecting all the working hypotheses.

Now, we pass to the fourth order of perturbation theory and solve equation (\ref{45}). In agreement with notations (\ref{39})--(\ref{41}) and by means of formulas (\ref{48}), (\ref{51})--(\ref{56}), (\ref{174})--(\ref{176}), (\ref{196})--(\ref{198}), and (\ref{202})--(\ref{204}), we infer that
\begin{align}
\Gamma =& \tfrac{1}{2} \bigg( \mu _{AC} (\varphi, \phi) \frac{\partial \bar{n} ^{C} (\varphi, \phi)}{\partial \phi ^{B}} + \mu _{BC} (\varphi, \phi) \frac{\partial \bar{n} ^{C} (\varphi, \phi)}{\partial \phi ^{A}} + \frac{\partial \mu _{AB} (\varphi, \phi)}{\partial \phi ^{C}} \bar{n} ^{C} (\varphi, \phi) \nonumber\\
& + \frac{\partial \mu _{AB} (\varphi, \phi)}{\partial \varphi} W(\varphi ) \bigg) \bar{n} ^{A} (\varphi, \phi) \bar{n} ^{B} (\varphi, \phi) A ^{\mu} A _{\mu} \eta , \label{205}
\end{align}
so it is identically vanishing since the parameterizing functions are solutions to the consistency conditions (\ref{193}) and thus equation (\ref{45}) takes the simple form
\begin{equation}
sd = \partial _{\mu} m ^{\mu}, \label{206}
\end{equation}
with $m ^{\mu}$ a local current. By virtue of (\ref{206}), the nonintegrated density of the fourth-order deformation of the solution to the master equation is spanned by the (nontrivial) elements from the cohomology $H^{0} (s|d)$ that comply with all the working hypotheses, which can be eliminated because they have already been considered once, at order one in the coupling constant, so we can choose
\begin{equation}
d = 0 \Leftrightarrow S _{4} =0 \label{207}
\end{equation}
without loss of new nontrivial deformations and also without further consistency conditions on the functions that parameterize the first-order deformation. Along the same line and using the results deduced until now, it is easy to see that all the remaining higher-order deformations can be taken to vanish
\begin{equation}
S _{k} =0,\qquad k>4. \label{208}
\end{equation}

Assembling the outcomes deduced so far via expansion (\ref{33}), we can state that the most general, nontrivial deformation of the solution to the master equation describing four-dimensional interactions among a topological BF theory and a (finite) set of massless real scalar fields that is consistent to all orders in the coupling constant and meanwhile displays all the required properties (analyticity in the deformation parameter, Lorentz covariance, spacetime locality, Poincar\'{e} invariance, and conservation of the number of derivatives on each field with respect to the free limit at the level of the deformed field equations), can be taken to stop at order three in the coupling constant
\begin{equation}
\bar{S} = S + \int d^4x [g(a ^{\mathrm{BF}}+ a ^{\mathrm{int}}) + g^2 b + g^3 c]. \label{209}
\end{equation}
In the above $S$ is the solution to the master equation in the absence of interactions, (\ref{32}), and the nonintegrated densities $a ^{\mathrm{BF}}$, $a ^{\mathrm{int}}$, $b$, and $c$ are expressed by formulas (\ref{51})--(\ref{56}), (\ref{174})--(\ref{176}), (\ref{196})--(\ref{198}), and respectively (\ref{202})--(\ref{204}). The smooth functions of the undifferentiated scalar fields from the theory involved in the deformation $\bar{S}$ are the solutions of the consistency equations (\ref{192}) and (\ref{193}), being understood that $\bar{n} ^{A}$ read like in (\ref{177}) and $\mu _{AB}$ are not trivial (see conditions (\ref{178})). Under these circumstances, in the sequel from (\ref{209}) we extract all the ingredients correlated with the Lagrangian formulation of the resulting interacting gauge theory and meanwhile emphasize some interesting solutions to the consistency equations together with their physical content.

\section{Main results: Lagrangian formulation of the interacting model(s)\label{inter}}

\subsection{General form of the Lagrangian action and gauge symmetries}

The Lagrangian action of the interacting gauge theory is recovered via those terms from (\ref{209}) that are both antifield- and ghost-independent
\begin{equation*}
\bar{S}^{\mathrm{L}}[\Phi^{\alpha_{0}}] = S^{\mathrm{L}}[\Phi^{\alpha_{0}}] + \int d^4x [g (a _{0} ^{\mathrm{BF}} + a _{0} ^{\mathrm{int}} ) + g^2 b _{0} + g^3 c _{0}],
\end{equation*}
with $S^{\mathrm{L}}[\Phi^{\alpha_{0}}]$ the free Lagrangian action (\ref{2}) and the interacting Lagrangian densities provided by formulas (\ref{52}), (\ref{175}), (\ref{197}), and respectively (\ref{203}), such that its concrete expression reads as
\begin{align}
\bar{S}^{\mathrm{L}}[\Phi^{\alpha_{0}}] =& \int d^4x \big\{ H^{\mu} \partial_{\mu }\varphi +\tfrac{1}{2}B^{\mu\nu}\partial_{[\mu }A_{\nu ]}+\tfrac{1}{2}k_{AB} \big( \partial_{\mu }\phi ^{A} \big) \big(\partial^{\mu }\phi ^{B} \big) \nonumber \\
& +g \big[ - W(\varphi) H^{\mu} A_{\mu} - \mathcal{V} (\varphi, \phi ) - k _{AB} \big( \partial ^{\mu } \phi ^{A} \big) \bar{n} ^{B}(\varphi, \phi) A_{\mu} \nonumber \\
& + \tfrac{1}{2} \mu _{AB} (\varphi, \phi ) \big( \partial _{\mu } \phi ^{A} \big) \big( \partial ^{\mu } \phi ^{B} \big) \big] + g^2 \big[ \tfrac{1}{2} k _{AB} A ^{\mu} \bar{n} ^{A} (\varphi, \phi) \nonumber\\
&- \mu _{AB} (\varphi, \phi) \big( \partial ^{\mu} \phi ^{A} \big) \big] \bar{n} ^{B} (\varphi, \phi) A _{\mu} \nonumber\\
&+ g^3 \tfrac{1}{2} \mu _{AB} (\varphi, \phi) \bar{n} ^{A} (\varphi, \phi) \bar{n} ^{B} (\varphi, \phi) A ^{\mu} A _{\mu} \big\} . \label{210}
\end{align}
We can alternatively write down the functional $\bar{S}^{\mathrm{L}}$ in a more compact form like
\begin{align}
\bar{S}^{\mathrm{L}}[\Phi^{\alpha_{0}}] = & \int d^4x \big[ H^{\mu} D _{\mu} \varphi +\tfrac{1}{2}B^{\mu\nu}\partial_{[\mu }A_{\nu ]} - g \mathcal{V} (\varphi, \phi ) \nonumber \\
&+ \tfrac{1}{2} (k _{AB} + g \mu _{AB} (\varphi, \phi)) \big( \hat{D} _{\mu} \phi ^{A} \big) \big( \hat{D} ^{\mu} \phi ^{B} \big) \big] , \label{211}
\end{align}
in terms of the ``covariant derivatives'' of the BF scalar field and respectively of the matter fields
\begin{align}
D _{\mu} \varphi &\equiv \partial _{\mu }\varphi - g W(\varphi ) A _{\mu} , \label{212} \\
\hat{D} _{\mu} \phi ^{A} &\equiv \partial_{\mu }\phi ^{A} - g \bar{n} ^{A} (\varphi, \phi) A _{\mu} . \label{213}
\end{align}
They are not standard covariant derivatives in the usual sense of field theory since they generate more that minimal couplings. In agreement with (\ref{210}) or (\ref{211}), the deformed Lagrangian contains: (a) a single class of vertices (derivative-free and of order $1$ in the coupling constant) that describes selfinteractions among the BF fields and is monitored by the function $W(\varphi)$ and (b) six families of vertices that couple the BF to the matter scalar fields, among which (b.1) three kinds without derivatives (one at each of orders $1$, $2$, and respectively $3$ of perturbation theory, with the last two types quadratic in the BF one-form $A _{\mu}$), (b.2) two types with a single derivative acting on the matter fields and simultaneously linear in the BF one-form $A _{\mu}$ (at orders $1$ and $2$ in $g$), and (b.3) one class with two derivatives acting only via terms that are quadratic in the first-order derivatives of the matter fields (at order $1$ in the deformation parameter). We remark that the cross-couplings between the BF and matter fields at the first order of perturbation theory are exhausted via the function $\mathcal{V} (\varphi, \phi )$, the nontrivial `kinetic' terms with respect to the matter fields, $(1/2)\mu _{AB} (\varphi, \phi ) \big( \partial _{\mu } \phi ^{A} \big) \big( \partial ^{\mu } \phi ^{B} \big)$, and also by the current-gauge field contribution induced by the presence of the nontrivial, one-dimensional rigid symmetry of the free action, $ - t _{0} ^{\mu} A_{\mu} = - k _{AB} \big( \partial ^{\mu } \phi ^{A} \big) \bar{n} ^{B}(\varphi, \phi) A_{\mu}$ (see formula (\ref{149}) from Appendix \ref{A}).

The gauge symmetries of action $\bar{S}^{\mathrm{L}}[\Phi^{\alpha_{0}}]$ are also deformed with respect to those corresponding to its free limit, $S^{\mathrm{L}}[\Phi^{\alpha_{0}}]$, due to the fact that functional (\ref{209}) collects (nontrivial) terms of antifield number $1$ in (all) the nonvanishing deformations of strictly positive orders, $d^4 x (g(a _{1} ^{\mathrm{BF}}+ a _{1} ^{\mathrm{int}})+ g^2 b _{1} + g^3 c_{1})$. Consequently, a generating set of gauge transformations for the coupled Lagrangian action is obtained by adding to (\ref{3})--(\ref{5}) the contributions resulting from the previously mentioned terms (see formulas (\ref{53}), (\ref{176}), (\ref{198}), and respectively (\ref{204})) via detaching the antifields and reverting the ghosts to the corresponding gauge parameters. Proceeding along this line, we find that $\bar{S}^{\mathrm{L}}[\Phi^{\alpha_{0}}]$ is invariant under the nontrivial, infinitesimal gauge transformations
\begin{align}
\bar{\delta} _{\Omega ^{\alpha_{1}}} \varphi =& g W(\varphi) \epsilon , \qquad \bar{\delta} _{\Omega ^{\alpha_{1}}} A _{\mu} = \partial _{\mu } \epsilon , \label{220} \\
\bar{\delta} _{\Omega ^{\alpha_{1}}} H^{\mu} =& -2 \partial _{\lambda} \xi ^{\lambda\mu} - g \frac{d W (\varphi)}{d \varphi} ( 2 A_{\lambda} \xi ^{\lambda \mu} + H ^{\mu} \epsilon ) \nonumber \\
&-g  (k _{AB} + g \mu _{AB} (\varphi, \phi)) \big( \hat{D} _{\mu} \phi ^{A} \big) \frac{\partial \bar{n} ^{B} (\varphi, \phi)}{\partial \varphi } \epsilon ,\label{221} \\
\bar{\delta} _{\Omega ^{\alpha_{1}}} B^{\mu\nu} =& -3 \partial _{\lambda} \epsilon ^{\lambda\mu\nu} + 2g W(\varphi) \xi ^{\mu \nu}, \label{222}\\
\bar{\delta} _{\Omega ^{\alpha_{1}}} \phi ^{A} =& g \bar{n} ^{A} (\varphi, \phi ) \epsilon . \label{223}
\end{align}
The previous gauge transformations exhibit several nice properties. Thus, only those of the BF vector field $A _{\mu}$ are not affected by the deformation procedure and reduce to the original $U(1)$ gauge transformation of parameter $\epsilon $. Meanwhile, the remaining BF fields, including the original, gauge-invariant scalar $\varphi$, gain nontrivial gauge transformations due to their selfinteractions (controlled by the function $W(\varphi)$ and its first-order derivative) strictly in the first order of perturbation theory. At the same time, the cross-couplings add nontrivial contributions (only via the gauge parameter $\epsilon $) to the gauge transformations of the BF vector field $H ^{\mu}$ at orders $1$, $2$, and $3$ and, most important, induce nontrivial gauge transformations of the matter fields (at order $1$). The terms generated in this context by the functions $\bar{n} ^{A}$ in the first order of perturbation theory, namely, $\bar{n} ^{A} \epsilon $ in (\ref{223}) and $-k _{AB} \big( \partial ^{\mu } \phi ^{A} \big) (\partial \bar{n} ^{B} / \partial \varphi ) \epsilon$ in (\ref{221}), are obtained precisely by gauging the nontrivial, one-parameter rigid symmetry constructed during the computation of the cross-coupling first-order deformation $a ^{\mathrm{int}}$ (see results (\ref{147}) and (\ref{148}) from Appendix \ref{A}). The associated conserved current, (\ref{149}), which is present in the interacting Lagrangian action at order one in $g$ via the term $-t_{0} ^{\mu} A_{\mu}$, is \emph{not} gauge invariant under these transformations. This result has two main consequences \emph{at the second order of perturbation theory}: the appearance of the term $(1/2) k _{AB}  \bar{n} ^{A} \bar{n} ^{B} A ^{\mu} A _{\mu}$ in the Lagrangian action (\ref{210}) and the introduction of the quantity $k _{AB} \bar{n} ^{A} (\partial \bar{n} ^{B} / \partial \varphi ) A ^{\mu} \epsilon $ into the gauge transformations (\ref{221}) of the one-form $H _{\mu}$. Moreover, there appears a rather unusual behavior related to the presence of $\mu _{AB}$: these functions cannot be involved in any gauge transformation at order $1$ since they stem from a first-order deformation of the solution to the master equation that is both antifield- and ghost-independent, but instead modify the gauge transformations of $H ^{\mu}$ at orders $2$ and $3$ and also contribute to cross-coupling vertices at the same orders.

The main properties of the deformed generating set of gauge transformations (\ref{220})--(\ref{223}), namely, the accompanying gauge algebra and reducibility, are investigated in Appendix \ref{B}.

\subsection{Solutions to the consistency equations: mass terms for the $U(1)$ vector field}

We recall that the entire Lagrangian formulation of the interacting theory is controlled by the functions $W(\varphi)$, $T ^{AB} (\varphi)$, $n ^{A} (\varphi)$, $\mathcal{V} (\varphi, \phi )$, and $\mu _{AB} (\varphi, \phi )$, which are restricted to satisfy the consistency equations (\ref{192}) and (\ref{193}), written in detail like
\begin{align}
\frac{\partial \mathcal{V} (\varphi, \phi )}{\partial \phi ^{A}} \big( n ^{A} (\varphi) + T ^{AB} (\varphi) k _{BC} \phi ^{C} \big) + \frac{\partial \mathcal{V} (\varphi, \phi )}{\partial \varphi} W(\varphi ) &=0, \label{246}\\
\mu _{AC} (\varphi, \phi) T^ {CD} (\varphi) k _{DB} + \mu _{BC} (\varphi, \phi) T^ {CD} (\varphi) k _{DA}& \nonumber \\
+ \frac{\partial \mu _{AB} (\varphi, \phi)}{\partial \phi ^{C}} \big( n ^{C} (\varphi) + T ^{CD} (\varphi) k _{DE} \phi ^{E} \big) + \frac{\partial \mu _{AB} (\varphi, \phi)}{\partial \varphi} W(\varphi ) &=0. \label{247}
\end{align}
Thus, our procedure is consistent provided these equations possess solutions. We give below two classes of solutions, in terms of which the deformed Lagrangian action, (\ref{211}), displays a mass term for the BF $U(1)$ vector field $A _{\mu}$.

\subsubsection{Type I solutions}

A first class of solutions to (\ref{246}) and (\ref{247}) is given by
\begin{gather}
W(\varphi) = \mathrm{arbitrary} \neq 0, \qquad T^{AB} (\varphi) = t ^{AB}, \qquad n ^{A} (\varphi) = m ^{A}, \label{I.1} \\
\mathcal{V} (\varphi, \phi) = \mathcal{V} (q), \qquad \mu _{AB} (\varphi, \phi) = k _{AB} \theta (\bar{q}), \label{I.2}
\end{gather}
where $t ^{AB} = - t ^{BA}$ as well as $m ^{A}$ are some nonvanishing, real constants, while $q$ and $\bar{q}$ read as
\begin{align}
q & = \tfrac{1}{2} k _{AB} \big( t _{\hphantom{A} C}^{A} \phi ^{C} + m ^{A}\big) \big( t _{\hphantom{B} D}^{B} \phi ^{D} + m ^{B} \big) , \label{I.3}\\
\bar{q} & \equiv q - \tfrac{1}{2} k _{AB} m ^{A} m ^{B} = k _{AB} t _{\hphantom{A} C}^{A} \phi ^{C} \big(  \tfrac{1}{2} t _{\hphantom{B} D}^{B} \phi ^{D} + m ^{B} \big), \label{I.4}
\end{align}
with $t _{\hphantom{A} C}^{A} = t ^{AE} k _{EC}$. In (\ref{I.2}) $\mathcal{V} (q)$ and $\theta (\bar{q})$ are some arbitrary, smooth functions of their arguments and, in addition, $\theta$ is constrained to satisfy
\begin{equation}
\theta (0) = 0. \label{I.5}
\end{equation}
Condition (\ref{I.5}) ensures that the function $\theta (\bar{q})$ contains no additive constants and, as a consequence, none of the functions $\mu _{AB}$ exhibits trivial components. Based on the above solutions and taking into account result (\ref{177}), we have that formula (\ref{213}) takes the particular form
\begin{equation}
\hat{D} _{(\mathrm{I})\mu} \phi ^{A} = \mathcal{D} _{(\mathrm{I})\mu} \phi ^{A} - g m ^{A} A _{\mu} , \label{I.6}
\end{equation}
in terms of the notation
\begin{equation}
\mathcal{D} _{(\mathrm{I})\mu} \phi ^{A} \equiv \partial _{\mu} \phi ^{A} - g t _{\hphantom{A} C}^{A} \phi ^{C} A _{\mu} . \label{I.7}
\end{equation}
Substituting solutions (\ref{I.1}) and (\ref{I.2}) together with (\ref{I.6}) in (\ref{211}), we arrive at
\begin{align}
\bar{S}^{\mathrm{L}} _{(\mathrm{I})} [\Phi^{\alpha_{0}}] = & \int d^4x \big[ H^{\mu} D _{\mu} \varphi +\tfrac{1}{2}B^{\mu\nu}\partial_{[\mu }A_{\nu ]} - g \mathcal{V} (q) \nonumber \\
&+ \tfrac{1}{2} k _{AB} (1 + g \theta (\bar{q})) \big( \mathcal{D} _{(\mathrm{I})\mu} \phi ^{A} -2g m ^{A} A _{\mu} \big) \mathcal{D} _{(\mathrm{I})} ^{\mu} \phi ^{B} \nonumber \\
&+ \tfrac{1}{2} g ^{2} k _{AB} m ^{A} m ^{B} A _{\mu} A ^{\mu} + \tfrac{1}{2} g ^{3} k _{AB} \theta (\bar{q}) m ^{A} m ^{B} A _{\mu} A ^{\mu} \big] . \label{I.8}
\end{align}
The same procedure applied to formulas (\ref{220})--(\ref{223}) reveals that action (\ref{I.8}) is now invariant under the gauge transformations
\begin{align}
\bar{\delta} _{(\mathrm{I})\Omega ^{\alpha_{1}}} \varphi =& g W(\varphi) \epsilon , \qquad \bar{\delta} _{(\mathrm{I})\Omega ^{\alpha_{1}}} A _{\mu} = \partial _{\mu } \epsilon , \label{I.9} \\
\bar{\delta} _{(\mathrm{I})\Omega ^{\alpha_{1}}} H^{\mu} =& -2 \partial _{\lambda} \xi ^{\lambda\mu} - g \frac{d W (\varphi)}{d \varphi} ( 2 A_{\lambda} \xi ^{\lambda \mu} + H ^{\mu} \epsilon ) ,\label{I.10} \\
\bar{\delta} _{(\mathrm{I})\Omega ^{\alpha_{1}}} B^{\mu\nu} =& -3 \partial _{\lambda} \epsilon ^{\lambda\mu\nu} + 2g W(\varphi) \xi ^{\mu \nu}, \label{I.11}\\
\bar{\delta} _{(\mathrm{I})\Omega ^{\alpha_{1}}} \phi ^{A} =& g \big( t _{\hphantom{A} B}^{A} \phi ^{B} + m ^{A} \big) \epsilon . \label{I.12}
\end{align}
Due to the fact that $k _{AB}$ was taken by assumption to be positively defined and $m ^{A}$ are nonvanishing, we find that
\begin{equation}
k _{AB} m ^{A} m ^{B} > 0. \label{I.13}
\end{equation}
As a result, the quantity
\begin{equation}
\tfrac{1}{2} g ^{2} k _{AB} m ^{A} m ^{B} A _{\mu} A ^{\mu} \equiv \tfrac{1}{2} g ^{2} M^{2} A _{\mu} A ^{\mu}  \label{I.14}
\end{equation}
from (\ref{I.8}) is precisely a mass term for the BF $U(1)$ vector field $A _{\mu}$. At the same time, we remark that the object
\begin{equation}
\tfrac{1}{2} g ^{3} k _{AB} \theta (\bar{q}) m ^{A} m ^{B} A _{\mu} A ^{\mu}  \label{I.15}
\end{equation}
cannot generate mass for $A _{\mu}$ due to the fact that $\theta (\bar{q})$ contains no additive constants (see requirement (\ref{I.5})), so (\ref{I.14}) is indeed the only mass term present in (\ref{I.8}).

We notice that in the case of type I solutions the mass term coexists with the BF selfinteractions (generated by the nonvanishing function $W(\varphi )$.

\subsubsection{Type II solutions}

The second class of solutions to the consistency equations (\ref{246}) and (\ref{247}) reads as
\begin{gather}
W(\varphi) = 0, \qquad T^{AB} (\varphi) = \mathrm{arbitrary}, \quad A,B = \overline{1,N}, \label{I.16}\\
n ^{A} (\varphi) = m ^{A} + \tilde{n} ^{A} (\varphi), \qquad \tilde{n} ^{A} (\varphi) = \mathrm{arbitrary}, \quad A = \overline{1,N}, \label{I.17} \\
\mathcal{V} (\varphi, \phi) = \mathcal{V} (p), \qquad \mu _{AB} (\varphi, \phi) = k _{AB} \tau(\bar{p}), \label{I.18}
\end{gather}
where $m ^{A}$ are some real constants with the same property like before and $p$ together with $\bar{p}$ take the form
\begin{align}
p = & \tfrac{1}{2} k _{AB} \big( T _{\hphantom{A} C}^{A} (\varphi) \phi ^{C} + \tilde{n} ^{A} (\varphi) + m ^{A}\big) \big( T _{\hphantom{B} D}^{B} (\varphi) \phi ^{D} + \tilde{n} ^{B} (\varphi) + m ^{B} \big) , \label{I.19}\\
\bar{p} \equiv & p - \tfrac{1}{2} k _{AB} \big( \tilde{n} ^{A} (\varphi) + m ^{A}\big) \big( \tilde{n} ^{B} (\varphi) + m ^{B} \big) \nonumber \\
= & k _{AB} T _{\hphantom{A} C}^{A} (\varphi) \phi ^{C} \big(  \tfrac{1}{2} T _{\hphantom{B} D}^{B} (\varphi) \phi ^{D} + \tilde{n} ^{B} (\varphi) + m ^{B} \big). \label{I.20}
\end{align}
In (\ref{I.18}) $\mathcal{V} (p)$ and $\tau (\bar{p})$ are some arbitrary, smooth functions of their arguments and, moreover, $\tau$ is asked to fulfill
\begin{equation}
\tau (0) = 0. \label{I.21}
\end{equation}
The last requirement grants in this context that $\tau (\bar{p})$ contains no additive constants, so $\mu _{AB}$ are nontrivial as well. With the help of the previous solutions and relying on result (\ref{177}), we then find that (\ref{213}) becomes in this case
\begin{equation}
\hat{D} _{(\mathrm{II})\mu} \phi ^{A} = \mathcal{D} _{(\mathrm{II})\mu} \phi ^{A} - g m ^{A} A _{\mu} , \label{I.22}
\end{equation}
where we employed the notation
\begin{equation}
\mathcal{D} _{(\mathrm{II})\mu} \phi ^{A} \equiv \partial _{\mu} \phi ^{A} - g \big( T _{\hphantom{A} B}^{A} (\varphi) \phi ^{B} + \tilde{n} ^{A} (\varphi) \big) . \label{I.23}
\end{equation}
Inserting (\ref{I.16})--(\ref{I.18}) and (\ref{I.22}) into (\ref{211}), we obtain the expression of the interacting Lagrangian action in this particular case
\begin{align}
\bar{S}^{\mathrm{L}} _{(\mathrm{II})} [\Phi^{\alpha_{0}}] = & \int d^4x \big[ H^{\mu} \partial _{\mu} \varphi +\tfrac{1}{2}B^{\mu\nu}\partial_{[\mu }A_{\nu ]} - g \mathcal{V} (p) \nonumber \\
&+ \tfrac{1}{2} k _{AB} (1 + g \tau (\bar{p})) \big( \mathcal{D} _{(\mathrm{II})\mu} \phi ^{A} -2g m ^{A} A _{\mu} \big) \mathcal{D} _{(\mathrm{II})} ^{\mu} \phi ^{B} \nonumber \\
&+ \tfrac{1}{2} g ^{2} k _{AB} m ^{A} m ^{B} A _{\mu} A ^{\mu} + \tfrac{1}{2} g ^{3} k _{AB} \tau (\bar{p}) m ^{A} m ^{B} A _{\mu} A ^{\mu} \big] . \label{I.24}
\end{align}
Acting along the same line with respect to relations (\ref{220})--(\ref{223}), we get that action (\ref{I.24}) displays the generating set of gauge transformations
\begin{align}
\bar{\delta} _{(\mathrm{II})\Omega ^{\alpha_{1}}} \varphi =& 0 , \qquad \bar{\delta} _{(\mathrm{II})\Omega ^{\alpha_{1}}} A _{\mu} = \partial _{\mu } \epsilon , \label{I.25} \\
\bar{\delta} _{(\mathrm{II})\Omega ^{\alpha_{1}}} H^{\mu} =& -2 \partial _{\lambda} \xi ^{\lambda\mu} \nonumber \\
&-g k _{AB} (1 + g \tau (\bar{p})) \big( \hat{D} _{(\mathrm{II})\mu} \phi ^{A} \big) \frac{\partial \hat{n} ^{B} (\varphi, \phi)}{\partial \varphi} \epsilon ,\label{I.26} \\
\bar{\delta} _{(\mathrm{II})\Omega ^{\alpha_{1}}} B^{\mu\nu} =& -3 \partial _{\lambda} \epsilon ^{\lambda\mu\nu}, \label{I.27}\\
\bar{\delta} _{(\mathrm{II})\Omega ^{\alpha_{1}}} \phi ^{A} =& g \big( \hat{n} ^{A} (\varphi, \phi) + m ^{A} \big) \epsilon , \label{I.28}
\end{align}
where
\begin{equation}
\hat{n} ^{A} (\varphi, \phi) \equiv T _{\hphantom{A} B}^{A} (\varphi) \phi ^{B} + \tilde{n} ^{A} (\varphi) . \label{I.29}
\end{equation}
Exactly like in the previous case, the component
\begin{equation}
\tfrac{1}{2} g ^{2} k _{AB} m ^{A} m ^{B} A _{\mu} A ^{\mu} \equiv \tfrac{1}{2} g ^{2} M^{2} A _{\mu} A ^{\mu}  \label{I.30}
\end{equation}
entering (\ref{I.24}) is nothing but a mass term for the BF $U(1)$ vector field $A _{\mu}$, while
\begin{equation}
\tfrac{1}{2} g ^{3} k _{AB} \tau (\bar{p}) m ^{A} m ^{B} A _{\mu} A ^{\mu}  \label{I.31}
\end{equation}
cannot generate mass for $A _{\mu}$ due to condition (\ref{I.21}).

The mass term specific to type II solutions appears in the absence of BF selfinteractions, dictated by the choice $W(\varphi )=0$.

\section{Conclusions and comments\label{concl}}

To conclude with, in this paper we have investigated the couplings between a topological BF model with a maximal field spectrum (a scalar field, two sorts of vector fields, and a two-form gauge field) and a set of massless real scalar fields by means of the deformation of the solution to the master equation with the help of local BRST cohomology. Initially, we constructed the concrete form of the deformed solution to the master equations (that can be taken to stop at order three in the coupling constant) and obtained that it is parameterized by five kinds of functions (depending on the undifferentiated scalar fields form the theory), which are restricted to satisfy some consistency equations. Next, from the above deformed solution we derived the general Lagrangian formulation of the interacting theory (Lagrangian action, gauge symmetries, gauge algebra, reducibility relations). Finally, we gave two types of solutions to the consistency equations, which led to two classes of gauge invariant interacting theories with a mass term for the BF $U(1)$ vector field $A_{\mu }$. We mention that the mass term emerged naturally and did not follow from the Higgs mechanism.

Actually, the mass term for the $U(1)$ vector field (see (\ref{I.8}) and (\ref{I.24})) originates in our approach from the quantity
\begin{equation}
\int d^{4}x \tfrac{1}{2} g^{2} k_{AB} n^{A}(\varphi ) n^{B}(\varphi ) A_{\mu } A^{\mu }  \label{C1}
\end{equation}
present in (\ref{211}) particularized to type I and type II solutions (see (\ref{I.1})--(\ref{I.5}) and respectively (\ref{I.16})--(\ref{I.21})). On the one hand, none of the functions that parameterize action (\ref{211}), excepting $n^{A}(\varphi )$, particularized to type I and type II solutions contribute
to the mass of the $U(1)$ vector field. On the other hand, the existence of the functions $n^{A}(\varphi )$ in (\ref{211}) is a consequence of the existence of the one-parameter global symmetry
\begin{align}
\Delta _{\Upsilon } \phi ^{A} & =\big[ n^{A} (\varphi ) + T^{AB} (\varphi ) k_{BC} \phi ^{C} \big] \Upsilon ,  \label{C2} \\
\Delta _{\Upsilon } H^{\mu }& =-k_{AB} \big( \partial ^{\mu } \phi ^{A} \big) \bigg[ \frac{\partial n^{B} (\varphi )}{\partial \varphi } + \frac{\partial T^{BC} (\varphi )}{\partial \varphi } k_{CD} \phi ^{D} \bigg]\Upsilon  \label{C3}
\end{align}
of (the free) action (\ref{2}) (see the discussion from Appendix \ref{A}). Therefore, the appearance of this mass term is a direct consequence of the deformation method employed here in the context of the free limit described by action (\ref{2}).

Let us consider now a free action involving massive matter scalar fields, of the form
\begin{align}
S^{{\rm L}}[\Phi ^{\alpha _{0}}] = & \int d^{4}x \big[ H^{\mu } \partial _{\mu } \varphi +\tfrac{1}{2} B^{\mu \nu } \partial _{[\mu } A_{\nu ]} + \tfrac{1}{2} k_{AB} \big( \partial _{\mu } \phi ^{A} \big) \big( \partial ^{\mu } \phi ^{B} \big)  \nonumber \\
&-\tfrac{1}{2} \mu ^{2} k_{AB} \phi ^{A} \phi ^{B} \big] \nonumber\\
\equiv & S ^{\mathrm{L,BF}} [\varphi , A_{\mu }, H^{\mu }, B^{\mu \nu }]+S ^{\mathrm{L, scalar}} [\phi ], \label{C4}
\end{align}
where $\mu ^{2}$ is a real, strictly positive constant. Action (\ref{C4}) admits the one-parameter global symmetry
\begin{align}
\Delta _{\Upsilon } \phi ^{A} & =T^{AB} (\varphi ) k_{BC}\phi ^{C}\Upsilon , \label{C5} \\
\Delta _{\Upsilon } H^{\mu }& =-k_{AB} \big( \partial ^{\mu }\phi ^{A}\big) \frac{\partial T^{BC} (\varphi )}{\partial \varphi } k_{CD}\phi ^{D} \Upsilon , \label{C6}
\end{align}
where the functions $n^{A}(\varphi )$ are no longer allowed. The presence of these functions in (\ref{C5}) and (\ref{C6}) is forbidden precisely by the mass term for the matter scalars $\phi ^{A}$. Thus, if we apply the deformation procedure starting from action (\ref{C4}), then we infer no mass term for the $U(1)$ vector field. These considerations justify once more the importance of the free limit (\ref{2}) in view of obtaining a mass term for $A_{\mu }$.

The fact that only the $U(1)$ vector field from the BF field spectrum gains mass is encouraging since it opens the perspective of a mass generation mechanism for gauge vector fields through a procedure similar to that applied here, but in the presence of a free limit describing a collection of (massless) Maxwell vectors and a set of massless real scalars. The successful solving of the last problem may enlighten certain aspects of the results following from the Higgs mechanism based on spontaneous symmetry breaking. The last issue will be reported elsewhere~\cite{inprep}.

\appendix

\section{Computation of $a ^{\mathrm{int}}$ \label{A}}

Here we generate $a ^{\mathrm{int}}$ as in (\ref{57}) via its components of antifield number $1$ and respectively $0$ computed as the general solutions to equations (\ref{61}) and (\ref{62}) that in addition fulfill the specific set of rules invoked in the preamble of section \ref{first}.

As it has been noticed before (see the paragraph following equation (\ref{62})), the solution $a _{1}^{\mathrm{int}}$ to equation (\ref{61}) is a nontrivial element of the cohomology of $\gamma $ in $\pgh = 1$ computed in the algebra of local nonintegrated densities in the framework of the working hypotheses, $H^{1}(\gamma )$. We remind that the entire cohomology algebra of the longitudinal exterior differential computed in a given algebra, $H(\gamma )$, is defined like the equivalence classes of $\gamma $-closed elements from that algebra modulo $\gamma $-exact ones and inherits from $\gamma $ the $\mathbf{N}$-grading in terms of the pure ghost number $\pgh$. Focusing on definitions (\ref{28})--(\ref{31}) plus the observation that there exist no objects of strictly negative pure ghost number constructed out of the BRST generators, we obtain that the general solution to equation (\ref{61}) in the above mentioned algebra reads as
\begin{equation}
a _{1}^{\mathrm{int}}=f _{1}\big( [\Phi^{\ast}_{\alpha_{0}}] _{\mathrm{lin}}, [\partial _{[\mu}A _{\nu ]}], [\partial _{\mu} H ^{\mu}], [\partial _{\nu} B ^{\nu \mu}],[\varphi ], [\phi ] \big)\eta . \label{63}
\end{equation}
The notation $f([y])$ signifies that $f$ depends on $y$ and its spacetime derivatives up to a finite order. All the antifields $\chi^{\ast}_{\Delta}$ and their spacetime derivatives of arbitrarily high, but finite, orders are nontrivial elements of $H(\gamma )$ in pure ghost number $0$ (see the latter notation from (\ref{18}), the first relation in (\ref{21}), and the first definition from (\ref{28})). Nevertheless, the antifield number of $a _{1}^{\mathrm{int}}$ is fixed to $1$, such that the dependence on $\chi^{\ast}_{\Delta}$ from (\ref{63}) is limited to a monomial of degree one simultaneously in both the antifields of the original fields $\Phi^{\ast}_{\alpha_{0}}$ and their spacetime derivatives up to a finite order since these are the only objects of antifield number equal to $1$ available here (see the latter notation in (\ref{14}) and the second relation from (\ref{20})). This monomial dependence of degree one is symbolized in (\ref{63}) by $[\Phi^{\ast}_{\alpha_{0}}] _{\mathrm{lin}}$. The objects
\begin{equation}
\big\{ \partial _{[\mu}A _{\nu ]}, \partial _{\mu} H ^{\mu}, \partial _{\nu} B ^{\nu \mu}, \varphi, \phi \big\} \equiv \omega ^{\Theta} \label{64}
\end{equation}
together with their spacetime derivatives stand for the only gauge-invariant quantities of the starting model (\ref{2})--(\ref{5}) and thus produce all the (obviously nontrivial) elements from $H(\gamma )$ constructed out of the original fields $\Phi^{\alpha_{0}}$ (see the former notation in (\ref{14})). Their pure ghost number is also equal to $0$, so actually $f _{1}\big( [\Phi^{\ast}_{\alpha_{0}}] _{\mathrm{lin}}, [\partial _{[\mu}A _{\nu ]}], [\partial _{\mu} H ^{\mu}], [\partial _{\nu} B ^{\nu \mu}],[\varphi ], [\phi ] \big)$ yields the most general representative of $H^{0}(\gamma )$ of definite antifield number, equal to $1$. We notice from (\ref{63}) that the sole allowed dependence of $a _{1}^{\mathrm{int}}$ on the ghosts is actually linear in the undifferentiated ghost $\eta $ of pure ghost number $1$ (and of $\agh $ equal to $0$), which corresponds to the $U(1)$ gauge symmetry of action (\ref{2}) due to the gauge transformation of the vector field $A _{\mu }$ (see the latter relation from (\ref{3})). Indeed, we established that $f _{1}$ exhibits $\pgh (f _{1})=0$ and $\agh (f _{1})=1$. Consequently, the entire nontrivial dependence of $a _{1}^{\mathrm{int}}$ on the ghosts ensuring (\ref{61}) remains to be a monomial of degree one in all nontrivial $\gamma $-closed linear combinations constructed out of the ghosts of pure ghost number equal to $1$, $\eta ^{\alpha _{1}}$, introduced in (\ref{15}). In agreement with the actions of $\gamma $ on $\eta ^{\alpha _{1}}$, expressed by definitions (\ref{30}), the $\gamma $-closed linear combinations of $\eta ^{\alpha _{1}}$ are spanned by $\eta $, $\partial _{\lambda } C^{\lambda\mu }$, $\partial _{\lambda } \eta ^{\lambda\mu \nu}$, and their spacetime derivatives. Nevertheless, $\partial _{\mu} \eta $, $\partial _{\lambda } C^{\lambda\mu }$, $\partial _{\lambda } \eta ^{\lambda\mu \nu}$, and their spacetime derivatives are trivial in $H^1(\gamma )$ respectively due to the last relation in (\ref{28}) and the first two definitions from (\ref{29}), which leaves us with the result that the nontrivial part of $a _{1}^{\mathrm{int}}$ is truly linear in the undifferentiated $U(1)$ ghost $\eta $.

We can still eliminate some trivial terms from expression (\ref{63}) and meanwhile bring it to a more accessible form. In fact, by simple manipulations, like integrations by parts, one can remove the spacetime derivatives to act on the antifields $\Phi^{\ast}_{\alpha_{0}}$ modulo adding some (trivial) divergences and (irrelevant) $\gamma $-exact terms. Indeed, assuming that the maximum derivative order of $f _{1}$ with respect to the antifields $\Phi^{\ast}_{\alpha_{0}}$ is equal to $l$ and employing notation (\ref{64}), we have that
\begin{align}
f _{1}\big( [\Phi^{\ast}_{\alpha_{0}}] _{\mathrm{lin}}, [\omega ^{\Theta}] \big) = & \Phi^{\ast}_{\alpha_{0}} f ^{\alpha_{0}}\big( [\omega ^{\Theta}] \big) + \big( \partial ^{\mu} \Phi^{\ast}_{\alpha_{0}} \big) f _{\mu} ^{\alpha_{0}} \big( [\omega ^{\Theta}] \big) + \cdots \nonumber \\
&+ \big( \partial ^{\mu _{1}\ldots \mu _{l}} \Phi^{\ast}_{\alpha_{0}} \big) f _{\mu _{1}\ldots \mu _{l}} ^{\alpha_{0}} \big( [\omega ^{\Theta}] \big). \label{65}
\end{align}
All the coefficients denoted by $f$ from the right-hand side of (\ref{65}) depend now only on the objects from (\ref{64}) and their spacetime derivatives up to a finite order, so they are gauge invariant or, in other words, nontrivial elements of $H (\gamma )$ with both $\pgh $ and $\agh $ equal to $0$. Inserting (\ref{65}) into (\ref{63}), moving the derivatives from the antifields, and using the last definition from (\ref{28}) together with the $\gamma $-closeness of all $f$'s, we finally arrive at
\begin{align}
a _{1}^{\mathrm{int}}= &\Phi^{\ast}_{\alpha_{0}} \bar{f} ^{\alpha_{0}}\big( [\omega ^{\Theta}] \big) \eta + \gamma \big(- w _{\mu} \big( [\Phi^{\ast}_{\alpha_{0}}] _{\mathrm{lin}}, [\omega ^{\Theta}] \big) A ^{\mu} \big) \nonumber \\
&+ \partial ^{\mu} \big( w _{\mu} \big( [\Phi^{\ast}_{\alpha_{0}}] _{\mathrm{lin}}, [\omega ^{\Theta}] \big) \eta \big) ,\label{66}
\end{align}
where
\begin{equation}
\bar{f} ^{\alpha_{0}}\big( [\omega ^{\Theta}] \big) = f ^{\alpha_{0}}\big( [\omega ^{\Theta}] \big) - \partial ^{\mu} f _{\mu} ^{\alpha_{0}} \big( [\omega ^{\Theta}] \big) + \cdots + (-) ^{l} \partial ^{\mu _{1}\ldots \mu _{l}} f _{\mu _{1}\ldots \mu _{l}} ^{\alpha_{0}} \big( [\omega ^{\Theta}] \big) \label{67}
\end{equation}
and $w _{\mu}$ is a local, $\gamma $-closed current with $\agh =1$ and $\pgh =0$, which starts like
\begin{equation}
w _{\mu} \big( [\Phi^{\ast}_{\alpha_{0}}] _{\mathrm{lin}}, [\omega ^{\Theta}] \big) = \Phi^{\ast}_{\alpha_{0}} f _{\mu} ^{\alpha_{0}}\big( [\omega ^{\Theta}] \big) + \cdots . \label{68}
\end{equation}
Analyzing (\ref{66}), it is clear that one can eliminate both the $\gamma $-exact term and respectively the divergence from its right-hand side without modifying either the cohomological class of $a _{1}^{\mathrm{int}}$ from $H ^{1}(\gamma)$ or respectively the cohomological class of $a ^{\mathrm{int}}$ (of the form (\ref{57})) from $H ^{0} (s|d)$. By virtue of this result, from now on we work with the general solution to equation (\ref{61}) in the form (obtained from (\ref{66}) without trivial terms and with $\bar{f} ^{\alpha_{0}}$ renamed by $f ^{\alpha_{0}}$)
\begin{equation}
a _{1}^{\mathrm{int}}= f _{1}\big( \Phi^{\ast}_{\alpha_{0}} , [\omega ^{\Theta}] \big) \eta, \qquad f _{1}\big( \Phi^{\ast}_{\alpha_{0}} , [\omega ^{\Theta}] \big) \equiv \Phi^{\ast}_{\alpha_{0}} f ^{\alpha_{0}}\big( [\omega ^{\Theta}] \big) , \label{69}
\end{equation}
where $f _{1}$, with $\agh =1$ and $\pgh =0$, is linear now only in the undifferentiated antifields corresponding to the original fields.

In order to solve equation (\ref{62}) we act with $\delta $ on (\ref{69}) and employ the last definition from (\ref{28}), which produces a necessary condition related to the existence of $a _{0}^{\mathrm{int}}$
\begin{equation}
\delta f _{1}\big( \Phi^{\ast}_{\alpha_{0}}, [\omega ^{\Theta}] \big) = \partial _{\mu} t _{0} ^{\mu}\big( [\Phi ^{\alpha_{0}}] \big) , \label{70}
\end{equation}
where the current $t _{0} ^{\mu}$ (with $\agh = 0$) should be local in the original fields and their spacetime derivatives. (There is no a priori reason to force this current to be gauge invariant, i.e., to depend on $[\omega ^{\Theta}]$.) We remark that condition (\ref{70}) does not depend on the ghosts or, equivalently, the pure ghost number of its both hand sides is equal to $0$.

Equation (\ref{70}) expressing the necessary condition on the existence of $a _{0}^{\mathrm{int}}$ has a precise cohomological content: it requires that the double $\{f _{1},t _{0} ^{\mu}\}$ defines (by a one-to-one correspondence) an element $\Big\{\overset{[4]}{f} _{1}, \overset{[3]}{t} _{0}\Big\}$ from the kernel of $\delta $ modulo $d$ in antifield number $1$ and in maximum form degree ($\deg $) computed in the algebra of local forms that are ghost-independent, $(\mathrm{Ker}(\delta |d))^{4} _{1}$:
\begin{gather}
\Big\{\overset{[4]}{f} _{1}, \overset{[3]}{t} _{0}\Big\} \in (\mathrm{Ker}(\delta |d))^{4} _{1} \Leftrightarrow  \delta \overset{[4]}{f} _{1} = d \overset{[3]}{t} _{0}, \label{71}\\
\overset{[4]}{f} _{1} \equiv f _{1}\big( \Phi^{\ast}_{\alpha_{0}}, [\omega ^{\Theta}] \big)d ^4 x, \qquad \overset{[3]}{t} _{0} \equiv \tfrac{1}{3!} \varepsilon _{\nu\rho\lambda\sigma} t _{0} ^{\sigma}\big( [\Phi ^{\alpha_{0}}] \big)dx ^{\nu} dx ^{\rho} dx ^{\lambda} . \label{72}
\end{gather}
In the above the overscript between brackets symbolizes the form degree and the lower numeric index is assigned, like before, to the antifield number. The operator $\delta $ is extended to the (supercommutative) algebra of local forms with coefficients that are ghost-independent by $\delta (dx^{\mu})=0$, so it still defines a differential, the spacetime differential $d$ is taken to act as a right derivation (such that it anticommutes with $s$, $\delta$, and $\gamma $), and $\varepsilon _{\nu\rho\lambda\sigma}$ signify the components of the four-dimensional Levi--Civita symbol. We omitted the wedge product symbol in the last formula from (\ref{72}) since there is no danger of confusion, i.e., it is understood that $dx ^{\nu} dx ^{\rho} dx ^{\lambda} \equiv dx ^{\nu} \wedge dx ^{\rho} \wedge dx ^{\lambda}$. It is the anticommuting property of the two differentials $\delta $ and $d$ on the above mentioned supercommutative algebra ($\delta ^2 =0$, $d^2=0$, $\delta d + d\delta =0$), endowed in this context with two main $\mathbb{N}$-gradings ($\agh$ with respect to $\delta$, $\agh (\delta ) =-1$ and $\deg$ in relation with $d$, $\deg (d) =1$) that do not interfere ($\agh (d) = 0 = \deg (\delta )$), which ensures the correct construction of the homology of $\delta$ modulo $d$, also known as the local homology of the Koszul--Tate differential and traditionally denoted by $H(\delta | d)$, and respectively the cohomology of $d$ modulo $\delta $, $H(d|\delta )$. We insist on the fact that here we work on the algebra of local forms that do not depend on the ghosts (at $\pgh =0$) since otherwise, if we allow the coefficients of the local forms to depend also on the ghost fields, then the homology $H(\delta | d)$ for both strictly positive values of $\agh $ \emph{and} $\pgh $ vanishes~\cite{CMP1995a,PR2000}.

On the other hand, the nontriviality of $a ^{\mathrm{int}}$ in $H^0(s|d)$ induces that $\Big\{\overset{[4]}{f} _{1}, \overset{[3]}{t} _{0}\Big\}$ should belong to a nontrivial class of the local homology of the Koszul--Tate differential in maximum form degree ($4$) and in antifield number $1$, $H _{1} ^{4}(\delta | d)$. A trivial element from $H _{1} ^{4}(\delta | d)$, $\Big\{\overset{[4]}{f} _{\mathrm{triv},1}, \overset{[3]}{t} _{\mathrm{triv},0} \Big\}$, is defined in the standard manner like an element from the image of $\delta $ modulo $d$ in antifield number $1$ and in maximum form degree, $(\mathrm{Im}(\delta |d))^{4} _{1}$:
\begin{equation}
\Big\{\overset{[4]}{f} _{\mathrm{triv},1}, \overset{[3]}{t} _{\mathrm{triv},0} \Big\} \in (\mathrm{Im}(\delta |d))^{4} _{1} \Leftrightarrow
\left\{
\begin{array}{l}
\overset{[4]}{f} _{\mathrm{triv},1} = \delta \overset{[4]}{g} _{2} + d \overset{[3]}{u} _{1},\\
\overset{[3]}{t} _{\mathrm{triv},0} = -\delta \overset{[3]}{u} _{1} + d \overset{[2]}{v} _{0} ,
\end{array}
\right.
\label{73}
\end{equation}
with $\overset{[4]}{g} _{2}$, $\overset{[3]}{u} _{1}$, and $\overset{[2]}{v} _{0}$ some ghost-independent, local forms of fixed form degrees and antifield numbers, such that $\delta \overset{[4]}{f} _{\mathrm{triv},1} \equiv d \overset{[3]}{t} _{\mathrm{triv},0}$. Going back to dual notations, the nontriviality of $\Big\{\overset{[4]}{f} _{1}, \overset{[3]}{t} _{0}\Big\}$ in $H _{1} ^{4}(\delta | d)$ is equivalent to the nontriviality of the double $\{f _{1},t _{0} ^{\mu}\}$. A double $\{f _{\mathrm{triv},1},t _{\mathrm{triv},0} ^{\mu}\}$ is said to be trivial in this context if and only if
\begin{equation}
f _{\mathrm{triv},1} = \delta g _{2} + \partial _{\mu} u _{1} ^{\mu}, \qquad t _{\mathrm{triv},0} ^{\mu} = \delta u _{1} ^{\mu} + \partial _{\nu} v _{0} ^{\nu\mu}, \label{74}
\end{equation}
where the nonintegrated density $g _2$, the current $u _{1} ^{\mu}$, and the two-tensor $v _{0} ^{\nu\mu}$ are local and ghost-independent, with $v _{0} ^{\nu\mu}$ antisymmetric, $v _{0} ^{\nu\mu}=-v _{0} ^{\mu\nu}$, such that $\delta f _{\mathrm{triv},1} \equiv \partial _{\mu} t _{\mathrm{triv},0} ^{\mu}$. The equivalence between expressions (\ref{73}) and (\ref{74}) follows immediately if we work in dual notations
\begin{gather}
\overset{[4]}{f} _{\mathrm{triv},1} \equiv f _{\mathrm{triv},1} d^4x, \quad \overset{[3]}{t} _{\mathrm{triv},0} \equiv \tfrac{1}{3!} \varepsilon _{\nu\rho\lambda\sigma} t _{\mathrm{triv},0} ^{\sigma} dx ^{\nu} dx ^{\rho} dx ^{\lambda}, \label{75}\\
\overset{[4]}{g} _{2} \equiv g _{2} d^4x, \quad \overset{[3]}{u} _{1} \equiv \tfrac{1}{3!} \varepsilon _{\nu\rho\lambda\sigma} u _{1} ^{\sigma} dx ^{\nu}  dx ^{\rho} dx ^{\lambda}, \quad \overset{[2]}{v} _{0} \equiv \tfrac{1}{4} \varepsilon _{\rho\lambda\sigma\theta}v _{0} ^{\sigma\theta} dx ^{\rho} dx ^{\lambda}. \label{76}
\end{gather}

The above discussion on the cohomological interpretation of equation (\ref{70}) implies two things: (i) its solution is unique only up to the addition of trivial elements
\begin{equation}
f _{1} \rightarrow f _{1} ^{\prime} = f _{1} + \delta g _{2} + \partial _{\mu} u _{1} ^{\mu}, \qquad t _{0} ^{\mu} \rightarrow t _{0} ^{\prime \mu} = t _{0} ^{\mu} + \delta u _{1} ^{\mu} + \partial _{\nu} v _{0} ^{\nu\mu}, \label{77}
\end{equation}
which does not change the class from $H _{1} ^{4}(\delta | d)$, $\delta f _{1} ^{\prime} - \partial _{\mu} t _{0} ^{\prime \mu} \equiv \delta f _{1} - \partial _{\mu} t _{0} ^{\mu} =0$, and (ii) if $\{f _{1},t _{0} ^{\mu}\}$ is found to be completely trivial, i.e., of the form (\ref{74}), then the corresponding $a _{1}^{\mathrm{int}}$ like in (\ref{69}) can be safely removed from the first-order deformation (\ref{57}).

There is also a crucial physical content of equation (\ref{70}). In agreement with the general results from~\cite{CMP1995a,PR2000}, the spaces $(\mathrm{Ker}(\delta |d))^{4} _{1}$ and respectively $(\mathrm{Im}(\delta |d))^{4} _{1}$ are in a bijective correspondence to the set of global symmetries and respectively of trivial global symmetries associated with the Lagrangian action (\ref{2}), such that the factor space $H _{1} ^{4}(\delta | d)\equiv (\mathrm{Ker}(\delta |d))^{4} _{1}/(\mathrm{Im}(\delta |d))^{4} _{1}$ is in a one-to-one correspondence with the inequivalent (nontrivial) rigid symmetries of action (\ref{2}). A global symmetry of a given action is said to be trivial if it coincides with a gauge symmetry (possibly modulo on-shell trivial gauge symmetries). The above correspondence can be easily exemplified in our setting by going back to dual notations, replacing $f _{1}$ with its expression from (\ref{69}), and acting with $\delta $ on it via definitions (\ref{24})--(\ref{25}) written compactly in terms of the EL derivatives of action (\ref{2}) as
\begin{equation}
\delta \Phi^{\ast}_{\alpha_{0}} = -\frac{\delta S^{\mathrm{L}}[\Phi^{\beta_{0}}]}{\delta \Phi^{\alpha_{0}}}. \label{78}
\end{equation}
In this manner we infer that condition (\ref{70}), which becomes equivalent to
\begin{equation}
f ^{\alpha_{0}}\big( [\omega ^{\Theta}] \big) \frac{\delta S^{\mathrm{L}}[\Phi^{\beta_{0}}]}{\delta \Phi^{\alpha_{0}}} + \partial _{\mu} t _{0} ^{\mu}\big( [\Phi ^{\alpha_{0}}] \big) =0, \label{79}
\end{equation}
is nothing but Noether's theorem requiring the invariance of action (\ref{2}), $\Delta _ {\Upsilon} S^{\mathrm{L}}[\Phi^{\alpha_{0}}] = 0 $, under a nontrivial, global one-parameter transformation
\begin{equation}
\Delta _ {\Upsilon} \Phi^{\alpha_{0}} = f ^{\alpha_{0}}\big( [\omega ^{\Theta}] \big) \Upsilon . \label{80}
\end{equation}
Thus, we replaced the necessary condition on the existence of $a ^{\mathrm{int}}$ set as existence of nontrivial elements from $H _{1} ^{4}(\delta | d)$ with the existence of nontrivial, one-dimensional rigid symmetries of the free Lagrangian action.

Before solving equation (\ref{79}), it is worth mentioning that the nontriviality of (\ref{80}) automatically induces the nontriviality of the conserved current $t _{0} ^{\mu}$ appearing in (\ref{79}), which, in turn, is a key point in ensuring the nontriviality of the coupled Lagrangian at the first order of perturbation theory, $a _{0} ^{\mathrm{int}}$. Indeed, the homology space $H _{1} ^{4}(\delta | d)$ in $\pgh = 0$ is known to be isomorphic to the cohomology space of $d$ modulo $\delta $ in antifield number $0$ and in form degree $3$ computed in the algebra of ghost-independent local forms, $H _{0} ^{3}(d | \delta)$ (for instance, see~\cite{CMP1995a,PR2000,PLB1991}). The cohomology $H _{0} ^{3}(d | \delta)$ can be analyzed even without introducing the antifields and is in a bijective correspondence with the space of inequivalent nontrivial conserved currents of action (\ref{2}). In view of the physical significance of its hand sides, the isomorphism $H _{1} ^{4}(\delta | d)\simeq H _{0} ^{3}(d | \delta)$ is a cohomological reformulation of Noether's theorem and stipulates the isomorphism between the space of inequivalent nontrivial global symmetries of action (\ref{2}) and the space of its inequivalent nontrivial conserved currents. According to equation (\ref{79}), $t _{0} ^{\mu}$ is precisely the conserved current of this action corresponding to the nontrivial rigid symmetry (\ref{80}) and hence, by the above isomorphism, it will also be nontrivial. We recall that a conserved current ($\partial _{\mu} t _{0} ^{\mu} \approx 0$) is said to be trivial if it coincides on-shell with an identically conserved current, $t _{\mathrm{triv},0} ^{\mu} \approx \partial _{\nu}v _{0} ^{\nu\mu}$, with $v _{0} ^{\nu\mu} = -v _{0} ^{\mu\nu}$. The on-shell or weak equality ``$\approx$'' means as usually an equality that holds on the (stationary) surface of field equations, $\delta S^{\mathrm{L}} / \delta \Phi ^{\alpha_{0}} \approx 0$.

Beside being nontrivial, the global symmetry (\ref{80}) of action (\ref{2}) has to meet several other requirements. Thus, the necessity of producing a \emph{one-dimensional} global invariance of the free Lagrangian action (there is a single constant parameter $\Upsilon$ in (\ref{80})) is precisely due to the sole presence of the scalar ghost $\eta $ which is allowed in $a _{1} ^{\mathrm{int}}$ of the form (\ref{69}). In addition, all the generators of this global symmetry, $f ^{\alpha_{0}}$, should be gauge invariant since they may depend \emph{only} on the gauge-invariant quantities $\omega ^{\Theta}$ introduced in (\ref{64}) and their derivatives up to a finite order. This dependence automatically ensures the spacetime locality and Poincar\'{e} invariance of deformations. Moreover, the coefficients $f ^{\alpha_{0}}$ are demanded to instate the remaining hypotheses, namely, the Lorentz covariance and the derivative order assumption. In order to analyze properly the implications of these features, we pass to the explicit form of relation (\ref{80})
\begin{gather}
\Delta _ {\Upsilon} \varphi = f \big( [\omega ^{\Theta}] \big) \Upsilon , \qquad \Delta _ {\Upsilon} A _{\nu} = f _{\nu} ^{\prime} \big( [\omega ^{\Theta}] \big) \Upsilon , \label{81}\\
\Delta _ {\Upsilon} H ^{\mu} = \bar{f} ^{\mu} \big( [\omega ^{\Theta}] \big) \Upsilon , \qquad \Delta _ {\Upsilon} B ^{\mu\nu} = \tilde{f} ^{\mu\nu} \big( [\omega ^{\Theta}] \big) \Upsilon , \label{82}\\
\Delta _ {\Upsilon} \phi ^{A} = \hat{f} ^{A} \big( [\omega ^{\Theta}] \big) \Upsilon ,\label{83}
\end{gather}
along with the concrete expressions of the EL derivatives of action (\ref{2})
\begin{gather}
\frac{\delta S^{\mathrm{L}}[\Phi^{\alpha_{0}}]}{\delta \varphi} \equiv - \partial _{\lambda }H^{\lambda }, \qquad \frac{\delta S^{\mathrm{L}}[\Phi^{\alpha_{0}}]}{\delta A _{\nu}} \equiv - \partial _{\lambda }B^{\lambda \nu}, \qquad \frac{\delta S^{\mathrm{L}}[\Phi^{\alpha_{0}}]}{\delta H ^{\mu}} \equiv \partial _{\mu }\varphi , \label{84}\\
\frac{\delta S^{\mathrm{L}}[\Phi^{\alpha_{0}}]}{\delta B ^{\mu\nu}} \equiv \tfrac{1}{2} \partial _{[\mu }A _{\nu ]}, \qquad \frac{\delta S^{\mathrm{L}}[\Phi^{\alpha_{0}}]}{\delta \phi ^{A}} \equiv - k _{AB} \Box \phi ^{B} = - \Box \phi _{A} ,\label{85}
\end{gather}
such that the detailed structure of equation (\ref{79}) is given by
\begin{equation}
-f \partial _{\lambda }H^{\lambda } - f _{\nu} ^{\prime} \partial _{\lambda }B^{\lambda \nu} + \bar{f} ^{\mu} \partial _{\mu }\varphi + \tfrac{1}{2} \tilde{f} ^{\mu\nu} \partial _{[\mu }A _{\nu ]} - \hat{f} ^{A} k _{AB} \Box \phi ^{B} + \partial _{\mu} t _{0} ^{\mu} =0 .\label{86}
\end{equation}
The Lorentz covariance attracts that $f$ from (\ref{81}) and $\hat{f} ^{A}$ from (\ref{83}) are some scalars, both $f _{\nu} ^{\prime}$ and $\bar{f} ^{\mu}$ appearing in (\ref{81})--(\ref{82}) stand for the components of some four-dimensional vector fields, while $\tilde{f} ^{\mu\nu}$ implied in (\ref{82}) defines an antisymmetric two-tensor. Along the same line, the conservation of the number of derivatives on each field with respect to the free Lagrangian in the corresponding $a _{0} ^{\mathrm{int}}$ as solution to equation (\ref{62}) limits the dependence on the derivatives of the fields allowed to enter each generator. Indeed, this hypothesis constrains the conserved current $t _{0} ^{\mu}$ involved in (\ref{79}) and (\ref{86}) to contain at most two derivatives and the terms with precisely two derivatives to be quadratic in the first-order derivatives of the matter fields. Inspecting (\ref{86}), it follows on the one hand that the BF generators $\{f, f _{\nu} ^{\prime}, \bar{f} ^{\mu}, \tilde{f} ^{\mu\nu}\}$ should respect the same rules like $t _{0} ^{\mu}$ and on the other hand that the generators related to the real scalar fields, $\{ \hat{f} ^{A} \}$, should be at most linear in the first-order derivatives of the matter fields. However, we show that we can \emph{relax this condition} to the requirement that \emph{the conserved current} $t _{0} ^{\mu}$ involved in (\ref{79}) and (\ref{86}) \emph{contains at most two derivatives acting on any of the fields} and $t _{0} ^{\mu}$ yet fulfills the conservation of the number of derivatives on each field with respect to the free limit. From (\ref{86}) we then deduce that under this weaker assumption the BF generators $\{f, f _{\nu} ^{\prime}, \bar{f} ^{\mu}, \tilde{f} ^{\mu\nu}\}$ may involve at most two derivatives, while those related to the real scalar fields, $\{ \hat{f} ^{A} \}$, at most a single one. In order to produce true cross-couplings between the BF and matter sectors, all the generators of the BF fields mandatorily depend on the real scalar fields and their derivatives, $[\phi]$.

Due to the fact that the field equations contain no Levi--Civita symbols, we can further split the generators of the searched one-parameter rigid transformations, (\ref{81})--(\ref{83}), into
\begin{gather}
f = f _{\mathrm{PT}} \big( [\omega ^{\Theta}] \big) + f _{\mathrm{nPT}} \big( [\omega ^{\Theta}] \big), \qquad f _{\nu} ^{\prime}  = f _{\mathrm{PT},\nu} ^{\prime} \big( [\omega ^{\Theta}] \big) + f _{\mathrm{nPT},\nu} ^{\prime} \big( [\omega ^{\Theta}] \big) , \label{87}\\
\bar{f} ^{\mu} = \bar{f} _{\mathrm{PT}} ^{\mu} \big( [\omega ^{\Theta}] \big) + \bar{f} _{\mathrm{nPT}} ^{\mu} \big( [\omega ^{\Theta}] \big) , \qquad \tilde{f} ^{\mu\nu} = \tilde{f} _{\mathrm{PT}} ^{\mu\nu} \big( [\omega ^{\Theta}] \big) + \tilde{f} _{\mathrm{nPT}} ^{\mu\nu} \big( [\omega ^{\Theta}] \big)  , \label{88}\\
\hat{f} ^{A} = \hat{f} _{\mathrm{PT}} ^{A} \big( [\omega ^{\Theta}] \big) + \hat{f} _{\mathrm{nPT}} ^{A} \big( [\omega ^{\Theta}] \big) ,\label{89}
\end{gather}
where each $f _{\mathrm{PT}} ^{\alpha _{0}}$ and respectively $f _{\mathrm{nPT}} ^{\alpha _{0}}$ contains  an even and respectively an odd number of Levi--Civita symbols. Consequently, (\ref{86}) becomes equivalent to two distinct equations
\begin{align}
-f _{\mathrm{PT}} \partial _{\lambda }H^{\lambda } - f _{\mathrm{PT},\nu} ^{\prime} \partial _{\lambda }B^{\lambda \nu} + \bar{f} _{\mathrm{PT}} ^{\mu} \partial _{\mu }\varphi + \tfrac{1}{2} \tilde{f} _{\mathrm{PT}} ^{\mu\nu} \partial _{[\mu }A _{\nu ]} & \nonumber \\
- \hat{f} _{\mathrm{PT}} ^{A} k _{AB} \Box \phi ^{B} + \partial _{\mu} t _{\mathrm{PT},0} ^{\mu} &=0 ,\label{90}\\
-f _{\mathrm{nPT}} \partial _{\lambda }H^{\lambda } - f _{\mathrm{nPT},\nu} ^{\prime} \partial _{\lambda }B^{\lambda \nu} + \bar{f} _{\mathrm{nPT}} ^{\mu} \partial _{\mu }\varphi + \tfrac{1}{2} \tilde{f} _{\mathrm{nPT}} ^{\mu\nu} \partial _{[\mu }A _{\nu ]} & \nonumber \\
- \hat{f} _{\mathrm{nPT}} ^{A} k _{AB} \Box \phi ^{B} + \partial _{\mu} t _{\mathrm{nPT},0} ^{\mu} &=0 ,\label{91}
\end{align}
where the currents $t _{\mathrm{PT},0} ^{\mu}$ and $t _{\mathrm{nPT},0} ^{\mu}$ should also involve an even and respectively an odd number of Levi--Civita symbols. The strategy goes as follows. We start from the general representations of the above generators in terms of $[\omega ^{\Theta}]$ that implement all the working hypotheses (in agreement with the discussion from the previous paragraph), solve equations (\ref{90}) and (\ref{91}), and then eliminate all the trivial terms from their general solutions.

In order to solve equation (\ref{90}), we begin with the most general representations of $f _{\mathrm{PT}} ^{\alpha _{0}}$ from (\ref{87})--(\ref{89}) in terms of (\ref{64}) and their spacetime derivatives that are \emph{covariant} and implement \emph{the more relaxed requirement regarding their maximum number of field derivatives: two} with respect to the \emph{BF generators} and respectively  \emph{one} at the level of the \emph{matter generators}
\begin{align}
f _{\mathrm{PT}} \big( [\omega ^{\Theta}] \big) =  & f _{1} (\varphi, \underline{\phi }) + f _{2} (\varphi, \underline{\phi }) \partial _{\rho} H^{\rho} + f _{3} (\varphi, \underline{\phi }) (\partial _{\rho} H ^{\rho}) (\partial _{\sigma} H ^{\sigma}) \nonumber \\
& + f _{4} (\varphi, \underline{\phi }) (\partial ^{\rho } B _{\rho \nu}) (\partial  _{\sigma } B ^{\sigma \nu}) + f _{5} (\varphi, \underline{\phi }) (\partial _{[\mu } A _{\nu ]}) (\partial ^{[\mu } A ^{\nu ]}) \nonumber \\
& + f _{6} (\varphi, \underline{\phi }) (\partial _{\mu } \varphi ) (\partial ^{\mu } \varphi ) + f _{7AB} (\varphi, \phi ) \big( \partial _{\mu } \phi ^{A} \big) \big( \partial ^{\mu } \phi ^{B} \big) \nonumber \\
& + f _{8} (\varphi, \underline{\phi }) (\partial ^{\rho } B _{\rho \nu}) (\partial ^{\nu } \varphi ) + f _{9A} (\varphi, \phi ) (\partial ^{\rho } B _{\rho \nu}) \big( \partial ^{\nu } \phi ^{A} \big) \nonumber \\
& + f _{10A} (\varphi, \phi ) (\partial _{\mu } \varphi ) \big( \partial ^{\mu } \phi ^{A} \big) + f_{11} (\varphi, \underline{\phi }) \Box \varphi \nonumber \\
&+ f_{12 A} (\varphi, \phi ) \Box \phi ^{A}, \label{92} \\
f _{\mathrm{PT},\nu} ^{\prime} \big( [\omega ^{\Theta}] \big) = & f _{1} ^{\prime} (\varphi, \underline{\phi }) \partial ^{\rho } B _{\rho \nu } + f _{2} ^{\prime} (\varphi, \underline{\phi }) \partial _{\nu } \varphi + f _{3A} ^{\prime} (\varphi, \phi ) \partial _{\nu } \phi ^{A} \nonumber \\
& + f _{4} ^{\prime} (\varphi, \underline{\phi }) \partial _{\nu } \partial _{\rho} H^{\rho} + f _{5} ^{\prime} (\varphi, \underline{\phi }) \partial ^{\rho} \partial _{[\rho} A _{\nu ]} \nonumber \\
&+ f _{6} ^{\prime} (\varphi, \underline{\phi }) (\partial _{\rho} H^{\rho})(\partial  ^{\sigma } B _{\sigma \nu})+ f _{7} ^{\prime} (\varphi, \underline{\phi }) (\partial _{\rho} H^{\rho})(\partial _{\nu } \varphi ) \nonumber \\
&+ f _{8A} ^{\prime} (\varphi, \phi ) (\partial _{\rho} H^{\rho}) \big( \partial _{\nu } \phi ^{A} \big) + f _{9} ^{\prime} (\varphi, \underline{\phi }) (\partial _{[\rho} A _{\nu ]}) (\partial _{\sigma } B ^{\sigma \rho })\nonumber \\
& + f _{10} ^{\prime} (\varphi, \underline{\phi }) (\partial _{[\rho} A _{\nu ]}) (\partial ^{\rho} \varphi ) + f _{11A} ^{\prime} (\varphi, \phi ) (\partial _{[\rho} A _{\nu ]}) \big( \partial ^{\rho} \phi ^{A} \big) , \label{93} \\
\bar{f} _{\mathrm{PT}} ^{\mu} \big( [\omega ^{\Theta}] \big) = & \bar{f} _{1} (\varphi, \underline{\phi }) \partial _{\rho } B ^{\rho \mu } + \bar{f} _{2} (\varphi, \underline{\phi }) \partial ^{\mu } \varphi + \bar{f} _{3A} (\varphi, \phi ) \partial ^{\mu } \phi ^{A} \nonumber \\
& + \bar{f} _{4} (\varphi, \underline{\phi }) \partial ^{\mu } \partial _{\rho} H^{\rho} + \bar{f} _{5} (\varphi, \underline{\phi }) \partial _{\rho} \partial ^{[\rho} A ^{\mu ]} \nonumber \\
&+ \bar{f} _{6} (\varphi, \underline{\phi }) (\partial _{\rho} H^{\rho})(\partial  _{\lambda } B ^{\lambda \mu}) + \bar{f} _{7} (\varphi, \underline{\phi }) (\partial _{\rho} H^{\rho})(\partial ^{\mu } \varphi ) \nonumber \\
&+ \bar{f} _{8A} (\varphi, \phi ) (\partial _{\rho} H^{\rho}) \big( \partial ^{\mu } \phi ^{A} \big) + \bar{f} _{9} (\varphi, \underline{\phi }) (\partial ^{\lambda } B _{\lambda \rho }) (\partial ^{[\rho} A ^{\mu ]}) \nonumber \\
& + \bar{f} _{10} (\varphi, \underline{\phi }) (\partial _{\rho} \varphi ) (\partial ^{[\rho} A ^{\mu ]}) + \bar{f} _{11A} (\varphi, \phi ) \big( \partial _{\rho} \phi ^{A} \big) (\partial ^{[\rho} A ^{\mu ]}), \label{94} \\
\tilde{f} _{\mathrm{PT}} ^{\mu\nu} \big( [\omega ^{\Theta}] \big) = & \tilde{f} _{1} (\varphi, \underline{\phi }) \partial ^{[\mu} A ^{\nu ]} + \tilde{f} _{2} (\varphi, \underline{\phi }) \partial ^{[\mu} \partial _{\rho} B ^{\nu ] \rho} + \tilde{f} _{3} (\varphi, \underline{\phi }) (\partial _{\rho} H^{\rho}) (\partial ^{[\mu} A ^{\nu ]}) \nonumber \\
&+ \tilde{f} _{4} (\varphi, \underline{\phi }) (\partial _{\rho} B ^{\rho [\mu }) (\partial ^{\nu ]} \varphi) + \tilde{f} _{5A} (\varphi, \phi ) (\partial _{\rho} B ^{\rho [\mu }) \big( \partial ^{\nu ]} \phi ^{A} \big) \nonumber \\
& + \tilde{f} _{6A} (\varphi, \phi ) (\partial ^{[\mu } \varphi) \big( \partial ^{\nu ]} \phi ^{A} \big) + \tilde{f} _{7} ^{AB} (\varphi, \phi ) \big( \partial ^{[\mu } \phi _{A} \big) \big( \partial ^{\nu ]} \phi _{B} \big), \label{95}\\
\hat{f} _{\mathrm{PT}} ^{A} \big( [\omega ^{\Theta}] \big) = & \hat{f} _{1} ^{A}(\varphi, \phi ) + \hat{f} _{2} ^{A}(\varphi, \phi )\partial _{\rho} H^{\rho}. \label{96}
\end{align}
In the above all the coefficients denoted by $f$, $f ^{\prime}$, $\bar{f}$, $\tilde{f}$, or $\hat{f}$ stand for some unknown smooth functions of the undifferentiated BF scalar field $\varphi $ and undifferentiated matter scalar fields $\{\phi ^A\}\equiv \phi$. In addition, all the coefficients containing an underlined $\phi$ mandatorily depend on the matter fields in order to ensure true cross-couplings among the BF and matter sectors in $a _{1} ^{\mathrm{int}}$ (see (\ref{69})). Also, the functions $f _{7AB}$ together with $\tilde{f} _{7} ^{AB}$ possess definite symmetry/antisymmetry properties ($f _{7AB} (\varphi, \phi ) = f _{7BA} (\varphi, \phi )$, $\tilde{f} _{7} ^{AB} (\varphi, \phi ) = - \tilde{f} _{7} ^{BA} (\varphi, \phi )$) generated by the expressions of the corresponding terms from (\ref{92}) and respectively (\ref{95}). We note that even if the derivative order assumption allows larger classes of terms in (\ref{92})--(\ref{96}), like for instance linear in the first-order derivatives of the matter fields with respect to $f _{\mathrm{PT}}$, $\tilde{f} _{\mathrm{PT}} ^{\mu\nu}$, and $\hat{f} _{\mathrm{PT}} ^{A}$ or quadratic in the first-order derivatives of the matter fields at the level of $f _{\mathrm{PT},\nu} ^{\prime}$ and $\bar{f} _{\mathrm{PT}} ^{\mu}$, it is the Lorentz covariance that actually kills them.

Some of the above expressions of the global PT generators can be still simplified given their uniqueness up to adding trivial contributions obtained by two mechanisms: either combinations of true gauge transformations with the gauge parameters replaced by functions of fields and their derivatives or trivial gauge transformations, i.e., antisymmetric combinations of free field equations. Initially, we show that we can remove the terms involving the functions $f _{4} ^{\prime}$, $f _{5} ^{\prime}$, $\bar{f} _{4}$, and $\bar{f} _{5}$ from (\ref{93}) and (\ref{94}). Indeed, let us transform $f _{\mathrm{PT},\nu} ^{\prime}$ and respectively $\bar{f} _{\mathrm{PT}} ^{\mu}$ by adding to each of them some combinations of true gauge transformations of the vector fields $A _{\lambda}$ and $H^{\lambda}$ (see formulas (\ref{3}) and (\ref{4}))
\begin{align}
f _{\mathrm{PT},\nu} ^{\prime} &\rightarrow f _{\mathrm{PT},\nu} ^{\prime} + \delta _{\nu} ^{\lambda}\delta_{\Omega_{1}^{\alpha_{1}}}A_{\lambda} + \sigma _{\nu\lambda}\delta_{\Omega_{1}^{\alpha_{1}}}H^{\lambda}, \label{97} \\
\bar{f} _{\mathrm{PT}} ^{\mu} &\rightarrow \bar{f} _{\mathrm{PT}} ^{\mu} + \sigma ^{\mu\lambda}\delta_{\Omega_{2}^{\alpha_{1}}}A_{\lambda} + \delta ^{\mu} _{\lambda}\delta_{\Omega_{2}^{\alpha_{1}}}H^{\lambda}, \label{98}
\end{align}
defined like
\begin{align}
\delta_{\Omega _{1} ^{\alpha_{1}}} A _{\lambda} &= \partial_{\lambda }\epsilon _{1}, & \epsilon _{1} &\equiv - f _{4} ^{\prime} \partial _{\rho} H^{\rho}, \label{99} \\
\delta_{\Omega _{1} ^{\alpha_{1}}} H ^{\lambda} &= - 2\partial _{\rho}\xi _{1}^{\rho\lambda}, & \xi _{1}^{\rho\lambda} &\equiv \tfrac{1}{2} f _{5} ^{\prime}\partial ^{[\rho} A ^{\lambda ]}, \label{100} \\
\delta_{\Omega _{2} ^{\alpha_{1}}} A _{\lambda} & = \partial _{\lambda }\epsilon _{2}, & \epsilon _{2} &\equiv - \bar{f} _{4} \partial _{\rho} H^{\rho}, \label{101}\\
\delta_{\Omega _{2} ^{\alpha_{1}}} H ^{\lambda} &= - 2\partial _{\rho}\xi _{2}^{\rho\lambda}, & \xi _{2}^{\rho\lambda} &\equiv \tfrac{1}{2} \bar{f} _{5} \partial ^{[\rho} A ^{\lambda ]}. \label{102}
\end{align}
After simple computations, we find that (\ref{99})--(\ref{102}) ensure that transformations (\ref{97}) and (\ref{98}) affect only the terms from (\ref{93}) and respectively (\ref{94}) depending on $f _{4} ^{\prime}$, $f _{5} ^{\prime}$, $\bar{f} _{4}$, and $\bar{f} _{5}$ in the following manner
\begin{align}
f _{\mathrm{PT},\nu} ^{\prime} : f _{4} ^{\prime} \partial _{\nu } \partial _{\rho} H^{\rho} + f _{5} ^{\prime} \partial ^{\rho} \partial _{[\rho} A _{\nu ]} \rightarrow & - (\partial _{\rho} H^{\rho}) \bigg( \frac{\partial f _{4} ^{\prime}}{\partial \varphi} \partial _{\nu } \varphi  + \frac{\partial f _{4} ^{\prime}}{\partial \phi ^{A}}  \partial _{\nu } \phi ^{A} \bigg) \nonumber \\
&- (\partial _{[\rho} A _{\nu ]}) \bigg(\frac{\partial f _{5} ^{\prime}}{\partial \varphi} \partial ^{\rho} \varphi + \frac{\partial f _{5} ^{\prime}}{\partial \phi ^{A}} (\partial ^{\rho} \phi ^{A} \bigg), \label{103}\\
\bar{f} _{\mathrm{PT}} ^{\mu} : \bar{f} _{4} \partial ^{\mu } \partial _{\rho} H^{\rho} + \bar{f} _{5} \partial _{\rho} \partial ^{[\rho} A ^{\mu ]} \rightarrow &- (\partial _{\rho} H^{\rho}) \bigg( \frac{\partial \bar{f} _{4}}{\partial \varphi} \partial ^{\mu } \varphi + \frac{\partial \bar{f} _{4}}{\partial \phi ^{A}} \partial ^{\mu } \phi ^{A} \bigg) \nonumber\\
&- (\partial ^{[\rho} A ^{\mu ]}) \bigg( \frac{\partial \bar{f} _{5}}{\partial \varphi} \partial _{\rho} \varphi + \frac{\partial \bar{f} _{5}}{\partial \phi ^{A}} \partial _{\rho} \phi ^{A} \bigg). \label{104}
\end{align}
But all these contributions can be absorbed in similar terms already present in (\ref{93}) and (\ref{94}) by means of the redefinitions
\begin{align}
f _{7} ^{\prime} - \frac{\partial f _{4} ^{\prime}}{\partial \varphi} &\rightarrow f _{7} ^{\prime}, & f _{8A} ^{\prime} - \frac{\partial f _{4} ^{\prime}}{\partial \phi ^{A}} &\rightarrow f _{8A} ^{\prime}, \label{105}\\
f _{10} ^{\prime} - \frac{\partial f _{5} ^{\prime}}{\partial \varphi} &\rightarrow f _{10} ^{\prime},  & f _{11A} ^{\prime} - \frac{\partial f _{5} ^{\prime}}{\partial \phi ^{A}} &\rightarrow f _{11A} ^{\prime}, \label{106}\\
\bar{f} _{7} - \frac{\partial \bar{f} _{4}}{\partial \varphi} &\rightarrow \bar{f} _{7}, & \bar{f} _{8A} - \frac{\partial \bar{f} _{4}}{\partial \phi ^{A}} &\rightarrow \bar{f} _{8A}, \label{107}\\
\bar{f} _{10} - \frac{\partial \bar{f} _{5}}{\partial \varphi} &\rightarrow \bar{f} _{10}, & \bar{f} _{11A} - \frac{\partial \bar{f} _{5}}{\partial \phi ^{A}} &\rightarrow \bar{f} _{11A}, \label{108}
\end{align}
which are allowed since all the (smooth) functions of $\varphi $ and $\phi $ from (\ref{92})--(\ref{96}) are arbitrary at this stage. In conclusion, we can indeed take
\begin{equation}
f _{4} ^{\prime} = f _{5} ^{\prime} = \bar{f} _{4} = \bar{f} _{5} =0 \label{109}
\end{equation}
in formulas (\ref{93}) and (\ref{94}) without loss of potentially nontrivial terms. On the other hand, the terms depending on the functions $f _{9} ^{\prime}$ from (\ref{93}) and $\bar{f} _{10}$  in (\ref{94}) are already trivial since they are involved in the rigid transformations of the vector field $A _{\nu}$ and respectively $H ^{\mu}$ and reduce to antisymmetric combinations of precisely the associated free field equations ($\delta S^{\mathrm{L}}/ \delta A _{\rho}$ and respectively $\delta S^{\mathrm{L}}/ \delta H ^{\rho}$)
\begin{align*}
f _{9} ^{\prime} (\partial _{[\rho} A _{\nu ]}) (\partial _{\sigma } B ^{\sigma \rho }) &= f _{9\nu \rho} ^{\prime} \frac{\delta S^{\mathrm{L}}}{\delta A _{\rho}}, & f _{9\nu \rho} ^{\prime} &= - f _{9\rho \nu} ^{\prime}, &  f _{9\nu \rho} ^{\prime} &\equiv f _{9} ^{\prime} \partial _{[\nu} A _{\rho ]}, \\
\bar{f} _{10} (\partial _{\rho} \varphi ) (\partial ^{[\rho} A ^{\mu ]}) &= \bar{f} ^{10\mu \rho} \frac{\delta S^{\mathrm{L}}}{\delta H ^{\rho}}, & \bar{f} ^{10\mu \rho} &= - \bar{f} ^{10\rho \mu}, &  \bar{f} ^{10\mu \rho} &\equiv - \bar{f} _{10} \partial ^{[\mu} A ^{\rho ]} ,
\end{align*}
so they can be safely removed by setting
\begin{equation}
f _{9} ^{\prime} = \bar{f} _{10} =0. \label{110}
\end{equation}

Inserting relations (\ref{109}) and (\ref{110}) into formulas (\ref{92})--(\ref{96}) and the resulting expressions in equation (\ref{90}), we find that it becomes equivalent to thirteen (independent) equations obtained by projection on the total number of derivatives from the corresponding current component ($0$, $1$, and respectively $2$) and then, for each distinct value of the derivative order, on all independent combinations of fields and their derivatives:
\begin{enumerate}
\item at \emph{zero} derivatives in $t _{\mathrm{PT},0} ^{\mu}$ there is a single equation
\begin{equation}
- f _{1} \partial _{\lambda} H^{\lambda} + \partial _{\mu} t _{\mathrm{PT},0,0} ^{\mu} =0; \label{111}
\end{equation}
\item at \emph{one} derivative there appear five distinct equations
\begin{align}
- f _{2} (\partial _{\rho} H^{\rho}) (\partial _{\lambda} H^{\lambda}) + \partial _{\mu} t _{\mathrm{PT},0,1} ^{1,\mu} &=0, \label{112}\\
- f _{1} ^{\prime} (\partial ^{\rho } B _{\rho \nu }) (\partial _{\lambda } B ^{\lambda \nu }) + \partial _{\mu} t _{\mathrm{PT},0,1} ^{2,\mu} &=0, \label{113}\\
\tfrac{1}{2} \tilde{f} _{1} (\partial _{[\mu} A _{\nu ]}) (\partial ^{[\mu} A ^{\nu ]}) + \partial _{\mu} t _{\mathrm{PT},0,1} ^{3,\mu} &=0, \label{114}\\
\big[ (\bar{f} _{1} - f _{2} ^{\prime}) \partial ^{\mu} \varphi - f _{3A} ^{\prime} \partial ^{\mu } \phi ^{A} \big] (\partial ^{\rho } B _{\rho \mu }) + \partial _{\mu} t _{\mathrm{PT},0,1} ^{4,\mu} &=0, \label{115}\\
\big( \bar{f} _{2} \partial ^{\mu} \varphi + \bar{f} _{3A} \partial ^{\mu} \phi ^{A} \big) (\partial _{\mu} \varphi ) - \hat{f} _{1} ^{A} k _{AB} \Box \phi ^{B} + \partial _{\mu} t _{\mathrm{PT},0,1} ^{5,\mu} &=0; \label{116}
\end{align}
\item at \emph{two} derivatives there occur seven independent equations
\begin{align}
- f _{3} (\partial _{\rho} H^{\rho}) (\partial _{\sigma} H^{\sigma}) (\partial _{\lambda} H^{\lambda}) + \partial _{\mu} t _{\mathrm{PT},0,2} ^{1,\mu} &=0, \label{117}\\
- (f _{4} + f _{6} ^{\prime}) (\partial ^{\rho } B _{\rho \nu }) (\partial _{\sigma } B ^{\sigma \nu })(\partial _{\lambda} H^{\lambda}) + \partial _{\mu} t _{\mathrm{PT},0,2} ^{2,\mu} &=0, \label{118}\\
\big( \tfrac{1}{2} \tilde{f} _{3} - f _{5} \big) (\partial _{[\mu } A _{\nu ]}) (\partial ^{[\mu} A ^{\nu ]}) (\partial _{\lambda} H^{\lambda}) + \partial _{\mu} t _{\mathrm{PT},0,2} ^{3,\mu} &=0, \label{119}\\
- \big[ (f _{8} + f _{7} ^{\prime} - \bar{f} _{6}) \partial ^{\mu} \varphi  + (f _{9A} + f _{8A} ^{\prime}) \partial ^{\mu} \phi ^{A} \big] (\partial ^{\rho } B _{\rho \mu })(\partial _{\lambda} H^{\lambda}) & \nonumber \\
+ \partial _{\mu} t _{\mathrm{PT},0,2} ^{4,\mu} &=0, \label{120}\\
- \big\{ \tilde{f} _{2} \partial ^{\mu} \partial _{\rho } B ^{\rho \nu } + \big[ (f _{10} ^{\prime} + \bar{f} _{9} + \tilde{f} _{4}) \partial ^{\mu} \varphi & \nonumber \\
+ (f _{11A} ^{\prime} + \tilde{f} _{5A}) \partial ^{\mu} \phi ^{A} \big] (\partial _{\rho } B ^{\rho \nu }) \big\} (\partial _{[\mu } A _{\nu ]}) + \partial _{\mu} t _{\mathrm{PT},0,2} ^{5,\mu} &=0, \label{121}\\
- \big[ f _{11} \Box \varphi + (f _{6} - \bar{f} _{7}) (\partial _{\mu} \varphi ) (\partial ^{\mu} \varphi ) & \nonumber \\
+ (f _{10A} - \bar{f} _{8A}) (\partial _{\mu} \varphi ) \big( \partial ^{\mu} \phi ^{A} \big) + f _{7AB} \big( \partial _{\mu} \phi ^{A} \big) \big( \partial ^{\mu} \phi ^{B} \big)  & \nonumber \\
+ (f _{12A} + k _{AB}\hat{f} _{2} ^{B}) \Box \phi ^{A} \big] (\partial _{\lambda} H^{\lambda}) + \partial _{\mu} t _{\mathrm{PT},0,2} ^{6,\mu} &=0, \label{122}\\
\big[ (\bar{f} _{11A} - \tilde{f} _{6A}) \big( \partial ^{\mu} \phi ^{A} \big) (\partial ^{\nu} \varphi ) + \tilde{f} _{7} ^{AB} \big( \partial ^{\mu} \phi _{A} \big) \big( \partial ^{\nu} \phi _{B} \big) \big] (\partial _{[\mu } A _{\nu ]}) & \nonumber \\
+ \partial _{\mu} t _{\mathrm{PT},0,2} ^{7,\mu} &=0. \label{123}
\end{align}
\end{enumerate}
The general solutions to these equations, organized according to their order of appearance in (\ref{92})--(\ref{96}), can be expressed like:
\begin{gather}
f _{1} (\varphi, \phi ) = c, \quad  f _{2} (\varphi, \phi ) = 0, \quad f _{3} (\varphi, \phi ) = 0, \quad f _{4} (\varphi, \phi ) = - f _{6} ^{\prime} (\varphi, \phi ), \label{124}\\
f _{5} (\varphi, \phi ) = \tfrac{1}{2} \tilde{f} _{3} (\varphi, \phi ), \quad f _{6} (\varphi, \phi ) = \bar{f} _{7} (\varphi, \phi ), \quad f _{7AB} (\varphi, \phi ) = 0, \label{125}\\
f _{8} (\varphi, \phi ) = - f _{7} ^{\prime} (\varphi, \phi ) + \bar{f} _{6} (\varphi, \phi ), \quad f _{9A} (\varphi, \phi ) = - f _{8A} ^{\prime} (\varphi, \phi ), \label{126}\\
f _{10A} (\varphi, \phi ) = \bar{f} _{8A} (\varphi, \phi ), \quad f _{11} (\varphi, \phi ) = 0, \label{127}\\
f _{12A} (\varphi, \phi ) = - k _{AB} \hat{f} _{2} ^{B} (\varphi, \phi ), \quad f _{1} ^{\prime} (\varphi, \phi ) = 0, \quad f _{3A} ^{\prime} (\varphi, \phi ) = \frac{\partial f _{3} ^{\prime} (\varphi, \phi )}{\partial \phi ^{A}}, \label{128}\\
f _{10} ^{\prime} (\varphi, \phi ) = - \bar{f} _{9} (\varphi, \phi ) - \tilde{f} _{4} (\varphi, \phi ), \quad f _{11A} ^{\prime} (\varphi, \phi ) = - \tilde{f} _{5A} (\varphi, \phi ),\label{129}\\
\bar{f} _{1} (\varphi, \phi ) = f _{2} ^{\prime} (\varphi, \phi ) - \frac{\partial f _{3} ^{\prime} (\varphi, \phi )}{\partial \varphi } + f ^{\prime} (\varphi ), \quad \bar{f} _{2} (\varphi, \phi ) = 0, \label{130}\\
\bar{f} _{3A} (\varphi, \phi ) = - k _{AB} \frac{\partial \hat{f} _{1} ^{B} (\varphi, \phi )}{\partial \varphi }, \quad k _{AC} \frac{\partial \hat{f} _{1} ^{C} (\varphi, \phi )}{\partial \phi ^{B}} + k _{BC} \frac{\partial \hat{f} _{1} ^{C} (\varphi, \phi )}{\partial \phi ^{A}} = 0,\label{131}\\
\tilde{f} _{1} (\varphi, \phi ) = 0, \quad \tilde{f} _{2} (\varphi, \phi ) = 0, \label{132}\\
\tilde{f} _{6A} (\varphi, \phi ) = \bar{f} _{11A} (\varphi, \phi ), \quad \tilde{f} _{7} ^{AB} (\varphi, \phi ) = 0. \label{133}
\end{gather}
The above solutions are parameterized by one constant ($c$), one arbitrary scalar smooth function of the undifferentiated BF scalar field ($f ^{\prime}$), nine arbitrary, smooth scalar functions  ($f _{2} ^{\prime}$, $f _{3} ^{\prime}$, $f _{6} ^{\prime}$, $f _{7} ^{\prime}$, $\bar{f} _{6}$, $\bar{f} _{7}$, $\bar{f} _{9}$, $\tilde{f} _{3}$, and $\tilde{f} _{4}$) and five arbitrary collections of smooth scalar functions ($f _{8A} ^{\prime}$, $\bar{f} _{8A}$, $\bar{f} _{11A}$, $\tilde{f} _{5A}$, and $\hat{f} _{2} ^{A}$) depending on all the undifferentiated scalar fields from the theory, supplemented by a set of smooth scalar functions of $\varphi $ and $\phi $, namely $\hat{f} _{1} ^{A}$, which are not arbitrary, but subject to the latter equations from (\ref{131}). In order to produce true cross-couplings at the level of the first-order deformation $a ^{\mathrm{int}}$, the nine parameterizing scalar functions should effectively depend on the undifferentiated matter fields and, in addition, we must take
\begin{equation}
c=0, \qquad f ^{\prime} (\varphi ) =0. \label{134}
\end{equation}

Substituting results (\ref{124})--(\ref{134}) together with choices (\ref{109}) and (\ref{110}) in formulas (\ref{92})--(\ref{96}) and employing (\ref{84}) and (\ref{85}), after some simple computations we arrive at
\begin{align}
f _{\mathrm{PT}} =  & \big( f _{6} ^{\prime} \partial ^{\lambda } B _{\lambda \nu} +  f _{7} ^{\prime} \partial _{\nu } \varphi + f _{8A} ^{\prime} \partial _{\nu } \phi ^{A} \big) \frac{\delta S^{\mathrm{L}}}{\delta A _{\nu}} + \tilde{f} _{3} \partial ^{[\mu } A ^{\nu ]} \frac{\delta S^{\mathrm{L}}}{\delta B ^{\mu\nu}} \nonumber \\
& + \big( \bar{f} _{6} \partial _{\lambda} B ^{\lambda \mu} + \bar{f} _{7} \partial ^{\mu } \varphi + \bar{f} _{8A} \partial ^{\mu } \phi ^{A} \big) \frac{\delta S^{\mathrm{L}}}{\delta H^{\mu}} + \hat{f} _{2} ^{A} \frac{\delta S^{\mathrm{L}}}{\delta \phi ^{A}}, \label{135} \\
f _{\mathrm{PT},\nu} ^{\prime} = & \partial _{\nu } f _{3} ^{\prime} - \big( f _{6} ^{\prime} \partial ^{\lambda } B _{\lambda \nu} +  f _{7} ^{\prime} \partial _{\nu } \varphi + f _{8A} ^{\prime} \partial _{\nu } \phi ^{A} \big) \frac{\delta S^{\mathrm{L}}}{\delta \varphi} + \bar{f} _{9} \partial _{[\nu} A _{\lambda ]} \frac{\delta S^{\mathrm{L}}}{\delta H _{\lambda}} \nonumber \\
& + \delta ^{\mu} _{\nu}\bigg( f _{2} ^{\prime} - \frac{\partial f _{3} ^{\prime}}{\partial \varphi } \bigg) \frac{\delta S^{\mathrm{L}}}{\delta H^{\mu}} + \delta _{\nu} ^{[\rho} \big( \tilde{f} _{4} \partial ^{\lambda]} \varphi + \tilde{f} _{5A} \partial ^{\lambda]} \phi ^{A} \big) \frac{\delta S^{\mathrm{L}}}{\delta B^{\rho\lambda}}, \label{136} \\
\bar{f} _{\mathrm{PT}} ^{\mu} = & - \big( \bar{f} _{6} \partial _{\lambda} B ^{\lambda \mu} + \bar{f} _{7} \partial ^{\mu } \varphi + \bar{f} _{8A} \partial ^{\mu } \phi ^{A} \big) \frac{\delta S^{\mathrm{L}}}{\delta \varphi} + \bar{f} _{9} \partial ^{[\mu} A ^{\lambda ]} \frac{\delta S^{\mathrm{L}}}{\delta A ^{\lambda}} \nonumber \\
& - \delta ^{\mu} _{\nu} \bigg( f _{2} ^{\prime} -\frac{\partial f _{3} ^{\prime}}{\partial \varphi } \bigg) \frac{\delta S^{\mathrm{L}}}{\delta A_{\nu}} - \bar{f} _{11A} \big( \delta _{[\rho} ^{\mu}\partial _{\lambda ]} \phi ^{A} \big) \frac{\delta S^{\mathrm{L}}}{\delta B _{\rho\lambda}} \nonumber \\
&- k _{AB} \big( \partial ^{\mu } \phi ^{A} \big) \frac{\partial \hat{f} _{1} ^{B}}{\partial \varphi } , \label{137} \\
\tilde{f} _{\mathrm{PT}} ^{\mu\nu} = & - \tilde{f} _{3} \partial ^{[\mu} A ^{\nu ]} \frac{\delta S^{\mathrm{L}}}{\delta \varphi} - \delta _{\lambda} ^{[\mu} \big( \tilde{f} _{4} \partial ^{\nu]} \varphi + \tilde{f} _{5A} \partial ^{\nu]} \phi ^{A} \big) \frac{\delta S^{\mathrm{L}}}{\delta A _{\lambda}} \nonumber \\
& + \bar{f} _{11A} \big( \delta _{\lambda} ^{[\mu} \partial ^{\nu ]} \phi ^{A} \big) \frac{\delta S^{\mathrm{L}}}{\delta H _{\lambda}} , \label{138}\\
\hat{f} _{\mathrm{PT}} ^{A} = & \hat{f} _{1} ^{A} - \hat{f} _{2} ^{A} \frac{\delta S^{\mathrm{L}}}{\delta \varphi} . \label{139}
\end{align}
Finally, we observe that all the contributions from (\ref{135})--(\ref{139}) excepting those depending on $\hat{f} _{1} ^{A}$ are trivial since they reduce either to a gauge transformation of the BF vector field $A _{\nu}$ with the $U(1)$ gauge parameter replaced by $f _{3} ^{\prime}$ (the first term from the right-hand side of (\ref{136})) or to purely trivial gauge transformations of action (\ref{2}) otherwise, so we will discard them via setting zero all the corresponding parameterizing functions
\begin{gather}
f _{2} ^{\prime} = f _{3} ^{\prime} = f _{6} ^{\prime} = f _{7} ^{\prime} = \bar{f} _{6} = \bar{f} _{7} = \bar{f} _{9} = \tilde{f} _{3} =  \tilde{f} _{4} = 0, \label{140} \\
f _{8A} ^{\prime} = \bar{f} _{8A} = \bar{f} _{11A} = \tilde{f} _{5A} = 0, \quad \hat{f} _{2} ^{A} = 0, \qquad A=\overline{1,N}. \label{141}
\end{gather}

In this manner, the general solutions to equation (\ref{90}), which is responsible for the PT-invariant part from the nontrivial one-parameter rigid transformations (\ref{81})--(\ref{83}), become
\begin{gather}
f _{\mathrm{PT}} = 0, \qquad f _{\mathrm{PT},\nu} ^{\prime} = 0, \qquad \tilde{f} _{\mathrm{PT}} ^{\mu\nu} = 0, \label{142}\\
\bar{f} _{\mathrm{PT}} ^{\mu} = - k _{AB} \big( \partial ^{\mu } \phi ^{A} \big) \frac{\partial \hat{f} _{1} ^{B} (\varphi, \phi)}{\partial \varphi } , \qquad \hat{f} _{\mathrm{PT}} ^{A} = \hat{f} _{1} ^{A}(\varphi, \phi). \label{143}
\end{gather}
The above generators depend now on a single set of smooth scalar functions (in number equal to the number of matter real scalar fields from the collection, $N$), $\hat{f} _{1} ^{A}(\varphi, \phi)$, which is subject to the latter equations from (\ref{131}). Their general solutions, to be denoted by $\bar{n} ^{A}$, are expressed by a linear dependence of the undifferentiated matter fields
\begin{equation}
\hat{f} _{1} ^{A} (\varphi, \phi) \equiv \bar{n} ^{A} (\varphi, \phi) = n ^{A} (\varphi) + T ^{AB} (\varphi) k _{BC} \phi ^{C}, \, \, T ^{AB} (\varphi) = - T ^{BA} (\varphi), \label{144}
\end{equation}
parameterized by an $N$-dimensional vector whose components depend arbitrarily (but yet smoothly) on the undifferentiated BF scalar field, $n(\varphi) \equiv \{ n ^{A} (\varphi),A=\overline{1,N} \}$, and by a skew-symmetric quadratic matrix of order $N$ with elements also arbitrary smooth functions of the same field, $T(\varphi) \equiv \{ T ^{AB} (\varphi),A,B=\overline{1,N} \}$.

An absolutely similar procedure developed with respect to equation (\ref{91}) can be shown to give rise to purely trivial terms only
\begin{equation}
f _{\mathrm{nPT}} = 0, \qquad f _{\mathrm{nPT},\nu} ^{\prime} = 0, \qquad \bar{f} _{\mathrm{nPT}} ^{\mu} =0, \qquad \tilde{f} _{\mathrm{nPT}} ^{\mu\nu} = 0, \qquad \hat{f} _{\mathrm{nPT}} ^{A} = 0. \label{145}
\end{equation}
Assembling now results (\ref{142})--(\ref{145}) via decompositions (\ref{87})--(\ref{89}), it follows that the general expressions of the generators of the one-parameter rigid transformations read as
\begin{align}
f \big( [\omega ^{\Theta}] \big) &= 0, \qquad f _{\nu} ^{\prime} \big( [\omega ^{\Theta}] \big) = 0, \qquad \tilde{f} ^{\mu\nu} \big( [\omega ^{\Theta}] \big) = 0, \label{146}\\
\bar{f} ^{\mu} \big( [\omega ^{\Theta}] \big) &= - k _{AB} \big( \partial ^{\mu } \phi ^{A} \big) \bigg( \frac{\partial n ^{B} (\varphi )}{\partial \varphi } + \frac{\partial T ^{BC} (\varphi )}{\partial \varphi } k _{CD} \phi ^{D} \bigg) \nonumber \\
&\equiv - k _{AB} \big( \partial ^{\mu } \phi ^{A} \big) \frac{\partial \bar{n} ^{B} (\varphi, \phi)}{\partial \varphi },  \label{147}\\
\hat{f} ^{A} \big( [\omega ^{\Theta}] \big) &= k _{AB} \big( n ^{B} (\varphi) + T ^{BC} (\varphi) k _{CD} \phi ^{D} \big) \equiv \bar{n} ^{A}(\varphi, \phi). \label{148}
\end{align}
The associated conserved current emerging from the conservation law (\ref{86}) is given by
\begin{align}
t _{0} ^{\mu} (\varphi,[\phi]) &= k _{AB} \big( \partial ^{\mu } \phi ^{A} \big) \big( n ^{B} (\varphi) + T ^{BC} (\varphi) k _{CD} \phi ^{D} \big) \nonumber \\
&\equiv k _{AB} \big( \partial ^{\mu } \phi ^{A} \big) \bar{n} ^{B} (\varphi, \phi), \label{149}
\end{align}
so it is clearly nontrivial, linear in the first-order spacetime derivatives of the matter fields, and, most important in what follows, is gauge-invariant under (\ref{3})--(\ref{5}), $\delta_{\Omega^{\alpha_{1}}}t _{0} ^{\mu} =0$, since it is allowed to depend only on the scalar fields from the theory. It is important to note that even if we worked with the more relaxed assumption that there may be at most two derivatives in $t _{0} ^{\mu}$ acting on \emph{any} fields (BF/matter), its nontrivial component, given by the right-hand side of (\ref{149}), comprises a single derivative (acting on the matter fields) and thus will indeed ensure the conservation of the number of derivatives on each field from the free limit at the level of the interacting model. The last observation enables us to state that so far we indeed determined the most general, nontrivial, one-parameter rigid symmetry of the free Lagrangian action (\ref{2}) that complies with all the working hypotheses and meanwhile couples the BF and the matter sectors.

Inserting results (\ref{146})--(\ref{148}) into (\ref{69}) we finally find the general nontrivial solution to equation (\ref{61}) that fulfills \emph{all} the working hypotheses and satisfies the necessary condition (\ref{70})
\begin{gather}
a _{1} ^{\mathrm{int}} = \bigg[ - H ^{\ast} _{\mu} k _{AB} \big( \partial ^{\mu } \phi ^{A} \big) \frac{\partial \bar{n} ^{B} (\varphi, \phi)}{\partial \varphi } + \phi ^{\ast} _{A}\bar{n} ^{A}(\varphi, \phi) \bigg] \eta, \label{154}\\
\delta \bigg[ - H ^{\ast} _{\mu} k _{AB} \big( \partial ^{\mu } \phi ^{A} \big) \frac{\partial \bar{n} ^{B} (\varphi, \phi)}{\partial \varphi } + \phi ^{\ast} _{A}\bar{n} ^{A}(\varphi, \phi) \bigg]= \partial _{\mu} t _{0} ^{\mu} (\varphi,[\phi]), \label{155}
\end{gather}
with $\bar{n} ^{A}$ and $t _{0} ^{\mu} (\varphi,[\phi])$ of the form (\ref{144}) and respectively (\ref{149}). Given the fact that from (\ref{154}) one determines all the deformed gauge transformations of the fields at order one in the coupling constant by detaching the antifields and replacing the ghosts with the corresponding gauge parameters (in our case $\eta \rightarrow \epsilon$), it follows that they are obtained simply by gauging the nontrivial, one-parameter rigid transformations of the fields obtained in the above. On behalf of (\ref{155}) and using the first definition in (\ref{24}) particularized to $\eta $ together with the last relation from (\ref{28}), it follows immediately that
\begin{align}
\delta a _{1} ^{\mathrm{int}} &= - \partial _{\mu} [ t _{0} ^{\mu} (\varphi,[\phi]) \eta ] + \gamma [ t _{0} ^{\mu} (\varphi,[\phi]) A_{\mu} ] - [ \gamma t _{0} ^{\mu} (\varphi,[\phi]) ] A _{\mu} \nonumber \\
&=  - \partial _{\mu} [ t _{0} ^{\mu} (\varphi,[\phi]) \eta ] + \gamma [ t _{0} ^{\mu} (\varphi,[\phi]) A_{\mu} ] \label{156}
\end{align}
since $\gamma t _{0} ^{\mu}=0$.  Taking into account formulas (\ref{156}) and (\ref{62}), we remark that it is precisely the gauge invariance of the conserved current (\ref{149}) under transformations (\ref{3})--(\ref{5}), equivalent to its $\gamma$-closeness, which turns the necessary condition (\ref{70}) for the existence of a nonvanishing solution $a _{0} ^{\mathrm{int}}$ to equation (\ref{62}) into a sufficient one. In this manner, (\ref{156}) renders that the general solution to (\ref{62}) can be written as
\begin{equation}
a _{0} ^{\mathrm{int}} = - t _{0} ^{\mu} (\varphi,[\phi]) A_{\mu} + \bar{a} _{0} ^{\mathrm{int}} \equiv - k _{AB} \big( \partial ^{\mu } \phi ^{A} \big) \bar{n} ^{B}(\varphi, \phi) A_{\mu} + \bar{a} _{0} ^{\mathrm{int}}. \label{157}
\end{equation}
Recalling that $a _{0} ^{\mathrm{int}}$ is the Lagrangian density of the coupled model at order one in the deformation parameter, we can partially synthesize (\ref{157}) by the standard result that the existence of a nontrivial rigid symmetry of action (\ref{2}) with a gauge-invariant current produces a minimal current-gauge field coupling at order one of perturbation theory. We observe that the nontrivial conserved current obtained in the above (as the right-hand side of (\ref{149})) under a weaker derivative order assumption implies that the associated interacting vertex at order one of perturbation theory, written as the first term from the right-hand side of (\ref{157}), truly conserves the number of derivatives on each field from the free limit since it is linear in the first-order derivatives of the fields.

The terms denoted by $\bar{a} _{0} ^{\mathrm{int}}$ stand for the general nontrivial solutions to the `homogeneous' equation associated with (\ref{62})
\begin{equation}
\gamma \bar{a} _{0} ^{\mathrm{int}} = \partial _{\mu} \bar{j} _{\mathrm{int},0} ^{\mu} \label{158}
\end{equation}
that should also comply with all the working hypotheses and couple the BF to the matter fields. Such solutions cannot deform the gauge transformations and depend only on the original fields and their spacetime derivatives, $\bar{a} _{0} ^{\mathrm{int}}= \bar{a} _{0} ^{\mathrm{int}} \big( [\Phi ^{\alpha_{0}}] \big)$. They provide solutions to the cross-coupling first-order deformation equation (\ref{50}) that are independent of the previous ones, which ended in antifield number $1$.  We stress that here we work in antifield number $0$, so we are not allowed to replace equation (\ref{158}) with its homogeneous version, $\gamma \bar{a} _{0} ^{\mathrm{int}} = 0$, like we did before, in antifield number $1$ (see the paragraph between formulas (\ref{60}) and (\ref{61})). Instead, we split $\bar{a} _{0} ^{\mathrm{int}}$ into
\begin{equation}
\bar{a} _{0} ^{\mathrm{int}} \big( [\Phi ^{\alpha_{0}}] \big) = \bar{a} _{0} ^{\prime \mathrm{int}} \big( [\Phi ^{\alpha_{0}}] \big) + \bar{a} _{0} ^{\prime \prime \mathrm{int}} \big( [\Phi ^{\alpha_{0}}] \big) , \label{159}
\end{equation}
where the first component is the general solution to the truly homogeneous equation corresponding to (\ref{158})
\begin{equation}
\gamma \bar{a} _{0} ^{\prime \mathrm{int}} \big( [\Phi ^{\alpha_{0}}] \big) = 0 \label{160}
\end{equation}
and the second corresponds to a nonvanishing current
\begin{equation}
\gamma \bar{a} _{0} ^{\prime \prime \mathrm{int}} \big( [\Phi ^{\alpha_{0}}] \big) = \partial _{\mu} \bar{j} _{\mathrm{int},0} ^{\mu}, \qquad \partial _{\mu} \bar{j} _{\mathrm{int},0} ^{\mu} \neq 0. \label{161}
\end{equation}
Actually, since the actions of the Koszul--Tate differential on all fields are vanishing (see the first relation in (\ref{24}) for $\chi ^{\Delta} = \Phi ^{\alpha_{0}}$) and the BRST differential reduces to (\ref{22}), equation (\ref{158}) defines by itself an element from $H ^{0} (s|d)$ computed in the space of local nonintegrated densities that satisfy the working hypotheses, which is both ghost- and antifield-independent
\begin{equation}
\gamma \bar{a} _{0} ^{\mathrm{int}} \big( [\Phi ^{\alpha_{0}}] \big) = \partial _{\mu} \bar{j} _{\mathrm{int},0} ^{\mu} \Leftrightarrow s\bar{a} _{0} ^{\mathrm{int}} \big( [\Phi ^{\alpha_{0}}] \big) = \partial _{\mu} \bar{j} _{\mathrm{int},0} ^{\mu}. \label{162}
\end{equation}
In view of the above equivalence, by trivial solution to (\ref{158}) we understand any $s$-exact object modulo a full divergence. In the sequel we determine the general nontrivial solutions to equations (\ref{160}) and (\ref{161}) that satisfy the general assumptions imposed on the deformations.

Due to the fact that $\bar{a} _{0} ^{\prime \mathrm{int}}$ depends only on the fields and their spacetime derivatives, it follows that equation (\ref{160}) is completely equivalent to the gauge invariance condition $\delta_{\Omega^{\alpha_{1}}} \bar{a} _{0} ^{\prime \mathrm{int}}=0$ and hence, by virtue of a previous result (see the paragraph containing formula (\ref{64})), its solutions depend (locally) only on the gauge-invariant quantities introduced in (\ref{64}) and their derivatives up to a finite order
\begin{equation}
\bar{a} _{0} ^{\prime \mathrm{int}} \big( [\Phi ^{\alpha_{0}}] \big) = h \big( [\omega ^{\Theta}] \big) , \label{163}
\end{equation}
so $h$ is actually a polynomial in all the quantities from (\ref{64}) and their spacetime derivatives excepting the undifferentiated scalar fields $\varphi $ and $\phi$, which are allowed to enter $h$ via a smooth dependence. We start from a general representation of $h$ in terms of $[\omega ^{\Theta}]$ that is local, covariant, and Poincar\'{e} invariant, but \emph{relax again the conservation of the number of derivatives on each field from the free limit} to the requirement that $h$ contains maximum two derivatives that may act on any of the fields. Subsequently, we eliminate the trivial contributions and show that the remaining (nontrivial) terms satisfy the stronger derivative order assumption as well. Under these circumstances, we begin with
\begin{align}
h \big( [\omega ^{\Theta}] \big) = & h _{1} (\varphi, \phi ) + h _{2} (\varphi, \phi ) \partial _{\rho} H^{\rho} + h _{3} (\varphi, \phi ) (\partial _{\rho} H ^{\rho}) (\partial _{\lambda} H ^{\lambda}) \nonumber \\
& + h _{4} (\varphi, \phi ) (\partial ^{\rho } B _{\rho \mu}) (\partial  _{\lambda } B ^{\lambda \mu}) + \tfrac{1}{2} h _{5} (\varphi, \phi ) (\partial _{[\mu } A _{\nu ]}) (\partial ^{[\mu } A ^{\nu ]}) \nonumber \\
& + h _{6} (\varphi, \phi ) (\partial _{\mu } \varphi ) (\partial ^{\mu } \varphi ) + \tfrac{1}{2} h _{7AB} (\varphi, \phi ) \big( \partial _{\mu } \phi ^{A} \big) \big( \partial ^{\mu } \phi ^{B} \big) \nonumber \\
& + 2 h _{8} (\varphi, \phi ) (\partial _{\mu } \varphi ) (\partial _{\rho } B ^{\rho \mu}) + h _{9A} (\varphi, \phi ) \big( \partial _{\mu } \phi ^{A} \big) (\partial _{\rho } B ^{\rho \mu}) \nonumber \\
& + h _{10A} (\varphi, \phi ) (\partial _{\mu } \varphi ) \big( \partial ^{\mu } \phi ^{A} \big) + h_{11} (\varphi, \phi ) \Box \varphi \nonumber \\
&+ h_{12 A} (\varphi, \phi ) \Box \phi ^{A} + \tfrac{1}{2} \varepsilon ^{\mu\nu\rho\lambda} h _{13} (\varphi, \phi ) (\partial _{[\mu } A _{\nu ]}) (\partial _{[\rho } A _{\lambda ]}), \label{164}
\end{align}
where all the coefficients denoted by $h$ stand for some arbitrary smooth functions of the undifferentiated scalar fields from the theory, with $h _{7AB}$ symmetric. With the help of definitions (\ref{24}) and (\ref{25}) and decomposing $h _{7AB}$ into
\begin{align}
h _{7AB} (\varphi, \phi ) &= \bigg( \frac{\partial h _{7A} (\varphi, \phi )}{\partial \phi ^{B}} + \frac{\partial h _{7B} (\varphi, \phi )}{\partial \phi ^{A}} \bigg) + \mu _{AB} (\varphi, \phi ), \label{165}\\
\mu _{AB} (\varphi, \phi ) &= \mu _{BA} (\varphi, \phi ), \qquad \mu _{AB} (\varphi, \phi ) \neq \frac{\partial u _{A} (\varphi, \phi )}{\partial \phi ^{B}} + \frac{\partial u _{B} (\varphi, \phi )}{\partial \phi ^{A}}, \label{166}
\end{align}
we find that
\begin{align}
h \big( [\omega ^{\Theta}] \big) = & h _{1} (\varphi, \phi ) + \tfrac{1}{2} \mu _{AB} (\varphi, \phi ) \big( \partial _{\mu } \phi ^{A} \big) \big( \partial ^{\mu } \phi ^{B} \big) \nonumber \\
&+ s \bigg\{ \varphi ^{\ast} \big( h _{2} (\varphi, \phi ) + h _{3} (\varphi, \phi ) \partial _{\lambda} H ^{\lambda} \big) + A ^{\ast \mu} \big( h _{4} (\varphi, \phi ) \partial ^{\rho } B _{\rho \mu} \nonumber \\
&+ h _{8} (\varphi, \phi ) \partial _{\mu } \varphi + h _{9A} (\varphi, \phi ) \partial _{\mu } \phi ^{A} \big) + H _{\mu} ^{\ast} \bigg[ - h _{6} (\varphi, \phi ) \partial ^{\mu } \varphi \nonumber \\
&+ \bigg( \frac{\partial h _{7A} (\varphi, \phi )}{\partial \varphi} - h _{10A} \bigg) \partial ^{\mu } \phi ^{A} - h _{8} (\varphi, \phi ) \partial _{\rho } B ^{\rho \mu} + \partial ^{\mu} h_{11} (\varphi, \phi ) \bigg] \nonumber \\
&- B _{\mu\nu} ^{\ast} \big( h _{5} (\varphi, \phi ) \partial ^{[\mu } A ^{\nu ]} + \varepsilon ^{\mu\nu\rho\lambda} h _{13} (\varphi, \phi ) \partial _{[\rho } A _{\lambda ]} \big) \nonumber \\
&+ \phi _{A} ^{\ast} k ^{AB} ( - h _{7B} (\varphi, \phi ) + h _{12B} (\varphi, \phi )) \bigg\} \nonumber \\
&+ \partial _{\mu} \big( h _{7A} (\varphi, \phi ) \partial ^{\mu} \phi ^{A} + h _{11} (\varphi, \phi ) \partial ^{\mu} \varphi \big) . \label{167}
\end{align}
Removing the trivial ($s$-exact modulo divergences) terms by setting zero the associated parameterizing functions
\begin{gather}
h _{2} = h _{3} = h _{4} = h _{5} = h _{6} = h _{8} = h _{11} = h _{13} = 0, \label{168}\\
h _{7A} = h _{9A} = h _{10A} = h _{12A} = 0, \label{169}
\end{gather}
and denoting the coefficient $h _{1}$ by $-\mathcal{V}$ for further convenience, we obtain that (\ref{167}) reduces to a sum between two types of terms only
\begin{equation}
h \big( [\omega ^{\Theta}] \big) \equiv \bar{a} _{0} ^{\prime \mathrm{int}} \big( [\Phi ^{\alpha_{0}}] \big) = -\mathcal{V} (\varphi, \phi ) + \tfrac{1}{2} \mu _{AB} (\varphi, \phi ) \big( \partial _{\mu } \phi ^{A} \big) \big( \partial ^{\mu } \phi ^{B} \big) ,\label{170}
\end{equation}
so it is parameterized by an arbitrary scalar smooth function of the undifferentiated scalar fields ($\mathcal{V} (\varphi, \phi )$) and by a symmetric quadratic matrix of order $N$ ($\mu (\varphi, \phi ) \equiv \mu _{AB} (\varphi, \phi), A,B = \overline{1,N}$) with elements also smooth functions of the same fields, which are arbitrary up to the requirement (\ref{166}). It is clear now that the components from the right-hand side of (\ref{170}) satisfy all the working hypotheses including the conservation of the number of derivatives on each field with respect to free Lagrangian since they contain just terms with maximum two derivatives that are precisely quadratic in the first-order derivatives of the matter fields and, most important, they are truly nontrivial. This is because on the one hand $\mathcal{V}$ exhibits no spacetime derivatives while any divergence or $s$-exact term incorporates at least one (see definitions (\ref{24})--(\ref{31}) with nonvanishing right-hand sides and decomposition (\ref{22})), so $\mathcal{V} (\varphi, \phi )$ is trivial in $H^{0}(s|d)$ iff $\mathcal{V} (\varphi, \phi ) = 0$ and, on the other hand, the symmetric functions $\mu _{AB}$ are subject to (\ref{166}). Indeed, it easy to see that the terms from the right-hand side of (\ref{170}) quadratic in the first-order derivatives of the matter fields are in a trivial of class of $H^{0} (s|d)$ if and only if the elements of the symmetric quadratic matrix of order $N$ are written like the symmetric first-order derivatives of the components of an $N$-dimensional vector with respect to the matter fields
\begin{equation}
\mu _{AB} (\varphi, \phi ) = \mu _{AB} ^{\mathrm{triv}} (\varphi, \phi ) \equiv \frac{\partial u _{A} (\varphi, \phi )}{\partial \phi ^{B}} + \frac{\partial u _{B} (\varphi, \phi )}{\partial \phi ^{A}}, \label{171}
\end{equation}
in which case we have that
\begin{align}
\tfrac{1}{2} \mu _{AB} ^{\mathrm{triv}} (\varphi, \phi ) \big( \partial _{\mu } \phi ^{A} \big) \big( \partial ^{\mu } \phi ^{B} \big) =& s \bigg[ H _{\mu} ^{\ast} \big( \partial ^{\mu } \phi ^{A} \big) \frac{\partial u _{A} (\varphi, \phi )}{\partial \varphi} - k ^{AB} \phi _{A} ^{\ast} u _{B} (\varphi, \phi ) \bigg] \nonumber \\
&+ \partial _{\mu} \big( u _{A} (\varphi, \phi ) \partial ^{\mu} \phi ^{A} \big) . \label{172}
\end{align}
In view of the above result, we will call any symmetric matrix of the form (\ref{171}) involved in vertices of the type $\mu _{AB} ^{\mathrm{triv}} (\varphi, \phi )\big( \partial _{\mu } \phi ^{A} \big) \big( \partial ^{\mu } \phi ^{B} \big)$ to be \emph{trivial}. Such terms have already been considered in (\ref{164}) via decomposition (\ref{165}) of $h _{7AB}$ between a trivial and a nontrivial part (see the quantities from (\ref{167}) involving the functions $h _{7A}$). In conclusion, condition (\ref{166}) guarantees that the second kind of terms from the right-hand side of (\ref{170}) belongs to a nontrivial class from the cohomology $H^{0} (s|d)$ and therefore gives rise to allowed first-order deformations that do not modify the initial gauge transformations. It is interesting to regard the previous necessary and sufficient condition yet from an another perspective: if needed, one can always add a trivial part to $\mu (\varphi, \phi )$ and use the associated terms quadratic in the first-order derivatives of the matter fields as `counterterms' to similar quantities appearing in higher orders of perturbation theory (possibly up to an appropriate redefinition of the coupling constant).

Equation (\ref{161}) can be approached in a standard fashion (for instance, see~\cite{JPA2006,PRD2006b}) by decomposing $\bar{a} _{0} ^{\prime \prime \mathrm{int}}$ according to the number of derivatives and by solving the emerging equivalent equations via introducing a derivation in the algebra of the fields and of their derivatives that counts the powers of all fields and of their derivatives excepting the undifferentiated scalar fields (BF and matter). Proceeding along this line it is easy to see that all the solutions to (\ref{161}) that fulfill the working hypotheses are nonetheless trivial, so we can safely take
\begin{equation}
\bar{a} _{0} ^{\prime \prime \mathrm{int}} \big( [\Phi ^{\alpha_{0}}] \big) = 0. \label{173}
\end{equation}

Putting together the results given in formulas (\ref{154}), (\ref{157}), (\ref{159}), (\ref{163}), (\ref{170}), and (\ref{173}) via the former expansion in (\ref{57}), we conclude that the general, nontrivial expression of the nonintegrated density of the  first-order deformation that couples the BF to the matter fields and satisfies all the working hypotheses reads as
\begin{align*}
a ^{\mathrm{int}} =& a _{0} ^{\mathrm{int}} + a _{1} ^{\mathrm{int}}, \\
a _{0} ^{\mathrm{int}} = & - k _{AB} \big( \partial ^{\mu } \phi ^{A} \big) \bar{n} ^{B}(\varphi, \phi) A_{\mu} - \mathcal{V} (\varphi, \phi ) \\
&+ \tfrac{1}{2} \mu _{AB} (\varphi, \phi ) \big( \partial _{\mu } \phi ^{A} \big) \big( \partial ^{\mu } \phi ^{B} \big) ,\\
a _{1} ^{\mathrm{int}} =& \bigg[ - H ^{\ast} _{\mu} k _{AB} \big( \partial ^{\mu } \phi ^{A} \big) \frac{\partial \bar{n} ^{B} (\varphi, \phi)}{\partial \varphi } + \phi ^{\ast} _{A}\bar{n} ^{A}(\varphi, \phi) \bigg] \eta ,
\end{align*}
where $\bar{n} ^{A} (\varphi, \phi)$ is given by (\ref{144}) and the symmetric functions $\mu _{AB} (\varphi, \phi )$ are nontrivial (relations (\ref{166})).
This completes the proof regarding the general solutions to equations (\ref{61}) and (\ref{62}).

\section{Main properties of the deformed generating set of gauge transformations\label{B}}

In order to analyze the characteristic features of the deformed gauge transformations (\ref{220})--(\ref{223}) we need the explicit form of the interacting field equations resulting from action (\ref{211}), namely,
\begin{align}
\frac{\delta \bar{S}^{\mathrm{L}}[\Phi^{\alpha_{0}}]}{\delta \varphi} = & - D _{\mu} ^{\prime} H ^{\mu} - g \frac{\partial \mathcal{V} (\varphi, \phi )}{\partial \varphi} + \frac{g}{2} \frac{\partial \mu _{AB} (\varphi, \phi)}{\partial \varphi} \big( \hat{D} _{\mu} \phi ^{A} \big) \big( \hat{D} ^{\mu} \phi ^{B} \big) \nonumber \\
& - g (k _{AB} + g \mu _{AB} (\varphi, \phi)) \big( \hat{D} _{\mu} \phi ^{A} \big) A ^{\mu} \frac{\partial \bar{n} ^{B} (\varphi, \phi)}{\partial \varphi} \approx 0, \label{215}\\
\frac{\delta \bar{S}^{\mathrm{L}}[\Phi^{\alpha_{0}}]}{\delta A _{\mu}} =& - \partial _{\lambda} B ^{\lambda\mu} - g W (\varphi ) H ^{\mu} \nonumber \\
&- g (k _{AB} + g \mu _{AB} (\varphi, \phi)) \big( \hat{D} ^{\mu} \phi ^{A} \big) \bar{n} ^{B} (\varphi, \phi) \approx 0, \label{216}\\
\frac{\delta \bar{S}^{\mathrm{L}}[\Phi^{\alpha_{0}}]}{\delta H ^{\mu}} =& D _{\mu} \varphi \approx 0, \qquad \frac{\delta \bar{S}^{\mathrm{L}}[\Phi^{\alpha_{0}}]}{\delta B ^{\mu\nu}} = \frac{\delta S^{\mathrm{L}}[\Phi^{\alpha_{0}}]}{\delta B ^{\mu\nu}} \equiv \tfrac{1}{2} \partial _{[\mu }A _{\nu ]} \approx 0, \label{217}\\
\frac{\delta \bar{S}^{\mathrm{L}}[\Phi^{\alpha_{0}}]}{\delta \phi ^{A}} = & - \partial ^{\mu} \big[ (k _{AB} + g \mu _{AB} (\varphi, \phi)) \big( \hat{D} _{\mu} \phi ^{B} \big) \big] - g \frac{\partial \mathcal{V} (\varphi, \phi )}{\partial \phi ^{A}} \nonumber \\
& + \frac{g}{2} \frac{\partial \mu _{BC} (\varphi, \phi)}{\partial \phi ^{A}} \big( \hat{D} _{\mu} \phi ^{B} \big) \big( \hat{D} ^{\mu} \phi ^{C} \big) \nonumber \\
& - g (k _{BC} + g \mu _{BC} (\varphi, \phi)) \big( \hat{D} _{\mu} \phi ^{B} \big) A ^{\mu} \frac{\partial \bar{n} ^{C} (\varphi, \phi)}{\partial \phi ^{A}} \approx 0, \label{218}
\end{align}
written in terms of (\ref{212})--(\ref{213}) and of the additional ``covariant derivative''
\begin{equation}
D _{\mu} ^{\prime} \equiv \partial _{\mu} + g \frac{d W (\varphi)}{d \varphi} A _{\mu} , \label{219}
\end{equation}
that may act on \emph{any} object involved in the Lagrangian formulation of the coupled theory (fields, gauge parameters, etc.). It is easy to see from the above formulas that the field equations satisfy the derivative order assumption. Indeed, on the one hand the derivative order of all BF field equations is equal to one and that of the matter fields is equal to two (with respect to the matter fields themselves) via the term $- k _{AB} \Box \phi ^{B}$ following from the first kind of quantities on the right-hand side of (\ref{218})) and, on the other hand, each term from every field equation is strictly linear in the first-order derivatives of the BF fields. Excepting the EL derivatives of action (\ref{211}) with respect to the BF two-form $B ^{\mu\nu}$, which coincide with those from the free limit, the others are deformed by contributions due to both the selfinteractions among the BF fields (in the first order of perturbation theory) or to the cross-couplings between BF and matter field sectors (at orders $1$, $2$, and $3$).

In the sequel we investigate the main properties of the deformed generating set of gauge transformations, (\ref{220})--(\ref{223}). The associated gauge algebra is defined by the commutators among the above gauge transformations of the fields, which, in turn, result by retaining from the deformed solution to the master equation, (\ref{209}), the terms of antifield number $2$ that are quadratic (in the ghosts with the pure ghost number equal to $1$). If no such components were present in $\bar{S}$, then all the commutators would vanish, and therefore the corresponding gauge algebra would be Abelian. In general the pieces of antifield number $2$ quadratic in the ghosts fall in two possible classes: either linear in the antifields of the ghosts with the pure ghost number equal to $1$ or quadratic in the antifields of the original fields. The appearance only of elements from the first class signalizes that all the commutators among the gauge transformations of the fields close off-shell, but some are nonvanishing, so the gauge algebra is still closed, but non-Abelian. If there exists at least one term from the second class, i.e. quadratic in the antifields of the original fields, this means that at least one commutator among the gauge transformations of the fields closes on-shell via trivial gauge transformations, i.e. on the stationary surface through some antisymmetric combinations of field equations, and therefore the emerging gauge algebra is said to be open. Inspecting from this perspective the structure of (\ref{209}) by means of formulas (\ref{32}), (\ref{51})--(\ref{56}), (\ref{174})--(\ref{176}), (\ref{196})--(\ref{198}), and (\ref{202})--(\ref{204}), we notice that there appear nonvanishing elements from both classes, but only at order one in $g$, via the first-order deformation $a ^{\mathrm{BF}}$ (the last two terms on the right-hand side of (\ref{54}), quadratic in $\eta$ and $C ^{\mu\nu}$). The former is linear in the antifields $C _{\mu\nu} ^{\ast}$ and contains the first-order derivative of the smooth function $W(\varphi)$, while the latter is quadratic in the antifields $H _{\mu} ^{\ast}$ of the BF one-form $H ^{\mu}$ and includes the second-order derivative of $W$. Consequently, only the commutators among the gauge transformations that depend on the gauge parameters $\xi ^{\mu\nu}$ (since $ C _{\mu\nu} ^{\ast}\leftrightarrow C^{\mu\nu} \leftrightarrow \xi ^{\mu\nu}$ with the help of formulas (\ref{6}) and (\ref{15})) may be nonvanishing, i.e., those of the BF fields $H ^{\mu}$ and $B ^{\mu\nu}$ (see the first two terms on the second line of (\ref{221}) and the last from (\ref{222})), while the sole commutator that may close one-shell is that corresponding to $H ^{\mu}$ precisely by antisymmetric combinations of its own field equations, $M ^{\mu\lambda}\delta \bar{S} ^{\mathrm{L}} / \delta H^{\lambda} \approx 0$, with $M ^{\mu\lambda} =- M ^{\lambda\mu}$.  All this information extracted from the structure of the deformed solution to the master equation is translated at the level of the Lagrangian formulation of the interacting theory along the following formulas
\begin{align}
\big[ \bar{\delta} _{\Omega _{1}^{\alpha_{1}}}, \bar{\delta} _{\Omega _{2} ^{\alpha_{1}}} \big] \phi ^{A} = &0, \qquad \big[ \bar{\delta} _{\Omega _{1}^{\alpha_{1}}}, \bar{\delta} _{\Omega _{2} ^{\alpha_{1}}} \big] \varphi =0, \qquad \big[ \bar{\delta} _{\Omega _{1}^{\alpha_{1}}}, \bar{\delta} _{\Omega _{2} ^{\alpha_{1}}} \big] A _{\mu} = 0, \label{229} \\
\big[ \bar{\delta} _{\Omega _{1}^{\alpha_{1}}}, \bar{\delta} _{\Omega _{2} ^{\alpha_{1}}} \big] H ^{\mu} =& -2 D _{\lambda} ^{\prime} \bigg[ g \frac{dW (\varphi)}{d \varphi } (\epsilon _{1} \xi _{2} ^{\lambda\mu} - \epsilon _{2} \xi _{1} ^{\lambda\mu} ) \bigg] \nonumber\\
&+ 2g \frac{d ^{2} W (\varphi )}{d \varphi ^{2}} (\xi _{1} ^{\mu\lambda}\epsilon _{2} - \xi _{2} ^{\mu\lambda}\epsilon _{1}) D _{\lambda} \varphi, \label{230}\\
\big[ \bar{\delta} _{\Omega _{1}^{\alpha_{1}}}, \bar{\delta} _{\Omega _{2} ^{\alpha_{1}}} \big] B ^{\mu\nu} =& 2g W(\varphi) \bigg[ g \frac{dW (\varphi)}{d \varphi } (\epsilon _{1} \xi _{2} ^{\mu\nu} - \epsilon _{2} \xi _{1} ^{\mu\nu} ) \bigg] \label{231},
\end{align}
where the ``covariant derivatives'' $D _{\lambda} ^{\prime}$ and $D _{\lambda} \varphi$ are of the form (\ref{219}) and respectively (\ref{212}), while the commutators among the gauge transformations are considered with respect to two independent sets of gauge parameters organized in agreement with notation (\ref{6})
\begin{equation}
\Omega _{1}^{\alpha_{1}} \equiv \{ \epsilon _{1}, \xi _{1} ^{\lambda\mu}, \epsilon _{1} ^{\lambda\mu\nu}\}, \qquad \Omega _{2}^{\alpha_{1}} \equiv \{ \epsilon _{2}, \xi _{2} ^{\lambda\mu}, \epsilon _{2} ^{\lambda\mu\nu}\}. \label{232}
\end{equation}
Comparing the right-hand sides of (\ref{229})--(\ref{231}) with the corresponding gauge transformations from (\ref{220})--(\ref{223}) and using the former field equation from (\ref{217}), we find that the previous commutators can be written like
\begin{align}
\big[ \bar{\delta} _{\Omega _{1}^{\alpha_{1}}}, \bar{\delta} _{\Omega _{2} ^{\alpha_{1}}} \big] \phi ^{A} = &0, \qquad \big[ \bar{\delta} _{\Omega _{1}^{\alpha_{1}}}, \bar{\delta} _{\Omega _{2} ^{\alpha_{1}}} \big] \varphi =0, \qquad \big[ \bar{\delta} _{\Omega _{1}^{\alpha_{1}}}, \bar{\delta} _{\Omega _{2} ^{\alpha_{1}}} \big] A _{\mu} = 0, \label{233} \\
\big[ \bar{\delta} _{\Omega _{1}^{\alpha_{1}}}, \bar{\delta} _{\Omega _{2} ^{\alpha_{1}}} \big] H ^{\mu} =& \bar{\delta} _{\tilde{\Omega } ^{\alpha_{1}}} H ^{\mu} + M ^{\mu\lambda} (\Omega _{1}^{\alpha_{1}}, \Omega _{2}^{\alpha_{1}} ) \frac{\delta \bar{S}^{\mathrm{L}}[\Phi^{\alpha_{0}}]}{\delta H ^{\lambda}} , \label{234}\\
\big[ \bar{\delta} _{\Omega _{1}^{\alpha_{1}}}, \bar{\delta} _{\Omega _{2} ^{\alpha_{1}}} \big] B ^{\mu\nu} =& \bar{\delta} _{\tilde{\Omega } ^{\alpha_{1}}} B ^{\mu\nu} , \label{235}
\end{align}
in terms of a new set of gauge parameters
\begin{equation}
\tilde{\Omega } ^{\alpha_{1}} \equiv \bigg\{ \tilde{\epsilon} = 0 , \tilde{\xi} ^{\lambda\mu} = g \frac{dW (\varphi)}{d \varphi } (\epsilon _{1} \xi _{2} ^{\lambda\mu} - \epsilon _{2} \xi _{1} ^{\lambda\mu} ), \tilde{\epsilon} ^{\lambda\mu\nu} = 0 \bigg\} \label{236}
\end{equation}
and of the antisymmetric coefficients
\begin{equation}
M ^{\mu\lambda} (\Omega _{1}^{\alpha_{1}}, \Omega _{2}^{\alpha_{1}} ) = 2g \frac{d ^{2} W (\varphi )}{d \varphi ^{2}} (\xi _{1} ^{\mu\lambda}\epsilon _{2} - \xi _{2} ^{\mu\lambda}\epsilon _{1}), \qquad M ^{\mu\lambda} = - M ^{\lambda\mu}. \label{237}
\end{equation}
In conclusion, the gauge algebra associated with (\ref{220})--(\ref{223}) is open if and only if the selfinteractions among the BF fields are allowed and the second-order derivative of $W$ is nonvanishing (so the coefficients $M ^{\mu\lambda}$ are not equal to zero). If it is vanishing, but the first-order derivative of $W$ is not, then the gauge algebra is closed, but non-Abelian (the coefficients $M ^{\mu\lambda}$ become zero, but the gauge parameters $\tilde{\xi} ^{\lambda\mu}$ given in (\ref{236}) resist), so it is still deformed with respect to the initial, Abelian one. Finally, if the BF selfinteractions are excluded ($W=0$), then all the above commutators vanish off-shell, such that the gauge algebra of the cross-coupled theory remains Abelian, like that from the free limit.

The deformation procedure cannot alter either the maximum reducibility order of the generating set of gauge transformations or the number of reducibility relations at each stage since it preserves both the field sector and the number of physical degrees of freedom from the free limit. But it may modify the form of the reducibility functions (and hence the transformations of the gauge or reducibility parameters of a given order in terms of those corresponding to the next order) and of the associated reducibility relations; for instance, if the reducibility relations from the free limit held off-shell, then it is possible that the deformed ones take place on-shell, meaning on the deformed stationary surface. These features are dictated by the pieces linear in the ghosts with the pure ghost number strictly greater than $1$ from the deformed solution to the master equation in \emph{strictly positive orders of perturbation theory}. If no such components were present, then both the reducibility functions and relations would be those from the free limit. In the opposite situation, two kinds of elements linear in the ghosts with $\pgh >1$ are of interest: those likewise linear in the antifields (of some ghosts since their antifield number should also be at least equal to $2$) and respectively those simultaneously quadratic in the antifields and containing two antifields of the original fields if $\pgh = 2 = \agh$ or a single one if $\pgh = \agh >2$. Whenever only elements belonging to the first kind are allowed, then some of the reducibility functions are modified with respect those from the free limit, but all the deformed reducibility relations hold like in the free limit, namely off-shell here. If at least one element from the second class is detected, then the associated reducibility relations hold only on-shell. With these observations in mind, we notice in (\ref{209}) contributions to both classes at the first order of perturbation theory only, coming again from the purely BF deformation (\ref{51}) (all the terms from (\ref{54}) linear in the ghosts $C ^{\mu\nu\rho}$ and just five out of the seven components from (\ref{55}) linear in $C ^{\mu\nu\rho\lambda}$). Consequently, all the reducibility functions may be affected with respect to their free limit, but only at order one in $g$, and some of the reducibility relations may hold now on-shell (at the same order), depending on the choice and properties of the smooth function $W(\varphi)$. Their detailed analysis will be given below maintaining notations (\ref{1}), (\ref{6}), (\ref{8}), and (\ref{11}) and relying on the deformed gauge transformations (\ref{220})--(\ref{223}). Related to the first-order reducibility, we transform the gauge parameters from (\ref{6}) in terms of the first-order reducibility ones, (\ref{8}), by adding to relations (\ref{7}) some supplementary quantities induced by the corresponding elements from the first class present in (\ref{54})
\begin{equation}
\Omega ^{\alpha_{1}}=\Omega ^{\alpha_{1}} (\Omega ^{\alpha_{2}}) \Leftrightarrow \left\{
\begin{array}{l}
\epsilon (\Omega ^{\alpha_{2}}) =0,\\
\xi^{\mu\nu} (\Omega ^{\alpha_{2}}) =-3 D _{\lambda} ^{\prime} \xi^{\lambda\mu\nu},\\
\epsilon ^{\mu\nu\rho} (\Omega ^{\alpha_{2}}) =-4 \partial _{\lambda} \epsilon^{\lambda\mu\nu\rho} - 2g W (\varphi) \xi ^{\mu\nu\rho},
\end{array}
\right. \label{238}
\end{equation}
with $D _{\lambda} ^{\prime}$ like in (\ref{219}). Consequently, the deformed gauge transformations of the fields, (\ref{220})--(\ref{223}), are transformed like
\begin{align}
\bar{\delta} _{\Omega ^{\alpha_{1}}(\Omega ^{\alpha_{2}})} \phi ^{A} &= 0, \qquad \bar{\delta} _{\Omega ^{\alpha_{1}}(\Omega ^{\alpha_{2}})} \varphi =0, \qquad \bar{\delta} _{\Omega ^{\alpha_{1}}(\Omega ^{\alpha_{2}})} A _{\mu} =0, \label{239} \\
\bar{\delta} _{\Omega ^{\alpha_{1}}(\Omega ^{\alpha_{2}})} H ^{\mu} &= 6 g \xi^{\mu\rho\lambda} \bigg( \frac{d ^{2} W (\varphi )}{d \varphi ^{2}}  A _{\rho} \frac{\delta \bar{S}^{\mathrm{L}}[\Phi^{\alpha_{0}}]}{\delta H ^{\lambda}} - \frac{d W (\varphi )}{d \varphi } \frac{\delta \bar{S}^{\mathrm{L}}[\Phi^{\alpha_{0}}]}{\delta B ^{\rho\lambda}} \bigg), \label{240} \\
\bar{\delta} _{\Omega ^{\alpha_{1}}(\Omega ^{\alpha_{2}})} B ^{\mu\nu} &= 6g \xi^{\mu\nu\lambda} \frac{d W (\varphi )}{d \varphi } \frac{\delta \bar{S}^{\mathrm{L}}[\Phi^{\alpha_{0}}]}{\delta H ^{\lambda}} \label{241}
\end{align}
and represent the first-order reducibility relations of the interacting model. Regarding the second-order reducibility, we express the first-order reducibility parameters from (\ref{8}) in terms of the second-order reducibility ones (see (\ref{11})) by modifying relations (\ref{10}) with some additional pieces generated by the associated elements from the first class appearing in (\ref{55})
\begin{equation}
\Omega ^{\alpha_{2}} = \Omega ^{\alpha_{2}} (\Omega ^{\alpha_{3}}) \Leftrightarrow
\left\{
\begin{array}{l}
\xi^{\mu\nu\rho} (\Omega ^{\alpha_{3}}) =-4 D _{\lambda} ^{\prime} \xi^{\lambda\mu\nu\rho},\\
\epsilon ^{\lambda\mu\nu\rho} (\Omega ^{\alpha_{3}})= 2g W (\varphi) \xi^{\lambda\mu\nu\rho},
\end{array}
\right. \label{242}
\end{equation}
such that the transformed gauge parameters (\ref{238}) become
\begin{align}
\epsilon (\Omega ^{\alpha_{2}} (\Omega ^{\alpha_{3}})) =& 0, \label{243}\\
\xi ^{\mu\nu} (\Omega ^{\alpha_{2}} (\Omega ^{\alpha_{3}})) =& 12 g \xi^{\mu\nu\rho\lambda} \bigg( \frac{d ^{2} W (\varphi )}{d \varphi ^{2}}  A _{\rho} \frac{\delta \bar{S}^{\mathrm{L}}[\Phi^{\alpha_{0}}]}{\delta H ^{\lambda}} \nonumber \\
&- \frac{d W (\varphi )}{d \varphi } \frac{\delta \bar{S}^{\mathrm{L}}[\Phi^{\alpha_{0}}]}{\delta B ^{\rho\lambda}} \bigg) , \label{244}\\
\epsilon ^{\mu\nu\rho} (\Omega ^{\alpha_{2}} (\Omega ^{\alpha_{3}})) =& 8g \xi^{\mu\nu\rho\lambda} \frac{d W (\varphi )}{d \varphi } \frac{\delta \bar{S}^{\mathrm{L}}[\Phi^{\alpha_{0}}]}{\delta H ^{\lambda}}  \label{245}
\end{align}
and provide the second-order reducibility relations of the interacting theory. With the help of results (\ref{238})--(\ref{245}), we conclude that at least some of the reducibility functions for the generating set of gauge transformations corresponding to the interacting theory are deformed with respect to that of the free limit if and only if the selfinteractions among the BF fields are permitted. If in addition at least the first-order derivative of the function $W(\varphi)$ is nonvanishing, then both the reducibility relations of order one and two hold on-shell, by contrast to those associated with the free model.

The remaining terms from (\ref{209}), not taken in account so far, of antifield number $3$ and respectively $4$, are entirely contained in the components (\ref{55}) and (\ref{56}) of the purely BF first-order deformation $a ^{\mathrm{BF}}$. They bring contributions to the higher order tensor structure functions corresponding to the deformed generating set of gauge transformations (\ref{220})--(\ref{223}) and are fully manifested (always only at the first order of perturbation theory) if and only if all the derivatives of the function $W(\varphi)$ up to the fourth order inclusively are not vanishing.


\begin{thebibliography}{99}
\bibitem{birmingham91} D.~Birmingham, M.~Blau, M.~Rakowski, G.~Thompson, Topological field theory, Phys. Rep. \textbf{209(4--5)} (1991) 129--340

    http://dx.doi.org/10.1016/0370-1573(91)90117-5

\bibitem{stroblspec} P.~Schaller, T.~Strobl, Poisson structure induced (topological) field theories, Mod. Phys. Lett. A \textbf{9(33)} (1994) 3129--3136

    http://dx.doi.org/10.1142/S0217732394002951

    arXiv:hep-th/9405110, http://arxiv.org/abs/hep-th/9405110

\bibitem{ezawa} K.~Ezawa, Ashtekar's formulation for $N = 1,2$ supergravities as ``constrained'' BF theories, Prog. Theor. Phys. \textbf{95(5)} (1996) 863--882

    http://dx.doi.org/10.1143/PTP.95.863

     arXiv:hep-th/9511047, http://arxiv.org/abs/hep-th/9511047

\bibitem{freidel} L.~Freidel, K.~Krasnov, R.~Puzio, BF description of higher-dimensional gravity theories, Adv. Theor. Math. Phys. \textbf{3} (1999) 1289--1324

    arXiv:hep-th/9901069, http://arxiv.org/abs/hep-th/9901069

\bibitem{smolin} L.~Smolin, Holographic formulation of quantum general relativity, Phys. Rev. D \textbf{61(8)} (2000) 084007

    http://dx.doi.org/10.1103/PhysRevD.61.084007

    arXiv:hep-th/9808191, http://arxiv.org/abs/hep-th/9808191

\bibitem{ling} Y.~Ling, L.~Smolin, Holographic formulation of quantum supergravity, Phys. Rev. D \textbf{63(6)} (2001) 064010

    http://dx.doi.org/10.1103/PhysRevD.63.064010

    arXiv:hep-th/0009018, http://arxiv.org/abs/hep-th/0009018

\bibitem{bf1} C.~Bizdadea, C.~C.~Ciobirca, E.~M.~Cioroianu, S.~O.~Saliu, S.~C.~Sararu, Hamiltonian BRST deformation of a class of $n$-dimensional BF-type theories, J. High Energy Phys. JHEP \textbf{0301} (2003) 049

    http://dx.doi.org/10.1088/1126-6708/2003/01/049

    arXiv:hep-th/0302037, http://arxiv.org/abs/hep-th/0302037

\bibitem{bf2} E.~M.~Cioroianu, S.~C.~Sararu, Self-interactions in a topological BF-type model in $D=5$, J. High Energy Phys. JHEP \textbf{0507} (2005) 056

    http://dx.doi.org/10.1088/1126-6708/2005/07/056

    arXiv:hep-th/0508035, http://arxiv.org/abs/hep-th/0508035

\bibitem{bf3} E.~M.~Cioroianu, S.~C.~Sararu, PT-symmetry breaking hamiltonian interactions in BF models, Int. J. Mod. Phys. A \textbf{21(12)} (2006) 2573--2599

    http://dx.doi.org/10.1142/S0217751X06029089

    arXiv:hep-th/0606164, http://arxiv.org/abs/hep-th/0606164

\bibitem{bf4} E.~M.~Cioroianu, S.~C.~Sararu, Two-dimensional interactions between a BF-type theory and a collection of vector fields, Int. J. Mod. Phys. A \textbf{19(24)} (2004) 4101--4125

    http://dx.doi.org/10.1142/S0217751X04019470

    arXiv:hep-th/0501056, http://arxiv.org/abs/hep-th/0501056

\bibitem{bf5} C.~Bizdadea, E.~M.~Cioroianu, I.~Negru, S.~O.~Saliu, S.~C.~Sararu, On the generalized Freedman-Townsend model, J. High Energy Phys. JHEP \textbf{0610} (2006) 004

    http://dx.doi.org/10.1088/1126-6708/2006/10/004

    arXiv:0704.3407 [hep-th], http://arxiv.org/abs/0704.3407

\bibitem{bf6} C.~Bizdadea, E.~M.~Cioroianu, S.~C.~Sararu, Couplings between a collection of BF models and a set of three-form gauge fields, Int. J. Mod. Phys. A \textbf{21(31)} (2006) 6477--6490

    http://dx.doi.org/10.1142/S0217751X06034331

    arXiv:0704.2656 [hep-th], http://arxiv.org/abs/0704.2656

\bibitem{bf7} E.~M.~Cioroianu, S.~C.~Sararu, Consistent interactions between BF and massive Dirac fields. A cohomological approach, Rom. Rep. Phys. \textbf{57(2)} (2005) 189--203

\bibitem{PLB1993} G.~Barnich, M.~Henneaux, Consistent couplings between fields with a gauge freedom and deformations of the master equation, Phys. Lett. B \textbf{311(1--4)} (1993) 123--129

     http://dx.doi.org/10.1016/0370-2693(93)90544-R

     arXiv:hep-th/9304057, http://arxiv.org/abs/hep-th/9304057

\bibitem{CM1998} M.~Henneaux, Consistent interactions between gauge fields: the cohomological approach, Contemp. Math. \textbf{219} (1998) 93--110

    Proceedings of a Conference on Secondary Calculus and Cohomological Physics, August 24-31, 1997, Moscow, Russia, eds. M. Henneaux, J. Krasil'shchik, A. Vinogradov, American Mathematical Society, 1998

    arXiv:hep-th/9712226, http://arxiv.org/abs/hep-th/9712226

\bibitem{CMP1995a} G.~Barnich, F.~Brandt, M.~Henneaux, Local BRST cohomology in the antifield formalism: I. General theorems, Commun. Math.
    Phys. \textbf{174(1)} (1995) 57--91

    http://dx.doi.org/10.1007/BF02099464

    arXiv:hep-th/9405109, http://arxiv.org/abs/hep-th/9405109

\bibitem{CMP1995b} G.~Barnich, F.~Brandt, M.~Henneaux, Local BRST cohomology in the antifield formalism: II. Application to Yang--Mills theory, Commun. Math. Phys. \textbf{174(1)} (1995) 93--116

    http://dx.doi.org/10.1007/10.1007/BF02099465

    arXiv:hep-th/9405194, http://arxiv.org/abs/hep-th/9405194

\bibitem{PR2000} G.~Barnich, F.~Brandt, M.~Henneaux, Local BRST cohomology in gauge theories, Phys. Rep. \textbf{338(5)} (2000) 439--569

    http://dx.doi.org/10.1016/S0370-1573(00)00049-1

    arXiv:hep-th/0002245, http://arxiv.org/abs/hep-th/0002245

\bibitem{Englert} F.~Englert, R.~Brout, Broken symmetry and the mass of gauge vector mesons, Phys. Rev. Lett. \textbf{13(9)} (1964) 321--323

    http://dx.doi.org/10.1103/PhysRevLett.13.321

\bibitem{Higgs1} P.~W. Higgs, Broken symmetries, massless particles and gauge fields, Phys. Lett. \textbf{12(2)} (1964) 132--133

    http://dx.doi.org/10.1016/0031-9163(64)91136-9

\bibitem{Higgs2} P.~W. Higgs, Broken symmetries and the masses of gauge bosons, Phys. Rev. Lett. \textbf{13(16)} (1964) 508--508

    http://dx.doi.org/10.1103/PhysRevLett.13.508

\bibitem{Kibble} G.~S.~Guralnik, C.~R.~Hagen, T.~W.~B.~Kibble, Global conservation laws and massless particles, Phys. Rev. Lett. \textbf{13(20)} (1964) 585--587
    http://dx.doi.org/10.1103/PhysRevLett.13.585

\bibitem{BVPLB81} I.~A.~Batalin, G.~A.~Vilkovisky, Gauge algebra and quantization, Phys. Lett. B \textbf{102(1)} (1981) 27--31

    http://dx.doi.org/10.1016/0370-2693(81)90205-7

\bibitem{BVPLB83} I.~A.~Batalin, G.~A.~Vilkovisky, Feynman rules for reducible gauge theories, Phys. Lett. B \textbf{120(1--3)} (1983) 166--170

    http://dx.doi.org/10.1016/0370-2693(83)90645-7

\bibitem{BVPRD83} I.~A.~Batalin, G.~A.~Vilkovisky, Quantization of gauge theories with linearly dependent generators, Phys. Rev. D \textbf{28(10)} (1983) 2567--2582

    http://dx.doi.org/10.1103/PhysRevD.28.2567

\bibitem{BVNPB84} I.~A.~Batalin, G.~A.~Vilkovisky, Closure of the gauge algebra, generalized Lie equations and Feynman rules, Nucl. Phys. B \textbf{234(1)} (1984) 106--124

    http://dx.doi.org/10.1016/0550-3213(84)90227-X

\bibitem{CMP1990a} J.~M.~L.~Fisch, M.~Henneaux, Homological perturbation theory and the algebraic structure of the antifield-antibracket formalism for gauge theories, Commun. Math. Phys. \textbf{128(3)} (1990) 627--640

    http://dx.doi.org/10.1007/BF02096877

\bibitem{NPPB1990} M.~Henneaux, Lectures on the antifield-BRST formalism for gauge theories, Nucl. Phys. B Proc. Suppl. \textbf{18A(1)} (1990) 47--105

    http://dx.doi.org/10.1016/0920-5632(90)90647-D

\bibitem{Princeton1992} M.~Henneaux, C.~Teitelboim, Quantization of gauge systems (Princeton: Princeton Univ. Press, 1992)

ISBN: 0691037698, 9780691037691

\bibitem{PR1995} J.~Gomis, J.~Paris, S.~Samuel, Antibracket, antifields and gauge-theory quantization, Phys. Rep. \textbf{259(1--2)} (1995) 1--145

    http://dx.doi.org/10.1016/0370-1573(94)00112-G

    arXiv:hep-th/9412228, http://arxiv.org/abs/hep-th/9412228

\bibitem{IJGMMP1995} A.~Fuster, M.~Henneaux, A.~Maas, BRST-antifield quantization: a short review, Int. J. Geom. Meth. Mod. Phys. \textbf{02(05)} (2005) 939--964

    http://dx.doi.org/10.1142/S0219887805000892

    arXiv:hep-th/0506098, http://arxiv.org/abs/hep-th/0506098

\bibitem{NPPB1997} M.~Henneaux, The role of antifields in BRST theory, Nucl. Phys. B Proc. Suppl. \textbf{57(1--3)} (1997) 131--137

    http://dx.doi.org/10.1016/S0920-5632(97)00361-7

\bibitem{AP2003} C.~Bizdadea, C.~C.~Ciob\^{\i}rc\u{a}, E.~M.~Cioroianu, S.~O.~Saliu, S.~C.~S\u{a}raru, Four-dimensional couplings among BF and matter theories from BRST cohomology, Annalen Phys. \textbf{12(9)} (2003) 543--571

    http://dx.doi.org/10.1002/andp.200310026

    arXiv:hep-th/0310243, http://arxiv.org/abs/hep-th/0310243

\bibitem{CMP1990b} M.~Henneaux, Spacetime locality of the BRST formalism, Commun. Math. Phys. \textbf{140(1)} (1991) 1--13

    http://dx.doi.org/10.1007/BF02099287

\bibitem{NPB2001} N.~Boulanger, T.~Damour, L.~Gualtieri, M.~Henneaux, Inconsistency of interacting, multi-graviton theories, Nucl. Phys. B \textbf{597(1--3)} (2001) 127--171

    http://dx.doi.org/10.1016/S0550-3213(00)00718-5

    arXiv:hep-th/0007220, http://arxiv.org/abs/hep-th/0007220

\bibitem{PRD2003} X.~Bekaert, N.~Boulanger, M.~Henneaux, Consistent deformations of dual formulations of linearized gravity: A no-go result, Phys. Rev. D \textbf{67(4)} (2003) 044010

     http://dx.doi.org/10.1103/PhysRevD.67.044010

     arXiv:hep-th/0210278, http://arxiv.org/abs/hep-th/0210278

\bibitem{IJMPA2004} C.~C.~Ciob\^{\i}rc\u{a}, E.~M.~Cioroianu, S.~O.~Saliu, Cohomological BRST aspects of the massless tensor field with the mixed symmetry $(k,k)$, Int. J. Mod. Phys. A \textbf{19(27)} (2004) 4579--4620

     http://dx.doi.org/10.1103/10.1142/S0217751X04018488

     arXiv:hep-th/0403017, http://arxiv.org/abs/hep-th/0403017

\bibitem{IJGMMP2004} C. ~Bizdadea, C.~C.~Ciob\^{\i}rc\u{a}, E.~M.~Cioroianu, S.~O.~Saliu, S.~C.~S\u{a}raru, BRST cohomological results on the massless tensor field with the mixed symmetry of the Riemann tensor, Int. J. Geom. Meth. Mod. Phys. \textbf{01(04)} (2004) 335--366

    http://dx.doi.org/10.1142/S0219887804000174

    arXiv:hep-th/0402099, http://arxiv.org/abs/hep-th/0402099

\bibitem{JHEP2006} X.~Bekaert, N.~Boulanger, S.~Cnockaert, Spin three gauge theory revisited, J. High Energy Phys. JHEP \textbf{0601} (2006) 052

     http://dx.doi.org/10.1088/1126-6708/2006/01/052

     arXiv:hep-th/0508048, http://arxiv.org/abs/hep-th/0508048

\bibitem{PRD2006} N.~Boulanger, S.~Leclercq, S.~Cnockaert, Parity-violating vertices for spin-3 gauge fields, Phys. Rev. D \textbf{73(6)} (2006) 065019

    http://dx.doi.org/10.1103/PhysRevD.73.065019

    arXiv:hep-th/0509118, http://arxiv.org/abs/hep-th/0509118

\bibitem{inprep} C.~Bizdadea, E.~M.~Cioroianu, S.~O.~Saliu, in preparation

\bibitem{PLB1991} M.~Dubois-Violette, M.~Henneaux, M.~Talon, C.~M.~Viallet, Some results on local cohomologies in field theory, Phys. Lett. B \textbf{267(1)} (1991) 81--87

    http://dx.doi.org/10.1016/0370-2693(91)90527-W

\bibitem{JPA2006} C.~Bizdadea, D.~Cornea, S.~O.~Saliu, No cross-interactions among different tensor fields with the mixed symmetry $(3,1)$ intermediated by a vector field, J. Phys. A: Math. Theor. \textbf{41(28)} (2008) 285202 (12pp)

    http://dx.doi.org/10.1088/1751-8113/41/28/285202

    arXiv:0901.4059 [hep-th], http://arxiv.org/abs/0901.4059

\bibitem{PRD2006b} C. ~Bizdadea, C.~C.~Ciob\^{\i}rc\u{a}, I.~Negru, S.~O.~Saliu, Couplings between a single massless tensor field with the mixed symmetry $(3,1)$ and one vector field, Phys. Rev. D \textbf{74(4)} (2006) 045031 (15pp)

    http://dx.doi.org/10.1103/PhysRevD.74.045031

    arXiv:0705.1048 [hep-th], http://arxiv.org/abs/0705.1048
\end{thebibliography}
\end{document}